\documentclass{article}

\usepackage{arxiv}

\usepackage[utf8]{inputenc} 
\usepackage[T1]{fontenc}    
\usepackage{hyperref}       
\usepackage{url}            
\usepackage{booktabs}       
\usepackage{amsfonts}       
\usepackage{nicefrac}       
\usepackage{microtype}      
\usepackage{lipsum,enumerate,bm,mathrsfs,yfonts}
\usepackage{graphicx,amsmath,multicol,multirow}
\usepackage{url}
\usepackage{amsthm}
\usepackage[style=apa, backend=biber, natbib=true]{biblatex}
\usepackage{adjustbox,xr,comment} 
\usepackage{pdfpages}

\usepackage{amsmath}
\usepackage{amssymb}
\usepackage{amsthm,tikz}
\usepackage{colortbl,caption,xr}
\usepackage{adjustbox}  
\usepackage{graphicx}
\usepackage{lscape}
\usepackage{rotating}
\usepackage{graphicx}
\usepackage{float}
\usepackage{url}
\usepackage{subfigure}
\usepackage{dcolumn}
\usepackage{multirow}
\usepackage{verbatim,booktabs, array,arydshln}
\usepackage{color}
\usepackage{setspace}
\usepackage{bbm}
\usepackage{booktabs}
\usepackage{hhline}
\usepackage{bm}
\usepackage{dsfont,import}
\usepackage{enumerate}
\usepackage{threeparttable}
\usepackage{array}
\usepackage{caption}

\usepackage{moreverb}
\usepackage{bm}
\usepackage{hyperref,rotating,graphicx}
\usepackage{xr}
\usepackage{verbatim}
\usepackage{url}
\theoremstyle{definition}

\usepackage{tcolorbox}
\usepackage{amsmath}
\usepackage{geometry}
\usepackage{fancyhdr}
\usepackage{placeins}
\usepackage{xcolor}
\usepackage{booktabs,tabularx,array}
\newcolumntype{Y}{>{\centering\arraybackslash}X}

\theoremstyle{definition}

\theoremstyle{remark}

\makeatletter

\title{Leveraging machine learning to estimate individualized treatment effects in cluster-randomized trials}

\author{
 Changjun Li \\
    Department of Biostatistics \\
    Yale School of Public Health \\
    New Haven, CT, USA\\
   \And
Xi Fang\\
    Department of Biostatistics \\
    Yale School of Public Health \\
    New Haven, CT, USA\\
\And
Michael O. Harhay\\
Center for Clinical Trials Innovation\\ Department of Biostatistics, Epidemiology \& Informatics\\ Perelman School of Medicine \\University of Pennsylvania\\ Philadelphia, PA, USA
\And
Andrew B. Forbes \\
Division of Quantitative Research Methodology\\ School of Public Health and Preventive Medicine\\Monash University\\ Melbourne, VIC, Australia
  \And
F. Perry Wilson\\
Section of Nephrology\\ Department of Internal Medicine\\ Yale School of Medicine\\ New Haven, CT, USA
\And
Guangyu Tong\\
Department of Biostatistics \\
    Yale School of Public Health \\
    New Haven, CT, USA\\
\And
 Fan Li \\
Department of Biostatistics \\
Yale School of Public Health \\
New Haven, CT, USA\\
  \texttt{fan.f.li@yale.edu} \\
}

\begin{document}
\maketitle
\begin{abstract}
Cluster-randomized trials (CRTs) are widely used to evaluate interventions delivered at the clinic, practice, or community level. Although standard analyses typically target average treatment effects, such summaries mask potentially meaningful variation in treatment response across individuals and clusters. This work addresses the estimation of conditional average treatment effects (CATEs) for continuous outcomes in two-arm parallel CRTs by defining causal estimands that incorporate both individual- and cluster-level baseline covariates while marginalizing over unobserved cluster heterogeneity. To estimate these quantities, we develop a unified framework based on mixed-effects machine learning, integrating and extending a range of existing approaches, including Bayesian additive regression trees with random effects, multilevel Bayesian causal forests, mixed-effects random forests, several mixed-effects gradient boosting procedures, and generalized additive mixed models, while incorporating cluster-specific random intercepts to account for within-cluster dependence. We evaluate these methods across diverse simulation scenarios and demonstrate their use in the Task Shifting and Blood Pressure Control in Ghana CRT, which investigates strategies for improving hypertension management. Drawing on these investigations, we provide practical guidance for applying mixed-effects machine learning to quantify treatment-effect heterogeneity in CRTs, together with reproducible code that enables investigators to implement all methods within a coherent workflow.
\end{abstract}


\section{Introduction}\label{sec:introduction}

Cluster-randomized trials (CRTs) represent a widely used experimental design in biomedical and public health research, wherein entire clusters such as clinics, schools, or communities are randomized to intervention arms, but outcomes are typically measured at the individual level \citep{murray1998design}. Cluster randomization is often motivated by logistical or ethical considerations, or by the need to minimize treatment contamination. The hierarchical structure inherent to CRTs induces within-cluster correlation and between-cluster heterogeneity, leading to nontrivial implications for both the definition and the estimation of causal estimands \citep{eldridge2012practical,kahan2023estimands} . Traditional CRT analyses typically target the average treatment effect under cluster-level randomization, often using cluster-level summaries or model-based estimators \citep{li2025model,wang2024model}. Despite the advancement of the estimation methods for average treatment effect, a sole focus on population averages may obscure important variation in causal effects across individual participants. The resulting heterogeneity, if appropriately quantified, often carries important evidence base for precision medicine and underlies the individualized treatment rules. Recent developments in causal inference have thus emphasized the estimation of heterogeneous treatment effects (HTEs) or conditional average treatment effects (CATEs), defined as the expected contrast in potential outcomes given measured covariates \citep{wager2018estimation}. HTEs extend beyond global comparisons to characterize the functional relationship between treatment effects and baseline covariates, thereby informing individualized decision making and adaptive intervention strategies \citep{kent2018personalized}. However, the hierarchical dependence structure and cluster-level randomization in CRTs pose unique challenges for HTE estimation. The vast number of existing causal learners have been developed based on independent data and do not properly address clustering in the estimation and inference, and hence may not be directly applicable to the CRT context \citep{powers2018some,anoke2019approaches}. On the other hand, conventional multilevel models that have been routinely used in CRTs impose restrictive parametric forms that limit flexibility in capturing nonlinear or high-dimensional effect modification.  


In contemporary CRTs, investigators routinely collect a rich set of baseline features at both the individual and cluster level, and the potential effect modification may be nonlinear in nature and possibly driven by higher-order interactions. These features make parametric multilevel models less attractive for individualized treatment effect estimation, because explicitly specifying complex treatment by covariate interactions quickly becomes infeasible and risks substantial misspecification bias. This has motivated the development of flexible machine-learning approaches for estimating heterogeneous treatment effects in the multilevel data settings, where the regression function is learned in a data-adaptive manner while accounting for within-cluster dependence. Under the frequentist paradigm, generalized additive mixed models offer a structured compromise by combining smooth functions with cluster-level random effects and allowing treatment-by-covariate interactions through by-treatment smooth terms \citep{lin1999inference,cho2022modelling,cho2024modeling}. Traditional ensemble methods, such as random forests and gradient boosting, have also been extended to hierarchical data settings by incorporating cluster-specific random effects. Mixed-effects random forests embed a standard random forest algorithm within an expectation-maximization (EM) framework to account for cluster-level random effects \citep{hajjem2011mixed,hajjem2014mixed,capitaine2021random}. For gradient boosting, several extensions have been proposed for multilevel data, using different strategies to incorporate grouped random effects and to model the associated covariance structure while growing trees sequentially \citep{sigrist2022gaussian,olaniran2025mixed,knieper2025gradient}.

In parallel, Bayesian tree ensembles provide a flexible alternative for modeling complex response surfaces. Among them, the Bayesian additive regression trees (BART) and their mixed-effects extensions incorporate cluster-level random effects to capture residual heterogeneity while retaining nonparametric flexibility in the fixed-effects components, while the Bayesian causal forest further separates prognostic and treatment-effect components to regularize heterogeneous effect estimation \citep{chipman2010bart,spanbauer2021nonparametric,dorie2022stan,hahn2020bayesian,yeager2022synergistic}. From a causal machine learning perspective, many of these mixed-effects learners are naturally deployed within an S-learner framework, in which a single predictive model for the outcome includes treatment as one of the inputs and heterogeneous effects are obtained by contrasting model predictions under different treatment assignments. In contrast, R-learner and related orthogonalized approaches separate baseline outcome modeling from treatment-effect modeling, typically using residualization or doubly robust constructions to target heterogeneity more directly, although these approaches have not been generalized to accommodate clustered data \citep{nie2021quasi,chernozhukov2018double}. Finally, the causal forests have been developed and implemented as a mainstream causal machine learning method, combining honesty, local weighting, and orthogonalization to estimate heterogeneous effects while reducing bias from adaptive partitioning; in standard software, they have been adapted to clustered data through cluster-level subsampling and appropriate variance adjustments \citep{wager2018estimation,athey2019estimating}.

Despite these seemingly parallel methodological advancements, few prior studies have systematically compared how different flexible machine learning methods perform for HTE estimation in CRTs, where the clustered data structure, within-cluster dependence, and multilevel covariates create challenges that are not present in individually randomized settings. This paper addresses that gap by reporting on a comprehensive, simulation-based evaluation of off-the-shelf mixed-effects machine learning methods for HTE estimation in CRTs. Specifically, we compare frequentist and Bayesian approaches across a range of scenarios varying in intracluster correlation, sample size, and level of treatment effect heterogeneity, with a focus on their estimation accuracy, inference validity, and overall performance in the presence of clustering. 
We then present a case study that demonstrates how the potential insights can be gained through using causal machine learning methods to explore effect heterogeneity in CRTs. Our central objective is to offer some empirical guidance for estimating treatment effect heterogeneity in increasingly common CRT designs and to identify methods that perform reliably in the case of a continuous outcome.

The remainder of this paper is organized as follows. In Section 2, we introduce the notation and define the target estimands. We then describe the ten machine learning methods that have been implemented in R and will be compared, along with some implementation details. Section 3 reports a simulation study that evaluates these methods across diverse scenarios. Section 4 provides sensitivity analyses to assess the robustness of the findings. Section 5 illustrates an application by estimating heterogeneous treatment effects in the TASSH real data. Section 6 concludes with a discussion of the main results, practical implications, limitations, and directions for future research.

\section{Statistical Methods} 
\subsection{Notation and estimands}

We consider a two-arm parallel group cluster-randomized trial with $I$ clusters, where cluster $i$ contains $N_i$ individuals indexed by $j=1,\ldots,N_i$. Let $A_i \in \{0,1\}$ denote the treatment assignment for cluster $i$, with $A_i=1$ if cluster $i$ is assigned to the intervention and $A_i=0$ otherwise. Under the potential outcomes framework, let $Y_{ij}(a)$ denote the outcome that would be observed for individual $j$ in cluster $i$ if, possibly contrary to fact, cluster $i$ were assigned to treatment level $a \in \{0,1\}$. The pair $\{Y_{ij}(1), Y_{ij}(0)\}$ represents the individual-level potential outcomes under intervention and control, respectively.
In a given study, it is impossible to fully observe the pair $\{Y_{ij}(1), Y_{ij}(0)\}$ for each individual. We describe two typical assumptions that are plausible in CRTs for identification. First, under the cluster-level Stable Unit Treatment Value Assumption (SUTVA), the observed outcome for individual $j = 1,\dots, N_i$ in cluster $i=1,\dots,I$ can be expressed as $Y_{ij} = A_i Y_{ij}(1) + (1 - A_i) Y_{ij}(0)$. Second, we assume a cluster-level randomization mechanism such that $A_i \perp \{\bm{Y}_i(1), \bm{Y}_i(0)\} \mid (\bm{X}_i,\bm{V}_i)$, where $\bm{Y}_i(a) = (Y_{i1}(a), \ldots, Y_{iN_i}(a))^\top$ is the vector of potential outcomes in cluster $i$ under treatment $a$, $\bm{X}_i = (\bm{X}_{i1}, \ldots, \bm{X}_{iN_i})^\top$ denotes the collection of individual baseline covariates, and \(\bm{V}_i\) denotes the cluster-level covariates in cluster $i$. Cluster-level covariates $\bm{V}_i$ may include the geographical region of the cluster or the cluster-population size, whereas individual-level covariates $\bm{X}_{ij}$ often include demographics and baseline clinical characteristics. Although the above cluster randomization assumptions are described more generally by conditioning on $\bm{X}_i$ and \(\bm{V}_i\) to allow for stratified randomization, in most cases we have marginal cluster randomization such that $A_i \perp \{\bm{Y}_i(1), \bm{Y}_i(0)\}$. We will focus on the marginal cluster randomization in our subsequent development.

To describe the causal estimand of interest, we further assume that given baseline covariates $\bm{X}_i$ and \(\bm{V}_i\), the distribution of $\{Y_{ij}(1), Y_{ij}(0)\}$ is identical across individuals within each cluster; hence this can be viewed as the \emph{restricted} informative cluster size assumption as $\bm{V}_i$ includes cluster size $N_i$. This then permits us to define the CATE in CRTs as 
\begin{equation}\label{eq:CATE}
\tau(\bm{x},\bm{v}) = \mathbb{E}[Y_{ij}(1) - Y_{ij}(0) \mid \bm{X}_{ij} = \bm{x},\bm{V}_i=\bm{v}].
\end{equation}
Several remarks follow from expression \eqref{eq:CATE}. First, analogous to the individually randomized trial setting, $\tau(\bm{x},\bm{v})$ describes how treatment effects vary across individuals with baseline covariates. However, unlike in the individually randomized setting, $\tau(\bm{x},\bm{v})$ in \eqref{eq:CATE} is defined with respect to both individual-level and cluster-level characteristics, and is more precisely interpreted as the CATE for individuals with covariates $\bm{X}_{ij} = \bm{x}$ in clusters characterized by $\bm{V}_i = \bm{v}$. Second, the estimand in \eqref{eq:CATE} is both conditional and marginal. It is conditional on the observed individual- and cluster-level covariates, yet marginal with respect to any unmeasured cluster-level heterogeneity, such as latent random effects that are routinely used to conceptualize the distribution of clustered observations. This marginalization is deliberate and practical, as $\tau(\bm{x},\bm{v})$ is often used as the basis to guide subgroup identification or the development of individualized treatment rules---tasks that rely on measured baseline information rather than unmeasured latent variables. Thus, although machine learning models should ideally account for unmeasured heterogeneity (to more precisely approximate the unknown data-generating process), the CATE estimand must remain defined in terms of observed covariates to yield actionable insights in the context of CRTs. Third, although \eqref{eq:CATE} is defined on the difference scale, the framework is general and can be extended to other scales, such as the risk ratio or odds ratio. In this paper, we focus on estimating CATE on the difference scale, leaving ratio estimands to future work. We next describe a set of machine learning models that may be used to quantify the CATE estimand in CRTs.

\subsection{Using machine learning tools to quantify CATE in CRTs} \label{sec:ml_method}

\subsubsection{Mixed-Effects BART}

Bayesian Additive Regression Trees (BART) \citep{chipman2010bart} is a Bayesian nonparametric ensemble method that expresses an unknown regression function as a sum over multiple shallow binary trees. Let \( f(\bm{X}_{ij}, \bm{V}_i, A_i) \) denote the regression surface for the observed outcome \( Y_{ij} \). In BART, this surface is modeled as a sum of \( K \) tree-structured functions,
\begin{equation*} 
f_{\mathrm{BART}}(\bm{X}_{ij}, \bm{V}_i, A_i) = \sum_{k=1}^K g_k(\bm{X}_{ij}, \bm{V}_i, A_i; \mathcal T_k, \mathcal M_k),
\end{equation*}
where each \( g_k \) is a decision tree defined by structure \( \mathcal T_k \) (i.e., internal splitting rules) and a set of terminal node values \( \mathcal M_k \). Each tree partitions the covariate space into disjoint regions and assigns a constant value to each region. The additive sum across trees provides a flexible, nonlinear approximation to the regression function. BART imposes priors that regularize the ensemble by shrinking individual tree contributions toward zero and favoring shallow trees, yielding global shrinkage with local adaptivity. The model is fitted using a Bayesian backfitting Markov chain Monte Carlo (MCMC) algorithm, which iteratively samples each tree conditional on the rest.

To account for within-cluster correlations in CRTs, a convenient off-the-shelf extension is the random-intercept BART (riBART), which introduces a cluster-specific random intercept to capture the cluster structure. More specifically, the model is defined as
\begin{equation} \label{eq:mebart}
Y_{ij} = f_{\mathrm{BART}}(\bm{X}_{ij}, \bm{V}_i, A_i) + b_i + \varepsilon_{ij}, \quad b_i \sim \mathcal{N}(0, \sigma_b^2), \quad \varepsilon_{ij} \sim \mathcal{N}(0, \sigma^2),
\end{equation}
where \( b_i \) captures residual intra-cluster correlation not explained by the cluster-level covariates \( \bm{V}_i \), and \( \varepsilon_{ij} \) denotes an individual-level error term. Under this model, the CATE for individuals with covariates \( \bm{X}_{ij} = \bm{x} \) and \( \bm{V}_i = \bm{v} \) is estimated by contrasting posterior draws of the regression surface under treatment and control:
\begin{equation} \label{eq:post_mean}
  \widehat{\tau}(\bm{x}_{ij}, \bm{v}_i) = \frac{1}{S} \sum_{s=1}^S \left\{ \widehat f^{(s)}(\bm{x}_{ij}, \bm{v}_i, 1) - \widehat f^{(s)}(\bm{x}_{ij}, \bm{v}_i, 0) \right\},  
\end{equation}
where \( \widehat f^{(s)} \) is the \( s \)-th posterior draw for \(s=1,\dots, S\) of the BART surface from the fitted riBART model. These draws marginalize over posterior samples of the trees and random intercepts \( b_i \), providing a flexible estimate of treatment effect conditional on observed covariates. This method can be implemented as \texttt{rbart\_vi} in the \texttt{dbarts} R package \citep{tan2016predicting}. It augments the BART mean with a cluster-specific Gaussian intercept that is updated by conjugate Gaussian steps while the trees are fitted via the standard Bayesian backfitting sampler. 

In addition to this implementation, since model \eqref{eq:mebart} is a special case of the mixed-effect BART framework with random effects restricted to an intercept term, it can also be implemented using the general mixed BART package. \citet{spanbauer2021nonparametric} developed a mixed-effect BART (mixedBART) framework that extends standard BART to account for hierarchical structure through random intercept and slopes. 
The treatment variable and all covariates are incorporated into the BART surface, and inference is performed using a Gibbs sampler that alternates between conjugate updates for \( b_i \) and Metropolis-Hastings updates for the tree structures and node parameters. This enables flexible borrowing of information across clusters while capturing both within- and between-cluster heterogeneity. This method is implemented in the R package \texttt{mxbart} which builds on the \texttt{dbarts} engine and supports both clustered or repeated-measures data as well as simpler intercept-only specifications for CRTs.
In addition to this, the stan4bart (R package \texttt{stan4bart}) framework developed by \citet{dorie2022stan} also extends BART to include random effects, but does so through a modular specification, combining a nonparametric BART component with a parametric multilevel block (fixed and random effects with variance components) estimated in Stan using Hamiltonian Monte Carlo. The resulting model is essentially the same form as in Spanbauer and Sparapani \citep{spanbauer2021nonparametric}. The difference is that rather than placing the mixed-effects updates inside the tree backfitting loop as in mixedBART \citep{spanbauer2021nonparametric}, stan4bart alternates BART updates with Stan’s Hamiltonian Monte Carlo for the hierarchical block. Thus, both approaches integrate BART with random effects, but they differ in posterior computation.

\subsubsection{Multilevel Bayesian Causal Forest}
In mixed-effects BART, the treatment indicator enters the BART surface in the same way as the other predictors, together with the cluster-level random intercept. Bayesian causal forest (BCF) takes the same ingredients but models the response surface as the sum of a prognostic function and a treatment-effect function, learned by separate tree ensembles with separate regularization to stabilize heterogeneous effect estimation \citep{hahn2020bayesian}. The baseline BCF takes the form

\begin{equation*}
Y_{ij}= \mu\left\{\bm x_{ij} ,\bm v_i, \widehat{\pi}(\bm x_{ij},\bm v_i)\right\}+\tau(\bm x_{ij},\bm v_i)A_i + \varepsilon_{ij}.  \quad \varepsilon_{ij} \sim \mathcal{N}(0, \sigma^2),
\end{equation*}

Here $\mu(\cdot)$ denotes the prognostic component that captures baseline outcome variation under control, and $\tau(\cdot)$ denotes the conditional treatment effect component. 
The term $\pi_{ij}$ represents the treatment assignment probability included in the prognostic component, following the usual BCF parameterization. In our CRT setting with equal-probability randomization, this quantity is known and constant, so we set $\pi_{ij}=0.5$ for all individuals rather than estimating a propensity score. Therefore, this term does not contribute additional subject-specific information in our implementation and serves only as a constant offset. In designs with unequal allocation or stratified randomization, the known overall assignment probability or the stratum-specific assignment probability could be used analogously.

To accommodate intra-cluster correlation in CRTs, the model augments the same cluster-specific random intercept as \eqref{eq:mebart}, yielding the multilevel BCF (MBCF):
\begin{equation} \label{eq:mbcf}
Y_{ij}=\mu\left\{\bm x_{ij},\bm v_i,\widehat{\pi}(\bm x_{ij},\bm v_i) \right\}+\tau(\bm x_{ij},\bm{v}_i)A_{i}+b_i+\varepsilon_{ij},  \quad b_i \sim \mathcal{N}(0, \sigma_b^2), \quad \varepsilon_{ij} \sim \mathcal{N}(0, \sigma^2),
\end{equation}
where the random intercept $b_i$ captures unmeasured cluster-level heterogeneity that induces within-cluster dependence while leaving the estimand unchanged. Since $\tau(\bm x_{ij},\bm v_i)$ conditions only on observed baseline covariates and is marginal with respect to \(b_i\), it coincides with the CATE defined in \eqref{eq:CATE}. Both $\mu(\cdot)$ and $\tau(\cdot)$ are represented by two BART-type ensembles with distinct priors, typically a richer, and more weakly regularized forest for \(\mu(.)\), and a smaller, more strongly regularized forest for $\tau(\cdot)$, so that effect heterogeneity is expressed only when supported by the data, and the random intercept \(b_i\) can be updated by conjugate Gaussian steps exactly as those in BART with random intercept implementation. In practice, at each MCMC iteration, the model gives a draw of the \(\tau\) forest at \((\bm x_{ij}, \bm v_i)\) and averaging these draws over iterations gives a posterior estimator \(\widehat{\tau}(\bm x_{ij},\bm{v}_i)\) as \eqref{eq:post_mean}. This can be implemented in the \texttt{multibart} R package \citep{yeager2022synergistic} .

\subsubsection{Causal Forest}

Causal forests estimate heterogeneous treatment effects by learning an adaptive notion of local neighborhoods in covariate space and then comparing treated and control outcomes within those neighborhoods. A central design used in many causal forest implementations is ``honesty,'' meaning that the data used to choose splits are separated from the data used to estimate within-leaf quantities; for example, Wager and Athey's ``double-sample'' construction uses one subsample to select splits and a held-out subsample to estimate leaf-level effects, which reduces bias and supports valid large-sample inference \citep{wager2018estimation}. For covariate-adjusted and observational studies, causal forests are naturally motivated by the Robinson residualization \citep{robinson1988root} and orthogonal (Neyman-orthogonal) scores. Specifically, define nuisance functions
$m(\bm x,\bm v)=\mathbb{E}(Y_{ij}\mid \bm X_{ij}=\bm x,\bm V_i=\bm v)$ and $\pi(\bm x,\bm v)=\mathbb{P}(A_i=1\mid \bm X_{ij}=\bm x,\bm V_i=\bm v)$, and form the residualized outcome and treatment, $Y_{ij}-m(\bm X_{ij},\bm V_i)$ and $A_i-\pi(\bm X_{ij},\bm V_i)$, respectively. Then the CATE $\tau(\bm x,\bm v)$ can be characterized as the local solution to the conditional orthogonal moment equation \citep{athey2019estimating}
\begin{equation}
\mathbb{E}\!\left[
\{A_i-\pi(\bm X_{ij},\bm V_i)\}
\Big\{Y_{ij}-m(\bm X_{ij},\bm V_i)-\tau(\bm x,\bm v)\{A_i-\pi(\bm X_{ij},\bm V_i)\}\Big\}
\,\middle|\,
\bm X_{ij}=\bm x,\bm V_i=\bm v
\right]=0,
\end{equation}
so that $\tau(\bm x,\bm v)$ is the best local coefficient relating the residualized outcome to the residualized treatment. This same orthogonalization yields the R-learner viewpoint of \citet{nie2021quasi}, which defines $\tau(\cdot)$ as the minimizer of a residualized squared-error criterion (with regularization), and provides a principled target for tuning forests specifically for treatment effect heterogeneity rather than for outcome prediction. In the cluster-randomized trials considered here, treatment is randomized with equal allocation at the cluster level, so this probability is known by design and fixed at $\pi(\bm x,\bm v)\equiv 0.5$.

The generalized random forests (GRF) framework was further developed to formalize the view of forests as adaptive kernel methods \citep{athey2019generalized}, under which a fitted forest induces similarity weights $\alpha_\ell(\bm x,\bm v)$ that define the learned neighborhood around $(\bm x,\bm v)$. For each tree, the weight assigned to training unit $\ell$ is proportional to the indicator that $\ell$ falls in the same leaf as $(\bm x,\bm v)$ (typically scaled by the leaf size), and $\alpha_\ell(\bm x,\bm v)$ is obtained by averaging these tree-wise weights over trees, yielding nonnegative weights that sum to one. These weights are then used to solve a locally weighted version of the residual-on-residual estimating equation, typically with out-of-bag (cross-fitted) estimates of $m(\cdot)$ and $\pi(\cdot)$, producing an estimator of $\tau(\bm x,\bm v)$ and associated standard errors and confidence intervals via GRF’s asymptotic normal approximation and variance estimation machinery. In the general GRF template, the target may be any local parameter $\theta(\bm x,\bm v)$ defined by a conditional moment restriction $\mathbb{E}\!\left[\psi_{\theta(\bm x,\bm v),\nu(\bm x,\bm v)}(O)\mid \bm X=\bm x,\bm V=\bm v\right]=0$,
where $\nu(\bm x,\bm v)$ denotes additional nuisance or auxiliary parameters required to specify and solve the local moment equation (e.g., local intercept or other components estimated jointly with $\theta$ within each neighborhood). Estimation proceeds by using the honest forest weights $\alpha_\ell(\bm x,\bm v)$ to form and solve a weighted empirical analogue of this moment restriction. Causal forests arise as a special case with $\theta=\tau$ and $\psi$ chosen as an orthogonalized treatment-effect score based on residualization. These methods are implemented in R in the package \texttt{grf} \citep{athey2019generalized}, which accommodates within-cluster dependence by subsampling at the cluster level (i.e., sampling clusters first, then sampling observations within sampled clusters) and by defining out-of-bag predictions at the cluster level (i.e., an observation is out-of-bag only if its entire cluster was not sampled), aligning resampling, nuisance prediction, and uncertainty quantification with the dependence structure induced by clustering.

\subsubsection{Mixed-Effects Random Forest}

As a frequentist counterpart to Bayesian tree ensembles, the mixed-effects random forest extends the random forest framework to clustered data by combining a flexible nonparametric fixed-effects surface with a cluster-specific random intercept. The model is specified as
\begin{equation}\label{eq:merf-crt}
Y_{ij}
\;=\;
f_{\mathrm{RF}}(\bm{X}_{ij}, \bm{V}_i, A_i)
\;+\;
b_i
\;+\;
\varepsilon_{ij},
\qquad
b_i \sim \mathcal{N}(0,\sigma_b^2),\ \ 
\varepsilon_{ij}\sim \mathcal{N}(0,\sigma^2),
\end{equation}
where \(f_{\mathrm{RF}}(.)\) denotes the forest predictor trained on \((\bm{X}_{ij},\bm{V}_i, A_i)\), and \(b_i\) captures unmeasured cluster-level heterogeneity \citep{capitaine2021random,hajjem2011mixed}. The fixed-effects surface inherits the forest mechanism, which is an aggregation of many decorrelated trees grown on bootstrap resamples with feature subsampling so that averaging over trees stabilizes prediction while accommodating nonlinearities and interactions without parametric specification \citep{breiman2001random}. Estimation in MERF proceeds through an alternating iterative algorithm:
(i) initial values of the random intercepts $(b_i)$ are assigned, typically set to zero; (ii) the forest predictor $f_{\mathrm{RF}}(.)$ is fitted to the residualized outcomes $Y_{ij}-b_i$ that have been adjusted for these current intercepts; (iii) the random intercepts and variance components $(b_i,\sigma_b^2,\sigma^2)$ are updated using best linear unbiased prediction (BLUP) and EM (BLUP-EM) algorithm based on the residuals obtained from the previous forest fit, $Y_{ij}-f_{\mathrm{RF}}(\bm{X}_{ij}, \bm{V}_i, A_i)$ \citep{hajjem2011mixed, sela2012re} . 
Heterogeneous treatment effects by contrasting predictions under the two treatment assignments can be estimated as 
\[
\widehat{\tau}(\bm{x},\bm{v}) \;=\; \widehat f_{\mathrm{RF}}(\bm{x},\bm{v},1) - \widehat f_{\mathrm{RF}}(\bm{x},\bm{v},0),
\]
which is conditional on observed baseline covariates and is marginal with respect to $b_i$. This method can be implemented via the \texttt{LongituRF} R package.

\subsubsection{Mixed-Effects Gradient Boosting} \label{met:boost}
Unlike mixed-effects random forests, which estimate the fixed-effects surface by averaging many trees grown in parallel on bootstrap samples, boosting constructs an additive predictor sequentially. Each shallow tree is fit to the current residuals or to a second-order approximation, contributing a small corrective step to the fitted function. This stage-wise construction provides fine control of model complexity through shrinkage and early stopping, while flexibly capturing nonlinearities and interactions. For CRTs, we consider the mixed-effects boosting model
\begin{equation}\label{eq_boosting}
Y_{ij}=f_{\mathrm{GB}}(\bm X_{ij},\bm V_i,A_i)+b_i+\varepsilon_{ij},\qquad 
b_i\sim\mathcal N(0,\sigma_b^2),\ \ \varepsilon_{ij}\sim\mathcal N(0,\sigma^2),
\end{equation}
where the working covariance for cluster \(i\) is \(\bm\Omega_i=\sigma_b^2\,\bm J_{N_i}+\sigma^2\,\bm I_{N_i}\), with \(\bm J_{N_i}\) denoting the \(N_i\times N_i\) all-ones matrix and \(\bm I_{N_i}\) the identity. Estimation of boosting algorithm proceeds through a blockwise alternating scheme at each iteration \(m\), consisting of two coordinated updates: (i) a boosting step, which refines the fixed-effects surface by adding a weak learner (e.g., decision tree) 
\begin{equation} \label{eq:boosting_fixed}   
f_{\mathrm{GB}}^{(m)} = f_{\mathrm{GB}}^{(m-1)} + \nu h_m,
\end{equation}
where \(h_m\) is a shallow regression tree at \(m\)-th iteration and \(0<\nu\le1\) is the learning rate controlling the contribution of the new tree; and (ii) a variance step, which updates the random-effects estimates and variance components \((\sigma_b^2,\sigma^2)\). 
The variance parameters are typically obtained by minimizing a function related to the negative log-likelihood
\[
\mathcal L_{\mathrm{RE}} (f_{\mathrm{GB}},\sigma_b^2,\sigma^2) =\frac{1}{2}\sum_{i=1}^I \left\{
\log|\bm\Omega_i| +(\bm y_i-\bm f_i)^{\top}\bm\Omega_i^{-1}(\bm y_i-\bm f_i)\right\},\]
where \(\bm y_i=(Y_{i1},\ldots,Y_{iN_i})^\top\) and  \(\bm f_i=\left\{f_{\mathrm{GB}}(X_{i1},V_i,A_i),\ldots,f_{\mathrm{GB}}(X_{iN_i},V_i,A_i)\right\}^\top\). Three methods have been proposed regarding how these two steps are implemented.

Gaussian process boosting (GPBoost) \citep{sigrist2022gaussian} treats estimation as a joint problem by alternating coordinate-wise updates of the covariance and the fixed-effects surface under a single working objective. At iteration \(m\), GPBoost proceeds in two steps: (i) it updates the covariance parameters by
\((\widehat\sigma_b^{2\,(m)},\,\widehat\sigma^{2\,(m)})=\arg\min_{\sigma_b^2,\sigma^2}\ \mathcal L_{\mathrm{RE}}(f_{\mathrm{GB}}^{(m-1)},\,\sigma_b^2,\,\sigma^2)\),
and (ii) then updates the fixed part by adding a weak learner as in \eqref{eq:boosting_fixed}. In this second step, the tree is grown under the updated covariance with candidate splits, which are scored by their GLS reduction of the objective, and leaf values are computed using either a GLS gradient update, a Newton-GLS update, or a hybrid variant which combines gradient-based split selection with Newton leaf values. This can be implemented through the R package \texttt{gpboost} \citep{sigrist2022gaussian}.

Mixed-effects gradient boosting (MEGB) \citep{olaniran2025mixed} adopts a fixed-first least-squares update that learns each new tree from pseudo-outcomes. At iteration \(m\), a pseudo-response is formed by removing the current cluster intercept, \(Y_{ij}^{*(m)} = Y_{ij} - \widehat b_i^{(m-1)}\), and the pseudo-outcome to be learned is the prediction error on this pseudo-response, \(r_{ij}^{(m)} = Y_{ij}^{*(m)} - f_{\mathrm{GB}}^{(m-1)}(\bm X_{ij}, \bm V_i, A_i)\). A shallow regression tree \(h_m\) is then fitted by ordinary least squares to minimize \(\sum_{i=1}^{I}\sum_{j=1}^{N_i} \left\{ r_{ij}^{(m)} - h_m(\bm X_{ij}, \bm V_i, A_i) \right\}^2\), and the fixed-effects surface is updated as in equation \eqref{eq:boosting_fixed}. With the fixed part refined, cluster intercepts are recalculated via BLUP shrinkage, namely, shrinking each cluster's mean residual toward zero with weight \(\sigma_b^2 / (\sigma_b^2 + \sigma^2 / N_i)\), and the variance components \(\sigma_b^2\) and \(\sigma^2\) are refreshed through an EM update for numerical stability. Conceptually, these two steps alternate descent on the same Gaussian working objective \(\mathcal L_{\mathrm{RE}}(f_{\mathrm{GB}}, \sigma_b^2, \sigma^2)\). The tree step improves model fit by refining \(f_{\mathrm{GB}}\), while the BLUP-EM step reduces the same objective by updating the random effects and variance components. This method has been implemented in the R package \texttt{MEGB} \citep{olaniran2025mixed}.

Building on a similar alternating scheme, the mermboost method \citep{knieper2025gradient} treats the CRT model as a generalized additive mixed model and implements component-wise gradient boosting for the fixed effects while refitting the entire random-effects block at each iteration. The fixed part is represented as a sum of base learners \(f_{\mathrm{GB}}=\sum_{r=1}^{R} f_r\), with each learner equipped with its own penalty to control complexity. At boosting step \(s\), every candidate base learner \(h_r\) is fit to the negative gradient of the penalized objective
\(\mathcal{L}_{\mathrm{pen}}=\mathcal{L}_{\mathrm{RE}}(f_{\mathrm{GB}},\sigma_b^2,\sigma^2) +\sum_{r=1}^{R}\lambda_r\,\mathcal{J}_r(f_r)\) with respect to the current linear predictor \(f_{\mathrm{GB}}^{(m-1)}(\bm X_{ij},\bm V_i,A_i)+b_i\). Here \(\mathcal{J}_r(f_r)\) denotes the learner-specific penalty with tuning parameter \(\lambda_r\). The learner that reduces \(\mathcal{L}_{\mathrm{pen}}\) the most is selected, and the fixed predictor is updated according to equation \eqref{eq:boosting_fixed}. Next, given \(f_{\mathrm{GB}}^{(m)}\), the random effects \(b_i\) and variance components \((\sigma_b^2,\sigma^2)\) are re-estimated by minimizing \(\mathcal{L}_{\mathrm{RE}}\) via a Laplace (REML-style) step. An implementation of this method is available in the R package \texttt{mermboost}.

Within the present framework, each method can be used as an S-learner for heterogeneous effects. The model takes $(\bm{X}_{ij}, \bm{V}_i, A_i)$ as inputs to produce a prediction function $f_{\mathrm{GB}}(\bm{x}, \bm{v}, a)$, while the dependence structure is handled internally by the corresponding mixed-effects machinery. The CATE is then estimated by
\[
\widehat{\tau}(\bm{x}, \bm{v}) \;=\; \widehat f_{\mathrm{GB}}(\bm{x}, \bm{v}, 1) \;-\; \widehat f_{\mathrm{GB}}(\bm{x}, \bm{v}, 0),
\]
obtained as the difference of two predictions for the same covariate profile under the two treatment levels. 

These three methods target the same estimand, while differing in the parameterization of dependence, in the alternation between boosting and variance-component updates, and in mechanisms that separate fixed and random contributions.

\subsubsection{Generalized Additive Mixed Models}

Finally, as a semiparametric counterpart to tree-based methods, the generalized additive mixed model (GAMM) represents the fixed-effect surface as a structured sum of smooth functions of covariates, with random effects to capture cluster dependence. For CRTs with continuous outcomes, the GAMM with nonlinear treatment-by-covariate interactions can be written as 
\begin{equation}\label{eq:gamm-crt}
Y_{ij} = \beta_0 + A_i\,\beta_A + \sum_{r=1}^R f_r\!\left(\bm X_{r,ij},\bm V_{r,i}\right) + A_i \sum_{r=1}^R f^{\tau}_r\!\left(\bm X_{r,ij},\bm V_{r,i}\right)+b_i+\varepsilon_{ij},
\end{equation}
with $b_i\sim\mathcal{N}(0,\sigma_b^2)$ and $\varepsilon_{ij}\sim\mathcal{N}(0,\sigma^2)$. The function $f_r(\cdot)$ denotes prognostic smooth functions common to both treatment arms, and the ``by-treatment'' smooths $f_r^{\tau}(\cdot)$ allow the treatment effect to vary flexibly with prognostic characteristics. The vector $b_i$ and \(\varepsilon_{ij}\) introduce the same working covariance \(\bm{\Omega}_i\) for cluster \(i\) as in Section \ref{met:boost}.
Each smooth term is represented by a spline basis function. For a univariate \(x\) entering the $r$-th smooth term, the prognostic and by-treatment components are written as  $f_r(x)=\bm B_r(x)^\top\bm \theta_r$ and $f_r^{\tau}(x)=\bm B_r(x)^\top\bm \theta_r^{\tau}$, 
where \(\bm{B}_r(x)\) is the vector of predetermined basis functions (e.g., B-splines or P-splines) and \(\bm{\theta}_r\), \(\bm{\theta}_r^{\tau}\) are the associated basis coefficients. Smoothness is controlled by quadratic penalties  $ \lambda_r\,\bm \theta_r^\top \bm S_r\bm \theta_r$ and $\lambda_r^{\tau}\,{\bm\theta_r^{\tau}}^\top \bm S_r \bm \theta_r^{\tau}$, with penalty matrix \(\bm S_r\) and smoothing parameters \(\lambda_r\) and \(\lambda_r^{\tau}\). Under the mixed model representation, the penalized spline coefficients are treated as Gaussian random effects whose covariance depends on the smoothing parameters, and estimation proceeds under a Gaussian working model by restricted maximum likelihood, jointly estimating the fixed effects \(\beta_0\) and \(\beta_A\), the random intercept \(b_i\), and spline coefficients, along with variance components \(\sigma_b^2,\sigma^2\) and the smoothing parameters. 

Let $\widehat\eta(\bm{x},\bm{v},a)$ denote the fitted linear predictor from equation \eqref{eq:gamm-crt} evaluated at covariate profile $(\bm{x},\bm{v})$ and treatment level $a\in\{0,1\}$.  For continuous outcomes, 
the CATE can be estimated as
\begin{equation*}
\widehat{\tau}(\bm{x},\bm{v}) = \widehat \eta(\bm{x},\bm{v},1)-\widehat\eta(\bm{x},\bm{v},0).
\end{equation*}
This simplifies to 
\(\widehat\tau(\bm{x},\bm{v}) = \widehat\beta_A + \sum_{r=1}^R \widehat f_r^{\tau}\!\big(x_r,v_r\big)\) under model \eqref{eq:gamm-crt}.
The resulting treatment effect surface is a smooth, potentially non-linear function of the baseline covariates and marginal with respect to unobserved cluster effects. This method can be implemented via the \texttt{mgcv} R package \citep{wood2011fast}.

\section{Simulation Studies}\label{sec:simulation}

\subsection{Simulation Design}\label{sec:simulation design}
We conducted a simulation study to evaluate the performance of the machine learning methods discussed in Section \ref{sec:ml_method} for estimating the CATE in CRTs with continuous outcomes. We simulated data with \( I \in\{ 10, 30, 100\} \) clusters, fixing the total sample size \(I\cdot\mathbb{E}[N_i] = 3000\), where \(N_i\) follows a discrete uniform distribution. Clusters were randomized to treatment or control with a 1:1 allocation ratio. Two cluster-level covariates were generated and denoted as \( \bm{V}_i = (V_{i1}, V_{i2}) \), where \( V_{i1} \sim \mathcal{N}(0,1) \) and \( V_{i2} \sim \text{Bernoulli}(0.5) \). Within each cluster, eight individual-level covariates were generated and denoted as  \( \bm{X}_{ij} = (X_{ij1}, \ldots, X_{ij8}) \), with \( X_{ij1}, \ldots, X_{ij5} \sim \mathcal{N}(0,1) \) and  \( X_{ij6}, X_{ij7}, X_{ij8} \sim \text{Bernoulli}(0.5) \). The two potential outcomes were simulated from 
\begin{align*}
Y_{ij}(0) &= f_0(\bm{X}_{ij}, \bm{V}_i) + b_i + \epsilon_{ij}, \\
Y_{ij}(1) &= f_0(\bm{X}_{ij}, \bm{V}_i) + \tau(\bm{X}_{ij}, \bm{V}_i) + b_i + \epsilon_{ij},
\end{align*}
where \( f_0(\cdot) \) is a baseline outcome function, \( \tau(\cdot) \) is the conditional treatment effect, \( b_i \sim \mathcal{N}(0, \sigma_b^2) \) is a cluster-level random effect, and \( \epsilon_{ij} \sim \mathcal{N}(0, \sigma^2) \) is an individual-level error. We fixed the total variance at 1 and 
set \( \sigma_b^2 = 0.1 \) and \( \sigma^2 = 0.9 \) yielding an intraclass correlation (ICC) of 0.1 .
We also examined the alternative settings with ICCs of 0.01 and 0.05. For this investigation, we focused on $I=30$ clusters and the patterns of findings remained the same; these results are presented in the Appendix \ref{supp:icc_sensitivity}. 
The baseline function was held fixed across all scenarios in the form of 
\[
f_0(\bm{X}_{ij}, \bm{V}_i) = \sin(\pi X_{ij1}X_{ij2})  - \sqrt{|X_{ij4} - V_{i1}|} + \log(1 + V_{i2}) + 0.5X_{ij6}.
\]

We considered three different heterogeneity settings (HS), each defined by distinct conditional treatment effect functions $\tau(\bm{X}_{ij}, \bm{V}_i)$ with increasing levels of complexity. Scenario HS1 (low-complexity nonlinearity) introduced nonlinear transformations of covariates:
\[
\tau(\bm{X}_{ij}, \bm{V}_i) = 1 - \frac{0.8}{1 + \exp(-X_{ij2})} - 0.4 X_{ij3}^2 + 0.5 V_{i2}.
\]
Scenario HS2 (moderate-complexity nonlinearity with interactions) incorporated covariate interactions and non-additive terms:
\[
\tau(\bm{X}_{ij}, \bm{V}_i) = -0.5 + 0.6 \sin(\pi X_{ij1} X_{ij2}) + 0.3 \cos(X_{ij4}) - 0.4 X_{ij5}^2 + 0.4 V_{i2}.
\]
Scenario HS3 (high-complexity nonlinearity with higher-order interactions) incorporated more complex covariate interactions:
\[
\tau(\bm{X}_{ij}, \bm{V}_i) = 0.6 + 0.7 \log( \lvert X_{ij1} X_{ij2} + X_{ij8}\rvert) + \frac{0.4}{1 + \exp(-(X_{ij2}+X_{ij5}))} - 0.3\sin(\pi X_{ij4}X_{ij5}) + 0.3 V_{i2}.
\]
We controlled the signal-to-total variance ratio (STR) around 0.6--0.65 for all scenarios which is defined as
\[
\mathrm{STR}
=\frac{\text{Var}\left(f_0(\bm X_{ij},\bm V_i)+A_i\,\tau(\bm X_{ij},\bm V_i)+b_i\right)}
       {\text{Var}\left(f_0(\bm X_{ij},\bm V_i)+A_i\,\tau(\bm X_{ij},\bm V_i)+b_i+\varepsilon_{ij}\right)}.
\]

\subsection{Implementation}

\begin{sidewaystable}[htbp]
\centering
\caption{Implementation details for all methods, including R packages, key tuning parameters and interval estimation.}\label{sim:summary}
\newcolumntype{L}{>{\raggedright\arraybackslash}p{0.16\linewidth}}
\newcolumntype{S}{>{\raggedright\arraybackslash}p{0.14\linewidth}} 
\newcolumntype{X}{>{\raggedright\arraybackslash}m{}} 
\resizebox{\linewidth}{!}{%
\begin{tabularx}{\linewidth}{@{}L S X L@{}}
\toprule
\textbf{Method} & \textbf{R package} & \textbf{Key settings} & \textbf{Reference} \\
\midrule
Mixed-effects BART with STAN (stan4bart) &
\texttt{stan4bart}; Version: 0.0.12 &
200 trees; burn-in 5{,}000; posterior draws 5{,}000 &
\citet{dorie2022stan} \\
\addlinespace\midrule

Mixed-effects BART for longitudinal data (mixedBART) &
\texttt{mxbart}; Version: 1.1 (GitHub:\ \url{https://github.com/rsparapa/bnptools/tree/master/mxBART}) &
200 trees; burn-in 5{,}000; posterior draws 5{,}000; default hyperparameters &
\citet{spanbauer2021nonparametric} \\
\addlinespace\midrule

BART with random intercept (riBART) &
\texttt{dbarts}; Version: 0.9.33 &
200 trees; burn-in 5{,}000; posterior draws 5{,}000 &
\citet{tan2016predicting} \\
\addlinespace\midrule

Multilevel Bayesian Causal Forest (MBCF) &
\texttt{multibart}; Version: 0.3 (OSF:\ \url{https://osf.io/3zmqc/}) &
Prognostic forest 200 trees; treatment-effect forest 200 trees; burn-in 5{,}000; posterior draws 5{,}000; propensity = 0.5 &
\citet{hahn2020bayesian}; \citet{yeager2022synergistic} \\
\addlinespace\midrule

Causal Forest (CF) &
\texttt{grf}; Version: 2.6.1 &
2000 trees; propensity = 0.5 &
\citet{athey2019generalized};\citet{wager2018estimation} \\
\addlinespace\midrule

Mixed-effects random forest (MERF) &
\texttt{LongituRF}; Version: 0.9 &
500 trees; max interaction order 2; $m_{\mathrm{try}}=p/3$; cluster bootstrap with 200 replicates &
\citet{capitaine2021random} \\
\addlinespace\midrule

Gradient boosting with grouped random effects / Gaussian Process (GPBoost) &
\texttt{GPBoost}; Version: 1.6.6 &
2{,}000 boosting rounds; learning rate 0.1; max depth 4; $L_2$ regularization; minimum data in leaf 100; cluster bootstrap with 200 replicates &
\citet{sigrist2022gaussian} \\
\addlinespace\midrule

Mixed-effects gradient boosting (MEGB) &
\texttt{MEGB}; Version: 0.2 &
Shrinkage 0.05; max interaction order 2; 500 base learners; cluster bootstrap with 200 replicates &
\citet{olaniran2025mixed} \\
\addlinespace\midrule

Component-wise boosting for GAMMs (mermboost) &
\texttt{mermboost}; Version: 0.1.1 &
Learning rate 0.1; max iterations 5{,}000; cluster bootstrap with 200 replicates &
\citet{knieper2025gradient} \\
\addlinespace\midrule

Generalized additive mixed model (GAMM) &
\texttt{mgcv}; Version: 1.9.4 &
REML; random intercept by cluster; spline basis; by-treatment smooths as applicable &
\citet{wood2011fast}; \citet{cho2022modelling,cho2024modeling} \\
\bottomrule
\end{tabularx}
}
\end{sidewaystable}

We estimate CATEs using the 10 machine-learning methods described in Section \ref{sec:ml_method}. All models were trained on pooled individual-level data with covariates \((\bm{X}_{ij},\bm{V}_i)\) and the cluster-level treatment  \(A_i\). For BART-based methods, including stan4bart, mixedBART, riBART, and MBCF, we used ensembles of 200 trees and ran 5,000 burn-in iterations followed by 5,000 posterior draws under default prior and hyperparameter specifications. The MBCF specification employed separate prognostic and treatment-effect ensembles (200 trees each), and the propensity score was fixed at 0.5 reflecting the randomized assignment. For causal forest, we used 2,000 trees by default with the propensity score fixed at 0.5.

For the random forest- and boosting-based mixed-effects methods (MERF, GPBoost, MEGB, and mermboost), we used default settings with minor deviations for computational reasons. MERF was fitted with 500 trees, and a maximum interaction order of 2. GPBoost was trained using 2,000 boosting iterations with a learning rate of 0.1, maximum tree depth of 4, and $L_2$ regularization. MEGB was specified with 500 boosting iterations, and a maximum interaction order of 2. The mermboost model was implemented with spline-based learners and up to 5,000 iterations. For all methods, variance estimation was obtained via nonparametric bootstrap with 200 replicates. Finally, we fitted GAMMs with treatment and covariates as fixed effects and smooth spline terms for continuous covariates. 

For every method, the CATE was computed as the difference between fitted outcomes under \(A_i=1\) and \(A_i=0\). Interval estimation was method-specific. For the Bayesian methods (stan4bart, mixedBART, riBART, and MBCF), 95\% intervals were obtained from posterior draws and were therefore posterior credible intervals. For causal forest, we used the asymptotic variance estimates returned by \texttt{grf} and constructed 95\% confidence intervals as \(\hat\tau \pm 1.96\times \widehat{\mathrm{SE}}(\hat\tau)\). For GAMM, we constructed 95\% confidence intervals by simulating coefficient draws from the approximate sampling distribution \(\mathcal{N}(\hat{\bm\beta}, \widehat{\mathrm{Var}}(\hat{\bm\beta}))\) using the unconditional covariance matrix from \texttt{mgcv}, propagating these draws through the treatment-specific linear predictors, and taking empirical quantiles of the resulting simulated treatment-effect contrasts. For MERF, GPBoost, MEGB, and mermboost, which do not directly provide posterior draws or analytic CATE interval estimators in our implementation, we used a nonparametric cluster-level bootstrap with 200 replicates to construct 95\% confidence intervals. We considered 200 replicates since more replicates lead to prohibitive computations for our simulation studies.

Finally, the performance of each method in each scenario was assessed over 500 Monte Carlo replications. All hyperparameter choices and package implementations are summarized in Table \ref{sim:summary}, and fully reproducible code for the simulation study is available in our GitHub repository : \url{https://github.com/changjun-li/crt-hte-ml-continuous}. All simulations were conducted in R version 4.4.3.

\subsection{Performance Metrics}

We evaluated the performance of the CATE estimators from each method using five metrics: precision in estimation of heterogeneous effects (PEHE), absolute bias, expected regret, average interval length, and empirical coverage probability. 
The PEHE is defined as the root mean squared error between the estimated and true conditional treatment effects across all individuals:
\[ \text{PEHE} = \sqrt{\frac{1}{\sum_{i=1}^{I}N_i} \sum_{i=1}^{I} \sum_{j=1}^{N_i} \left\{\widehat{\tau}(\bm{X}_{ij},\bm{V}_i) - \tau(\bm{X}_{ij},\bm{V}_i)   \right\}^2    }. \]
To assess bias, we compute the absolute bias, defined as:
\[
\text{Absolute Bias}
=
\frac{1}{\sum_{i=1}^{I}N_i}
\sum_{i=1}^{I}\sum_{j=1}^{N_i}
\left|
\widehat{\tau}(\bm{X}_{ij},\bm{V}_i)-\tau(\bm{X}_{ij},\bm{V}_i)
\right|.
\]
To evaluate the utility of estimated effects for informing individualized treatment decisions, we computed the expected regret, defined as the average magnitude of the true effect among individuals for whom the estimated optimal treatment assignment (based on the sign of \(\widehat{\tau}_{ij}(\bm{X}_{ij},\bm{V}_i)\)) differed from the true optimal assignment:

\[
\text{Regret}
=
\frac{1}{\sum_{i=1}^{I}N_i}
\sum_{i=1}^{I}\sum_{j=1}^{N_i}
\left[
\left|
\tau(\bm{X}_{ij},\bm{V}_i)
\right|
\cdot
\mathbb{I}
\left\{
\mathrm{sign}\!\left(\widehat{\tau}(\bm{X}_{ij},\bm{V}_i)\right)
\neq
\mathrm{sign}\!\left(\tau(\bm{X}_{ij},\bm{V}_i)\right)
\right\}
\right].
\]

Finally, to quantify the uncertainty of estimators, we compute the average interval length, defined as mean width of the \(95\%\) confidence/credible intervals across all individuals. We also calculate the empirical coverage probability, defined as the proportion of cases in which the true value 
lies within the corresponding \(95\%\) confidence interval. Although the Bayesian methods yield posterior credible intervals, whereas the remaining methods yield frequentist confidence intervals, we evaluated all interval estimates on a common empirical scale through the Monte Carlo coverage of the nominal 95\% interval.

\section{Simulation Results}

\subsection{Performance under different scenarios}

\begin{figure}[htbp]
  \centering
  \includegraphics[width=\textwidth,keepaspectratio]{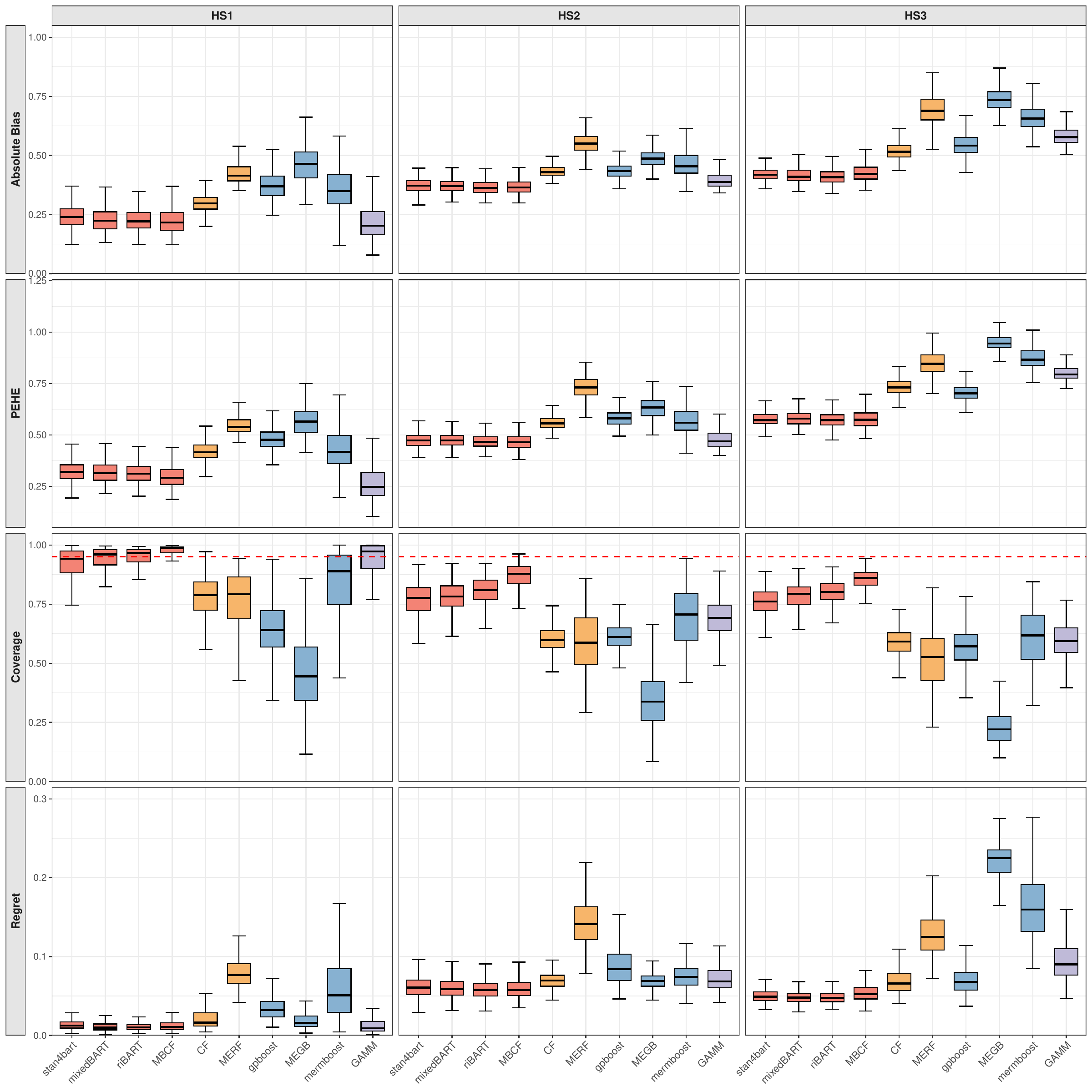}
  \caption{Absolute bias, PEHE, empirical coverage, and regret across methods for the setting with ICC \(=0.1\) and \(I=30\) clusters. Columns correspond to scenarios HS1--HS3. Rows show absolute bias (top), PEHE (second), empirical coverage of nominal \(95\%\) intervals (third), and expected regret (bottom). Boxplots summarize Monte Carlo replicates, and the red dashed line marks the nominal 0.95 frequentist coverage level.}
  \label{fig:pehe-coverage-30cluster}
\end{figure}

Figure \ref{fig:pehe-coverage-30cluster} summarizes results for the setting with \(I=30\) clusters. The four rows report, from top to bottom, absolute bias, PEHE, empirical coverage of nominal \(95\%\) intervals, and expected regret. Smaller values of absolute bias, PEHE, and regret indicate better performance, whereas coverage values close to the horizontal 0.95 reference line indicate better frequentist calibration. The three columns correspond to HS1--HS3, and within each panel, the same ten methods appear in a fixed left-to-right order (stan4bart, mixedBART, riBART, MBCF, CF, MERF, gpboost, MEGB, mermboost, GAMM). 

Under HS1 (column 1), the four BART-based methods form a clear low-bias group, with MBCF having slightly smaller absolute bias than stan4bart, mixedBART, and riBART. GAMM also performs favorably in this setting, with absolute bias comparable to the BART-based methods, which is plausible because the additive and smooth structure of the GAMM working model is well aligned with the HS1 data-generating mechanism. CF exhibits somewhat larger absolute bias than the BART-based methods and GAMM, whereas MERF and the boosting-based methods, especially MEGB, show less favorable performance. Under HS2 (column 2), absolute bias increases across all methods relative to HS1, but the overall ranking remains similar: the BART-based methods continue to perform best, with MBCF retaining a modest advantage, CF remains the strongest non-BART tree-based competitor, and GAMM loses the advantage it had under HS1. Under HS3 (column 3), where the heterogeneity structure is most complex, the separation becomes more pronounced. The BART-based methods continue to exhibit the smallest absolute bias, whereas GAMM performs substantially worse, reflecting the difficulty of capturing nonlinearities and higher-order interactions using a structured additive model. Among the remaining machine-learning approaches, CF and gpboost are generally more competitive than MERF, MEGB, and mermboost, but they are still generally worse than the BART-based methods.

The PEHE results (second row of Figure \ref{fig:pehe-coverage-30cluster}) are broadly consistent with the bias patterns. Under HS1, the four BART-based methods again form a low-error group. Among them, stan4bart, mixedBART, and riBART have nearly indistinguishable PEHE, while MBCF presents modestly lower PEHE than the other BART-based methods, indicating slightly better recovery of the CATE surface. CF exhibits higher PEHE than the BART group, but remains more accurate than MERF in this simplest setting. The boosting-style mixed-effects methods show more diverse performance under HS1: gpboost improves upon MEGB, while mermboost has the smallest PEHE among the three boosting methods, although it still does not outperform the BART-based methods. Among all frequentist methods, GAMM delivers the most favorable PEHE under HS1, with values as small as those of the BART-based methods. Taken together with the low absolute bias noted above, this suggests that the good HS1 performance of GAMM reflects genuine alignment between its working model and the true data-generating mechanism. In HS2, PEHE increases across all methods relative to HS1, with an upward shift in the PEHE distributions, but the overall ranking remains similar. The BART-based methods continue to yield the smallest PEHE, with MBCF remaining slightly better than the other BART methods, and CF again representing the best non-BART tree-based method. In this setting, however, the relative advantage of GAMM disappears, and it presents higher PEHE than the BART-based methods. In HS3, where the heterogeneity structure is the most complex, GAMM performs substantially worse than the BART-based methods, with markedly larger PEHE, indicating that it fails to capture nonlinearities and higher-order interactions between features. In this setting, the BART-based methods clearly dominate the other approaches. Among the remaining machine-learning methods, gpboost and CF show better performance than MERF, MEGB, mermboost, and GAMM, although their PEHE remains generally higher than that of the BART-based methods.

The coverage results (third row of Figure \ref{fig:pehe-coverage-30cluster}) summarize the frequentist calibration of the interval estimators across methods. Comparisons with interval length are provided in Appendix \ref{supp:interval_length}. Under HS1 (column 1), the four BART-based methods (stan4bart, mixedBART, riBART, MBCF) are the only methods whose coverage is consistently close to the frequentist nominal \(0.95\) level, with MBCF exhibiting the highest coverage within this class of methods. This slight conservatism is explained by its comparatively wider intervals, whereas the other BART methods achieve near-nominal coverage with narrower intervals. GAMM also attains near-nominal coverage in HS1, since the HS1 data-generating mechanism remains additive. Interval estimators from the remaining methods exhibit undercoverage even under HS1; that is, CF and MERF have coverage below \(0.95\), and the undercoverage is even more pronounced for boosting-based methods, especially MEGB. Interestingly, the bootstrap intervals for MEGB are noticeably narrower than those produced by the other methods, thereby contributing to undercoverage. The mermboost is comparatively less anti-conservative than gpboost and MEGB in HS1, with coverage closer to nominal, but its distribution is substantially more variable across Monte Carlo replications. Under HS2 (column 2), coverage declines across all methods relative to HS1. Nevertheless, the BART-based methods remain the closest to the nominal level, and MBCF continues to have the highest coverage within this group, again partially due to its wider intervals. By comparison, CF and MERF remain sub-nominal, the boosting-based methods continue to under-cover, and MEGB performs the worst. Under HS3 (column 3), the same ordering persists but becomes more pronounced. Even the BART-based methods can lead to undercoverage, although they remain the best-performing group. In this case, gpboost and MEGB again stand out for particularly low coverage and therefore lead to less credible uncertainty quantification for estimating CATE in CRTs.

The regret results (bottom row of Figure \ref{fig:pehe-coverage-30cluster}) are also broadly consistent with the patterns for absolute bias and PEHE. Under HS1, the BART-based methods and GAMM exhibit the smallest regret, CF shows intermediate performance, and MERF together with the boosting-based methods tends to have larger regret, especially MERF and MEGB. As the scenarios become more complex, regret increases across all procedures. Nevertheless, the BART-based methods continue to outperform the other approaches, with MBCF again tending to be the most favorable within this class. These findings reinforce the conclusion that better recovery of treatment-effect heterogeneity translates into improved treatment decision performance.

We also conducted additional simulations varying the number of clusters to \(I = 10\) and \(I = 100\), where the corresponding figures are reported in the Appendix \ref{supp:i10i100}. For \(I = 10\) clusters, the relative performance of methods is similar to that for \(I=30\), where the BART-based procedures (and, in HS1, GAMM) yield the smallest absolute bias, PEHE, and regret across HS1--HS3. Under HS1, GAMM becomes quite competitive with this top-performing group due to the additive data-generating process, but this advantage diminishes in HS2 and HS3. By contrast, CF, MERF and the boosting-type methods perform less favorably and exhibit pronounced undercoverage. The main effect of moving from \(I=30\) to \(I=10\) is therefore not a change in ranking for the performance across various methods, but increased variability of all performance metrics and amplifies the undercoverage of the frequentist machine learning methods. When the number of clusters increases to \(I = 100\), all procedures benefit from improved estimation accuracy, as reflected in lower absolute bias, PEHE, and regret, together with more stable coverage. However, the performance ranking of methods remains unchanged. BART-based methods still dominate, with MBCF retaining its advantage in frequentist coverage, i.e., more conservative and therefore closer to (or slightly above) the nominal \(95\%\) level. Methods such as CF, MERF, gpboost, MEGB, mermboost, and GAMM perform inadequately overall, particularly under the more complex scenarios HS2 and HS3.

\subsection{Sensitivity to the distribution of random intercepts}

We assessed the sensitivity of our findings to a more complex specification of the random intercepts. Specifically, we replicated the simulation for \(I = 30\) using the same data-generating process described in Section~\ref{sec:simulation design}, but assumed that the cluster-specific random intercepts \(b_i\) follow a gamma distribution rather than a normal distribution. In particular, we assumed \(b_i \sim \mathrm{Gamma}(k,\theta)\) with density \(f(b) = b^{k-1} e^{-b/\theta} \big/ \{\Gamma(k)\,\theta^{k}\}\) for \(b>0\), where \(k\) and \(\theta\) are the shape and scale parameters, respectively. In our setting, we took \(b_i \sim \mathrm{Gamma} \left(2, \sigma_b/\sqrt{2}\right)\), so that the skewness of the random intercepts is fixed at \(\sqrt{2}\). For ICC \(=0.1\), we set \(\sigma_b^2 = 0.1\), matching the variance used in the normal random-intercept specification. 
The results are shown in Appendix \ref{supp:sen_gamma}. Performance is qualitatively similar to that in Figure \ref{fig:pehe-coverage-30cluster} under normally distributed intercepts, indicating limited sensitivity to the random effect distribution misspecification.

\subsection{Performance when handling high-dimensional covariates}
In certain CRT applications, the number of baseline covariates may be more than a handful, and high-dimensional covariate spaces require methods to reliably separate true signal from substantial noise. To investigate performance in such settings, we conducted additional simulations with 20, 50, and 100 covariates. These simulations were based on the scenario with ICC=0.1 and \(I = 30\), and were designed as high-dimensional extensions of the HS3 mechanism: we increased the number of effect-modifying covariates entering $\tau$ (each governed by HS3-style nonlinearities and interactions) and also added additional covariates, including irrelevant noise variables, to raise the overall dimensionality. For the designs with 20, 50, and 100 covariates, we selected 10, 15, and 20 covariates, respectively, to enter either the baseline outcome function or the treatment effect function. We calibrated the variance components so that \(\sigma_b^2 + \sigma^2\) equals 1.5, 2, and 2.5 in the three designs, which yielded a signal-to-total variance ratio (STR) of approximately 0.65 in each case. Due to space constraint, the detailed data-generating mechanisms are provided in the Appendix \ref{supp::sen::high}.

Because the frequentist machine-learning methods were both less accurate and substantially more computationally demanding in these more complex scenarios, and given that we have already established that BART-based methods perform best overall,  we restricted the high-dimensional simulations to the BART-based approaches and examined how their performance degraded.

Here we focus on the top-performing class of methods based on BART. The simulation results for the high-dimensional scenarios are summarized in Figure \ref{fig:pehe-coverage-hd}, which displays absolute bias, PEHE, coverage and regret, and in Supplementary Figure~\ref{fig:hd_interval_length}, which reports interval length. As the dimensionality increases, the relative advantage of MBCF becomes more pronounced. With 20 covariates, the four BART-based methods perform broadly similarly in terms of absolute bias, PEHE, and regret, although MBCF already shows somewhat better coverage than the other three methods. When additional noise covariates are introduced, yielding 50 and 100 covariates, stan4bart, mixedBART, and riBART deteriorate more noticeably, with higher absolute bias, larger PEHE, higher regret, and persistent undercoverage. In contrast, MBCF remains comparatively robust in these higher-dimensional settings. For both 50 and 100 covariates, it achieves the smallest absolute bias, lowest PEHE, and lowest regret among the four BART-based methods, while also maintaining coverage much closer to the nominal 95\% level. These results indicate that the advantage of MBCF becomes increasingly substantial as the covariate space expands and irrelevant features are added, reinforcing its overall strong performance for estimating CATEs in CRTs.

\begin{figure}[h!]
  \centering
  \includegraphics[width=\textwidth,keepaspectratio]{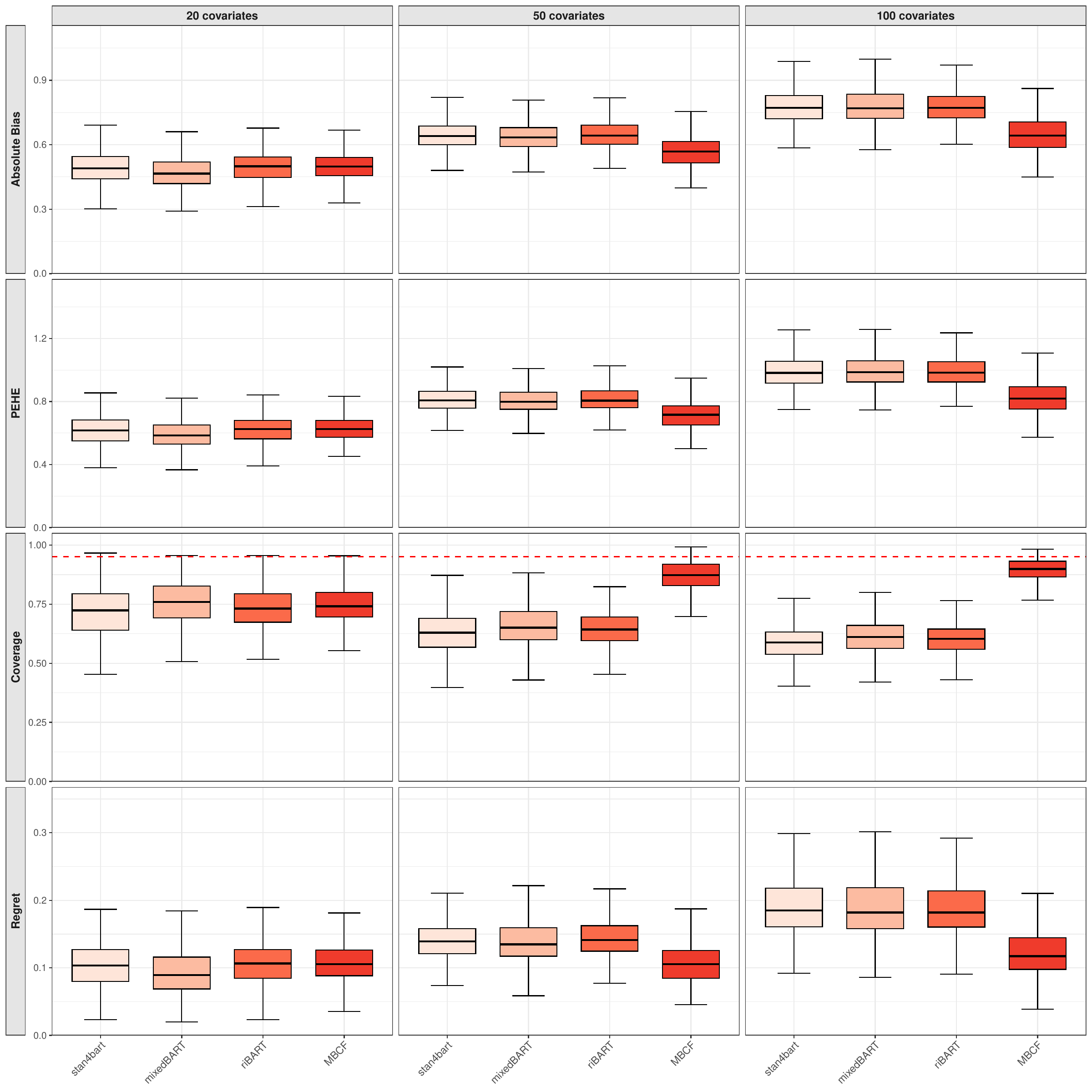}
  \caption{Absolute bias, PEHE, empirical coverage, and regret across BART-based methods in high-dimensional settings with ICC \(=0.1\) and \(I=30\) clusters. Columns correspond to 20, 50, and 100 covariates. Rows show absolute bias (top), PEHE (second), empirical coverage of nominal \(95\%\) intervals (third), and expected regret (bottom). Boxplots summarize Monte Carlo replicates, and the red dashed line marks the nominal 0.95 frequentist coverage level.}
  \label{fig:pehe-coverage-hd}
\end{figure}

\section{Estimating Conditional Average Treatment Effects in the TASSH cluster-randomized trial}\label{sec:data}
We analyzed data from the Task Shifting and Blood Pressure Control in Ghana (TASSH), a CRT that evaluated a nurse-led task shifting strategy for hypertension control \citep{ogedegbe2018health}. 
Thirty-two community health centers (clusters) in Ghana’s public healthcare system were randomized in a 1:1 ratio to receive either health insurance coverage (HIC) alone or HIC plus the TASSH intervention. TASSH recruited 389 patients in the TASSH plus HIC arm and 368 patients in the HIC alone arm, with the cluster sizes varying from 17 to 31 participants. Among the 757 enrolled patients, 641 had a well-defined 12-month outcome, corresponding to an overall completion rate of 85\% (88\% in the TASSH+HIC group and 82\% in the HIC group).

We estimated the CATEs on 12-month diastolic blood pressure (DBP) using the multilevel Bayesian causal forest, which was identified as the top performing approach in our simulation study. CATEs were defined as the difference between the fitted 12-month outcomes under intervention ($A_i=1$) and control ($A_i=0$), conditional on prespecified baseline covariates at both the individual and cluster levels. We used 200 trees for the prognostic forest and 200 trees for the treatment-effect forest, with 5{,}000 burn-in iterations followed by 5{,}000 posterior draws (default priors and initialization); the propensity score was fixed at 0.5 reflecting 1:1 cluster randomization. Individual-level features included baseline diastolic blood pressure and systolic blood pressure (SBP), demographics and behaviors (age, sex, education level, literacy, smoking status, and employment status); baseline physical activity measured as all activity in weighted metabolic equivalent minutes per week. Cluster-level features included staffing and setting characteristics (number of doctors and nurses on staff, annual patient volume, and rural versus urban location). All 13 baseline covariates had less than 10\% missingness, and, for illustrative purposes, missing covariate values were imputed using the mean for continuous variables and the mode for binary or categorical variables.

After estimating individual-level CATEs with MBCF, we assessed which baseline covariates contributed most to treatment effect heterogeneity using a leave-one-out treatment effect variable importance measure (TE-VIM) as introduced by \citet{hines2022variable}. Let \(\bm{Z}_{ij} = \{\bm{X}_{ij}^\top, \bm{V}_i^\top\}^\top\) denote the \(P\)-dimensional collection of individual- and cluster-level baseline covariates for individual \(j\) in cluster \(i\), and let \(\bm{Z}_{ij,-p}\) denote the same vector with the \(p\)-th covariate removed. For each covariate \(p \in \{1,\dots,P\}\), the population TE-VIM is defined as
\[
\mathrm{TE\text{-}VIM}_p=\mathbb{E}\!\left[\left\{\tau(\bm Z_{ij}) - \tau_{-p}(\bm Z_{ij})\right\}^2\right],
\]
where \(\tau(\bm Z_{ij}) = \tau(\bm x_{ij}, \bm v_i)\) is the CATE conditional on the full covariate vector, and
\[
\tau_{-p}(\bm Z_{ij})=\mathbb{E}\!\left[\tau(\bm Z_{ij}) \mid \bm Z_{ij,-p}\right]\]
is the best prediction of the treatment effect when covariate \(p\) is omitted from the conditioning set. Thus, \(\mathrm{TE\text{-}VIM}_p \ge 0\) quantifies the increase in mean squared error in predicting the CATE when covariate \(p\) is excluded, with larger values indicating greater importance of that covariate in explaining treatment effect heterogeneity.

In our analysis, we adopt a projection-based plug-in approach motivated by \citet{hines2022variable}, using the posterior mean CATE estimates from the fitted MBCF model as surrogates for the true CATEs. Specifically, let \(\widehat{\tau}(\bm x_{ij}, \bm v_i)\) denote the posterior mean CATE obtained from the full MBCF model fit including all \(P\) covariates. For each covariate \(p\), we estimate \(\tau_{-p}(\bm Z_{ij})\) by regressing the full-model posterior mean CATE estimates \(\widehat{\tau}(\bm x_{ij}, \bm v_i)\) on the reduced covariate set \(\bm Z_{ij,-p}\) using a BART regression model, yielding the projected estimate \(\widetilde{\tau}_{-p}(\bm z_{ij,-p})\). The TE-VIM for covariate \(p\) is then estimated as
\[
\widehat{\mathrm{TE\text{-}VIM}}_p=\frac{1}{\sum_{i=1}^{I} N_i}
\sum_{i=1}^{I}\sum_{j=1}^{N_i}
\left\{
\widehat{\tau}(\bm x_{ij}, \bm v_i)
-
\widetilde{\tau}_{-p}(\bm z_{ij,-p})
\right\}^2.
\]

We further characterized clinically interpretable subgroups with differential benefits from TASSH plus HIC versus HIC alone by adopting a ``fit-the-fit'' strategy \citep{foster2011subgroup, logan2019decision} . Under this strategy, we selected the six most important covariates identified by the TE-VIM analysis and fitted a classification and regression tree (CART) to the posterior mean CATE estimates from the MBCF model, using these baseline covariates as predictors, in order to identify key effect modifiers and define clinically interpretable subgroups.

\begin{figure}[htbp]
  \centering
  \includegraphics[width=\textwidth,keepaspectratio]{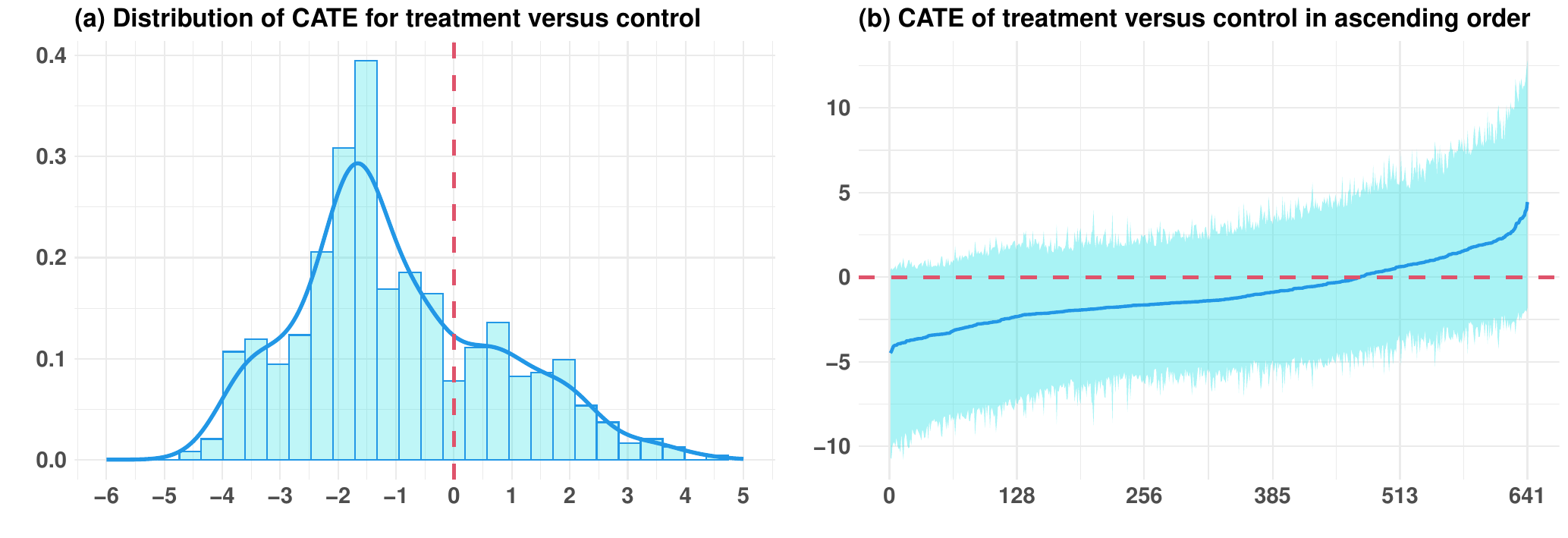}
  \caption{Estimated CATEs for DBP at 12 months.
  Panel (a) shows the posterior distribution of individual CATEs; values below zero indicate lower DBP under the intervention. The posterior mean ATE is $-1.00$ with a 95\% credible interval [$-3.48$, $1.40$]. Panel (b) plots participant specific CATEs in ascending order; the solid line is the posterior mean and the shaded band is the corresponding 95\% credible interval.}
  \label{fig:dbp_hist}
\end{figure}

Panel (a) of Figure \ref{fig:dbp_hist} shows the posterior distribution of the individual CATE estimates for 12-month DBP. The distribution is shifted markedly toward negative values, with most estimates indicating lower 12\text{-}month DBP under the intervention. This overall leftward pattern suggests that many individuals are expected to benefit from the treatment. Nonetheless, the posterior mean ATE was \(-1.00\) with a 95\% credible interval \([-3.48,\, 1.40]\), and the estimated intraclass correlation coefficient (ICC) 0.027 (95\% credible interval: 0.000 to 0.094); the average treatment effect hence is associated with a credible interval crossing zero. Panel (b) plots participant\text{-}specific CATEs for 12-month DBP ordered by their posterior mean. The solid curve traces the posterior mean CATE for each individual, and the shaded band gives the corresponding 95\% credible interval. The curve spans a wide range from strongly negative to clearly positive values, which shows substantial between-individual heterogeneity in the estimated treatment response. Although a large fraction of intervals cross zero, the overall gradient remains evident, with a sizable portion of participants exhibiting reductions in DBP, while others show little to no effect or even increases in DBP.

\begin{figure}[htbp]
  \centering
  \includegraphics[width=0.7\textwidth,keepaspectratio]{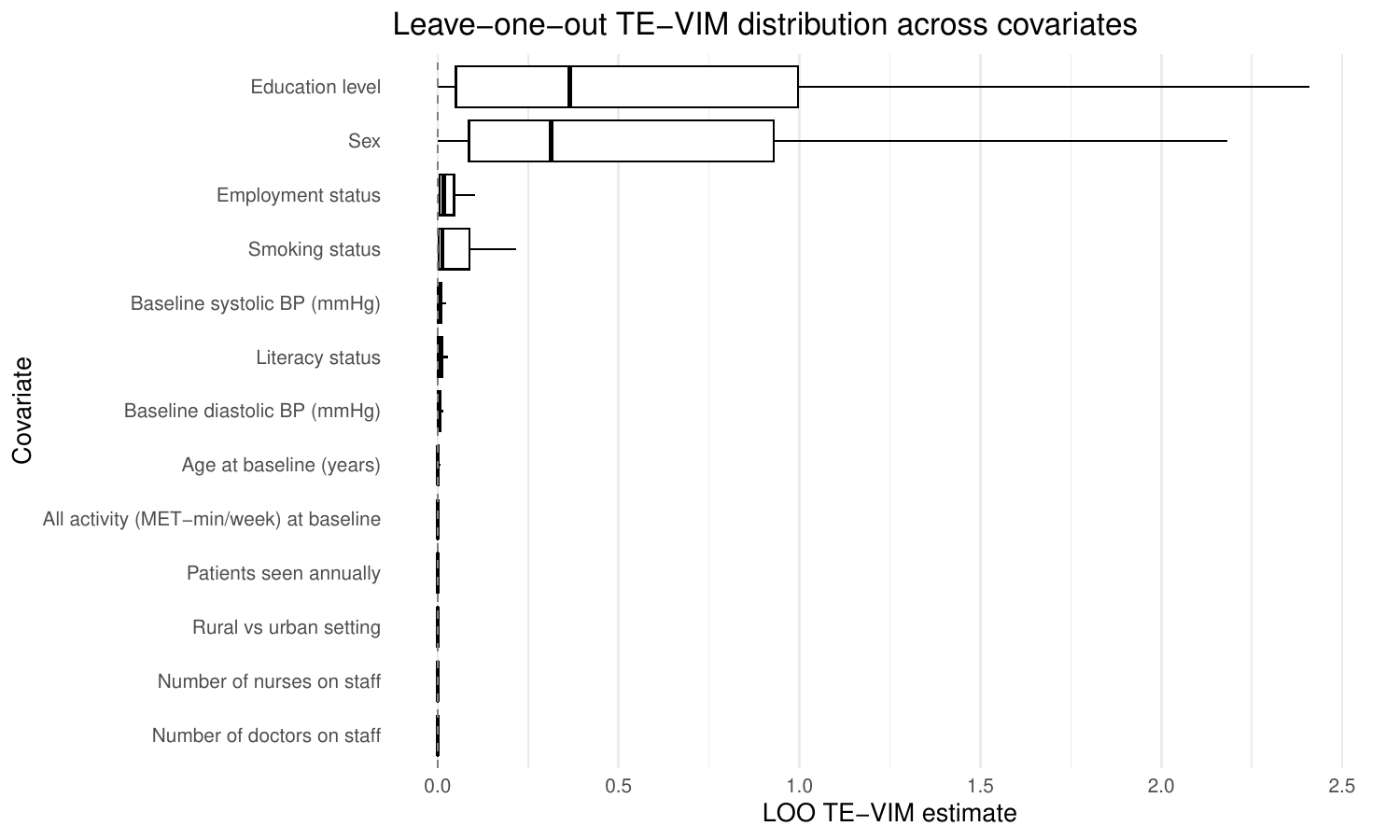}
  \caption{Leave-one-out TE-VIM across baseline covariates. For each covariate \(p\), we first obtained the full-model posterior mean CATE estimates \(\widehat{\tau}(\mathbf{x},\mathbf{v})\), and then regressed these estimates on the reduced covariate set excluding \(p\) using a BART projection model to obtain \(\widetilde{\tau}_{-p}(\mathbf{z}_{-p})\). Individual-level contributions to the TE-VIM were computed as \(\left\{\widehat{\tau}(\mathbf{x},\mathbf{v})-\widetilde{\tau}_{-p}(\mathbf{z}_{-p})\right\}^{2}\), and the boxplots summarize their distribution across participants. Larger values indicate that omitting the covariate leads to poorer prediction of the full-model CATEs and therefore greater importance for treatment-effect heterogeneity. Education level, sex, employment status, and smoking status show the largest TE-VIM values, whereas most other covariates display comparatively small contributions.}
  \label{fig:dbp_tevim}
\end{figure}

Figure \ref{fig:dbp_tevim} presents the TE-VIM results for each baseline covariate. For each covariate, the boxplots summarize the distribution of the individual-level squared differences contributing to the TE-VIM. Larger values indicate greater changes in the estimated CATEs when that covariate is omitted, suggesting that the covariate plays a more important role in explaining treatment-effect heterogeneity. The figure suggests that education level, sex, employment status, and smoking status have the greatest influence on the estimated heterogeneity, whereas the remaining covariates have comparatively smaller contributions.

\FloatBarrier

\begin{figure}[ht]
  \centering
  \includegraphics[width=0.8\textwidth,keepaspectratio]{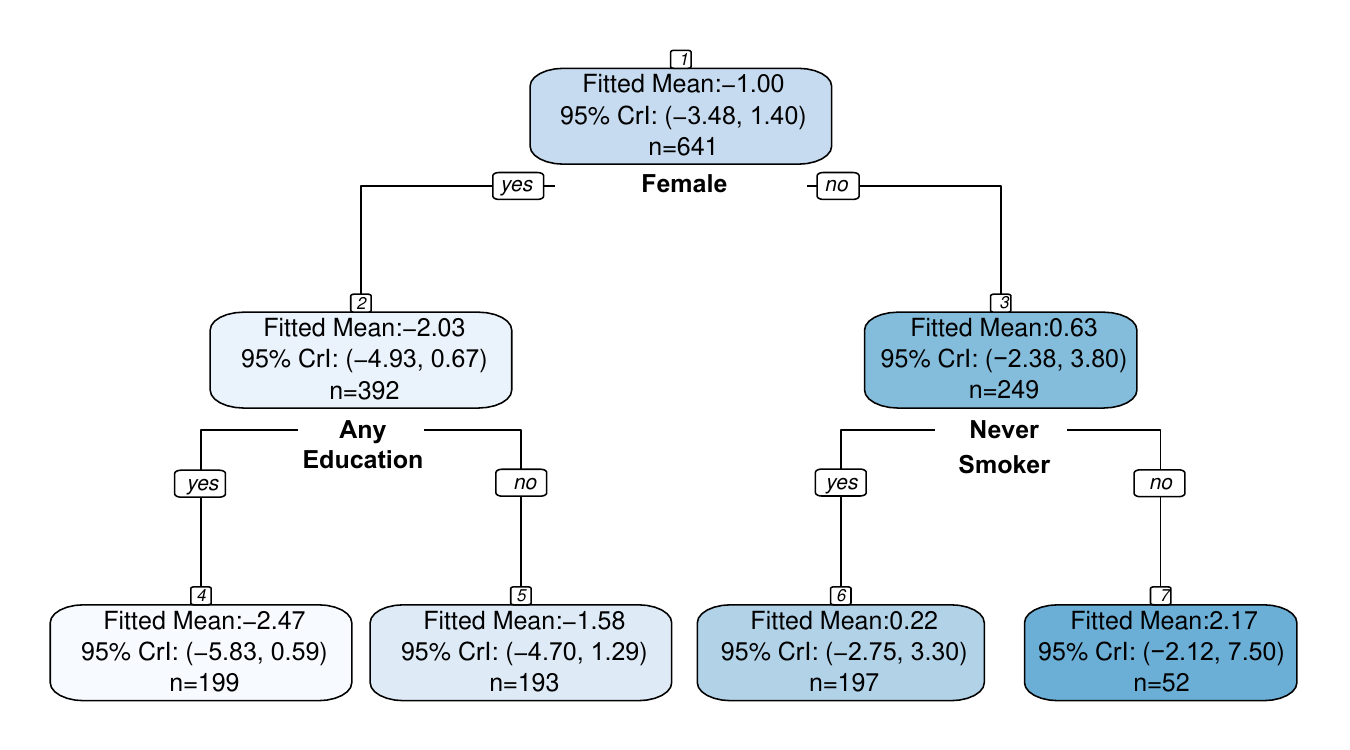}
  \caption{CART for posterior mean CATE on DBP at 12 months.
The tree regresses participant-specific posterior mean CATEs (from MBCF) on baseline covariates to summarize treatment effect heterogeneity between TASSH plus HIC and HIC. Internal nodes show splitting variables and cut points, and terminal nodes report the node-specific mean CATE and sample size. Negative values indicate lower 12-month DBP under TASSH plus HIC relative to HIC. Tree depth was limited to two for interpretability.}

  \label{fig:dbp_cart1}
\end{figure}

We then selected the six covariates with the largest TE-VIM values to fit a depth-2 classification and regression tree (CART). These six covariates were education level, sex, employment status, smoking status, baseline SBP, and literacy status. Figure \ref{fig:dbp_cart1} further summarizes heterogeneity by a two-level classification and regression tree fit to the posterior mean CATEs. Subgroup effects were obtained by averaging the individual CATE draws for participants in each terminal node, and 95\% credible intervals were computed from the corresponding posterior samples within each node. The first split is sex, with women exhibiting a more favorable average response to TASSH plus health insurance coverage than men. Among women, the next split is education level, indicating that those with any schooling show the largest average reduction in 12-month DBP, whereas women with no schooling have smaller estimated benefit. Among men, the next split is smoking status: non-smokers have more favorable average effects than those with any smoking. These patterns are consistent with the idea that factors related to health literacy, adherence, and risk behaviors may modify the intervention’s effect, although the exact mechanisms underlying effect modification cannot be established in this exploratory analysis alone. A three-level tree is reported in Appendix \ref{supp:MBCF_cart_3} as an additional exploration; employment status enter that more elaborate trees structure.

\FloatBarrier 

\begin{figure}[!htbp]
  \centering
  \includegraphics[width=\textwidth,keepaspectratio]{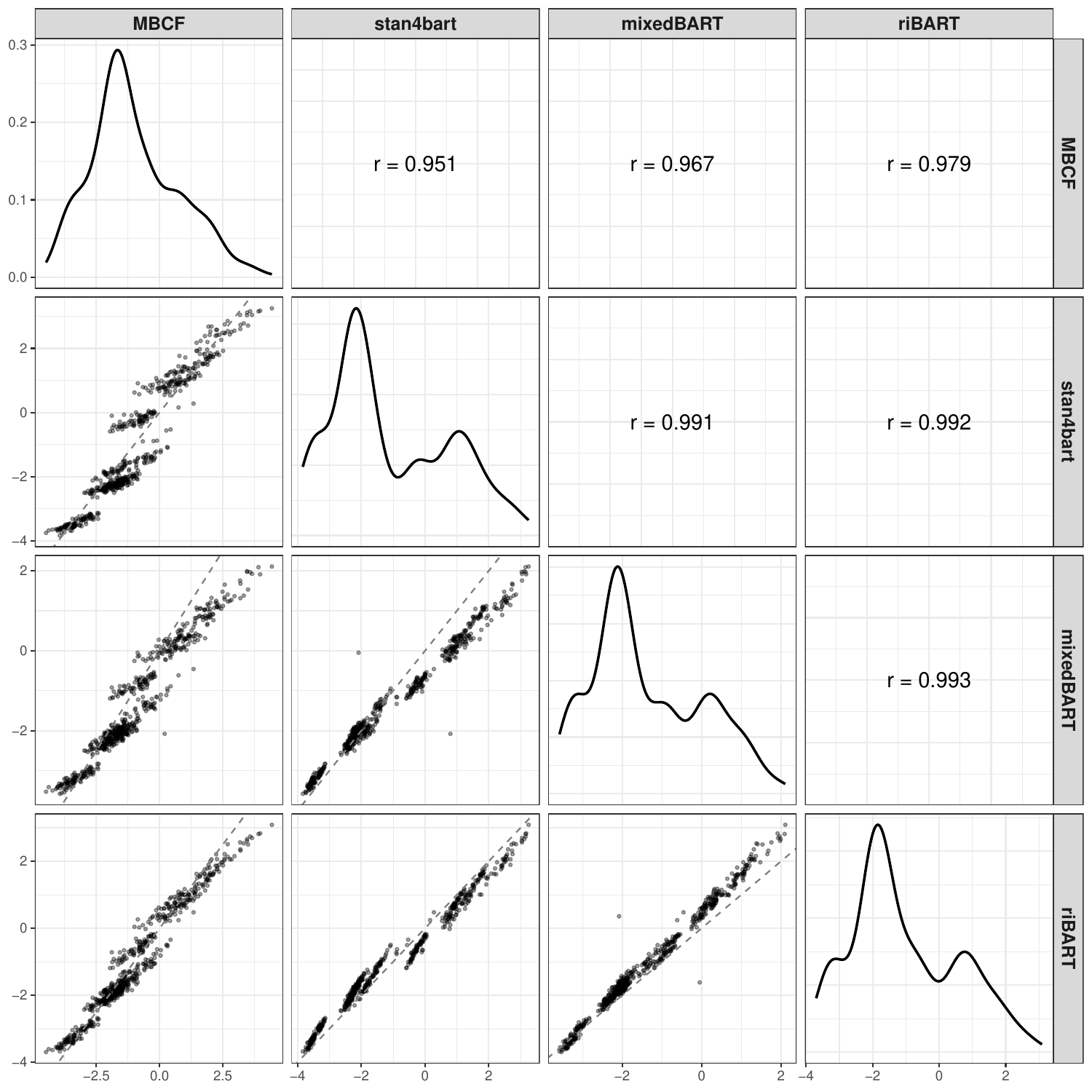}
  \caption{Pairwise scatterplot matrix of participant-specific posterior mean CATE estimates across the four BART-based methods in the TASSH analysis for 12-month DBP. Diagonal panels show marginal density estimates, upper panels report Pearson correlations, and lower panels display pairwise scatterplots.}
  \label{fig:cate_pairwise_matrix}
\end{figure}

As a supplementary analysis, we also repeated the analysis using three other BART-based methods, namely stan4bart, mixedBART, and riBART, to examine whether the findings were consistent across methods. The estimation and analysis strategy was the same as that used for MBCF. For each method, we used 200 trees, 5{,}000 burn-in iterations, and 5{,}000 posterior draws. Additional results from this supplementary analysis are presented in Appendix \ref{supp:TASSH_stan4bart}--\ref{supp:TASSH_comparison}. Overall, the four methods yielded highly consistent findings. Figure \ref{fig:cate_pairwise_matrix} summarizes the agreement of posterior mean CATE estimates across the four BART-based methods. Most points lie close to the 45-degree line, indicating close agreement in participant-specific CATE estimates across methods. In addition, all pairwise Pearson and Spearman correlations exceeded 0.95 across the four methods. The three mixed-effects BART methods were especially similar to one another, with correlations close to 0.99, whereas their correlations with MBCF were slightly lower, but still very high. The distributions of the posterior mean CATE estimates were also very similar across methods. The TE-VIM results were broadly consistent, with sex, education level, smoking status, and employment status consistently ranked among the most important covariates. Finally, the fitted classification and regression trees had the same overall structure across methods, differing only slightly in subgroup-specific point estimates and credible intervals.

\FloatBarrier

\section{Discussion}
Understanding treatment effect heterogeneity is crucial for generating scientifically meaningful hypotheses and discovering subgroups with differential benefit \citep{kent2020predictive, wang2007statistics,assmann2000subgroup}. A substantial literature has now grown to advise on methods for estimating heterogeneous treatment effects with independent data, and prior studies have compared their performance across randomized trials and observational settings with a variety of outcomes \citep{hu2021estimating}. However, these findings may not directly apply to CRTs where individual outcomes are nested within clusters that are randomized to conditions. As multilevel study designs have become more common, machine learning methods have also been extended to account for the clustering structure, including Bayesian Additive Regression Trees, gradient boosting, random forests, and generalized additive mixed models. In light of the growing interest in quantifying treatment effect heterogeneity in CRTs, this study is the first to evaluate the performance of these mixed-effects learning tools to estimate CATE in CRTs and provides supporting evidence to guide practice with continuous outcomes across various representative HTE scenarios. 

Our simulation results indicate that methods in the BART family were the most robust overall. Multilevel Bayesian causal forest showed consistently strong performance for estimating individual conditional average treatment effects, producing stable and interpretable rankings of effect modifiers, and conservative credible intervals with coverage close to nominal levels. Random intercept BART variants also performed well in low to moderate dimensional settings.
However, as the covariate dimension increased, their performance became more sensitive to noise variables, underscoring the need for explicit variable selection in high-dimensional CATE estimation. 
By contrast, gradient boosting and random forest implementations with random intercepts, such as mixed-effects gradient boosting, mixed-effects random forest, and GPBoost, were competitive for point estimation in simpler heterogeneity structures and under favorable signal-to-noise conditions, although uncertainty quantification was less reliable and variable importance was more sensitive to correlation among covariates. Generalized additive mixed models provided transparent smooth effects and interpretable modifier patterns in simpler settings but degraded as nonlinearity and higher-order interactions increased. All these results support using MBCF, or, more broadly, BART methods with cluster effect when estimating CATEs or conducting subgroup discovery is the primary goal. They also motivate sparsity-aware regularization in high-dimensional settings. For BART-based methods, one natural extension is to replace the uniform prior on splitting probabilities with a sparse Dirichlet prior, as in the Dirichlet additive regression trees (DART) framework. This prior concentrates probability mass on a small subset of covariates and shrinks the selection probabilities of noise variables toward zero, thereby improving robustness and efficiency in high-dimensional settings \citep{spanbauer2021nonparametric}.  In mixedBART, the \texttt{mxbart} package implements this sparse Dirichlet prior, and in additional simulations, we found that it can meaningfully improve the performance of mixedBART in high-dimensional settings. The corresponding simulation results are reported in Appendix \ref{supp:mxdart}.
We then apply the MBCF to the TASSH reanalysis, posterior mean CATEs for diastolic blood pressure at twelve months were mostly below zero, consistent with a tendency toward better control under the combined intervention, whereas the average effect was associated with a credible interval that crossed zero; leave one out treatment effect variable importance highlighted education level, sex, employment status and smoking as influential modifiers, and a shallow classification and regression tree on posterior mean CATEs yielded simple, clinically interpretable subgroups.

Our findings highlight several directions for further work. First, CRTs routinely feature outcomes beyond continuous measures, including binary endpoints, time to event outcomes, and outcomes truncated by death, each of which raises distinct estimands and modeling challenges. For binary outcomes, not all learners considered here have principled extensions, and future studies should therefore compare methods under clearly defined causal scales, for example, risk difference, risk ratio, and odds ratio, and evaluate their ability to recover CATE surface, rank effect modifiers, and support reliable inference. For survival outcomes, models must accommodate censoring and clustering; recent developments in random intercept accelerated failure time BART provide a promising foundation, but systematic evaluation across ICC levels, cluster size imbalance, and varying event rates is needed \citep{hu2022flexible,hu2024new}. Second, real-world CRTs can encounter intermediate outcomes such as patient compliance and death where treatment effect heterogeneity is only well defined among a subset of individuals \citep{tong2023bayesian, tong2025bayesian,cheng2025identification}.  In this setting, combining principal stratification with Bayesian machine learning offers a principled framework for subgroup-specific causal estimation. For clustered settings, \citet{li2026uncovering} recently developed a mixture of mixed-effects BART approaches for estimating conditional survivor average causal effects, highlighting the need for broader validation, guidance on estimand choice, and sensitivity analysis. Third, longitudinal data are common in CRTs; methods must account simultaneously for within-cluster and within individual correlation and allow time-varying effect modification. Extending mixed-effects learners to jointly model trajectories and treatment interactions, and benchmarking them in longitudinal CRT scenarios, would be valuable \citep{liang1986longitudinal,donner2000design}. Fourth, stepped wedge and cluster-randomized crossover designs are increasingly used. In stepped wedge designs, clusters cross over from control to intervention at prespecified time points so that all clusters eventually receive the intervention, whereas in cluster-randomized crossover designs, clusters switch back and forth between control and intervention according to designated schedules. Learners for these designs must handle period effects, staggered adoption or repeated crossover, and potential temporal confounding while preserving the unit of randomization. Flexible machine learners must simultaneously handle period effects, staggered adoption, and potential temporal confounding while preserving the unit of randomization.

\section*{Acknowledgments}
MOH and FL are supported by the United States National Institutes of Health (NIH), National Heart, Lung, and Blood Institute (NHLBI, grant numbers R01-HL168202 and R01-HL178513). All statements in this report, including its findings and conclusions, are solely those of the authors and do not necessarily represent the views of the NIH. The research of ABF is supported in part by the National Health and Medical Research Council of Australia (Ideas Grant number 2027218).

\section*{Supplementary Material}

The supplementary material includes additional simulation results, data analysis and the implementation code in R for all methods in Section \ref{sec:ml_method}.

\section*{Data availability}
The TASSH data can be obtained via \url{https://datadryad.org/dataset/doi:10.5061/dryad.16c9m51}, which is publicly available.

\section*{Conflict of interest}
The author(s) declared no potential conflicts of interest with respect to the research, authorship, and/or publication of this article.

\clearpage
\newpage
\printbibliography

\newpage
\appendix
\setcounter{section}{0}
\setcounter{subsection}{0}
\setcounter{subsubsection}{0}

\renewcommand{\thesection}{\arabic{section}}
\renewcommand{\thesubsection}{\thesection.\arabic{subsection}}
\renewcommand{\thesubsubsection}{\thesubsection.\arabic{subsubsection}}

\section{Additional simulation results} \label{supp:add_simu}

This section provides additional simulation results that complement the findings in the main text. We first examine how the relative performance of the competing methods changes when the number of clusters is reduced to \(I=10\) or increased to \(I=100\). We then report interval-length results, which complement the coverage findings in the main text. Finally, we present sensitivity analyses under lower intraclass correlation settings (\(\mathrm{ICC}=0.01\) and \(0.05\)).

\subsection{Simulation results for \(I=10\) and \(I=100\) clusters} \label{supp:i10i100}

\begin{figure}[htbp!]
  \centering
  \includegraphics[width=\textwidth,keepaspectratio]{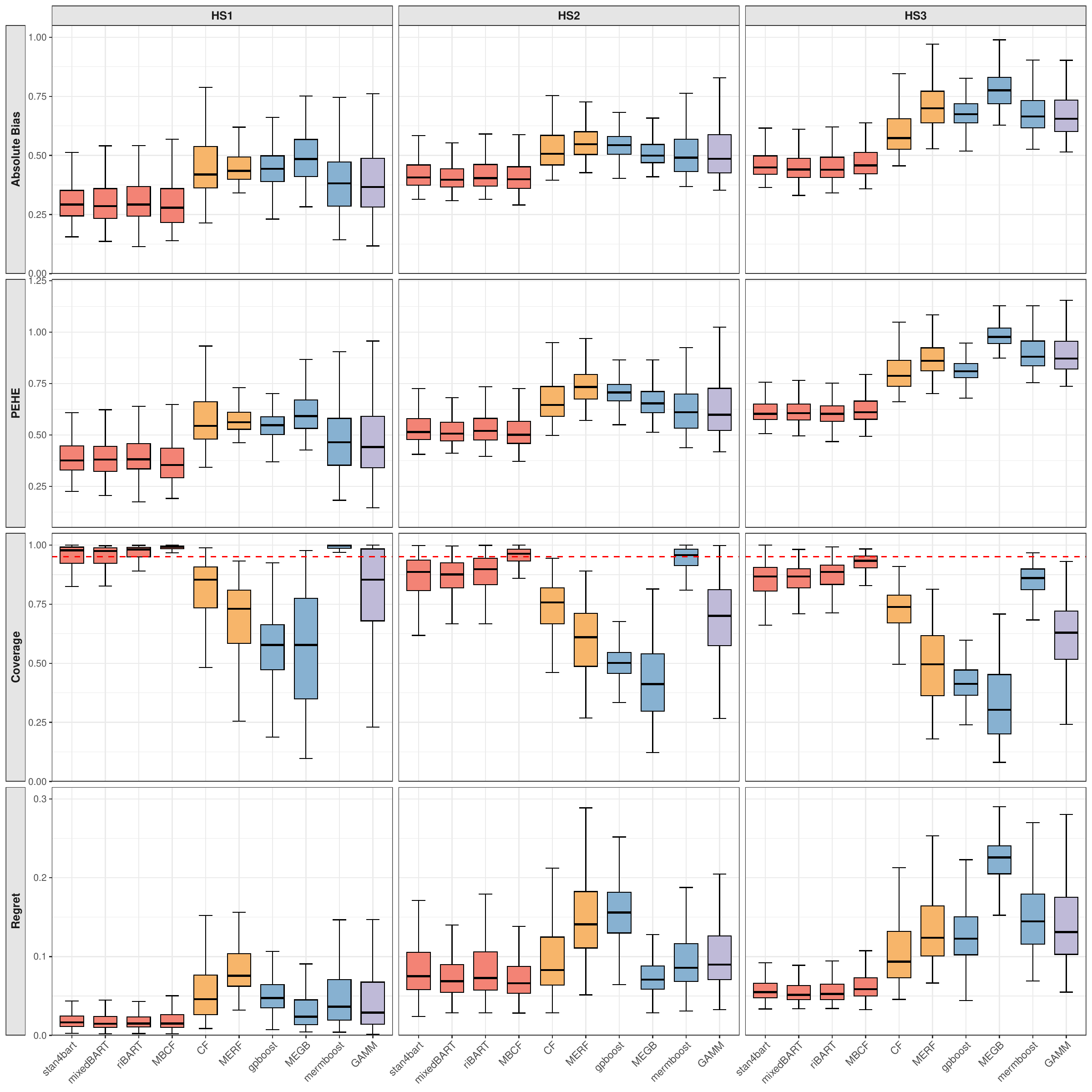}
  \caption{Absolute bias, PEHE, empirical coverage, and regret across methods for the setting with \(\mathrm{ICC}=0.1\) and \(I=10\) clusters. Columns correspond to scenarios HS1--HS3. Rows show absolute bias (top), PEHE (second), empirical coverage of nominal \(95\%\) intervals (third), and expected regret (bottom). Boxplots summarize Monte Carlo replicates, and the red dashed line marks the nominal 0.95 frequentist coverage level.}
  \label{fig:sim_i10}
\end{figure}

\begin{figure}[htbp!]
  \centering
  \includegraphics[width=\textwidth,keepaspectratio]{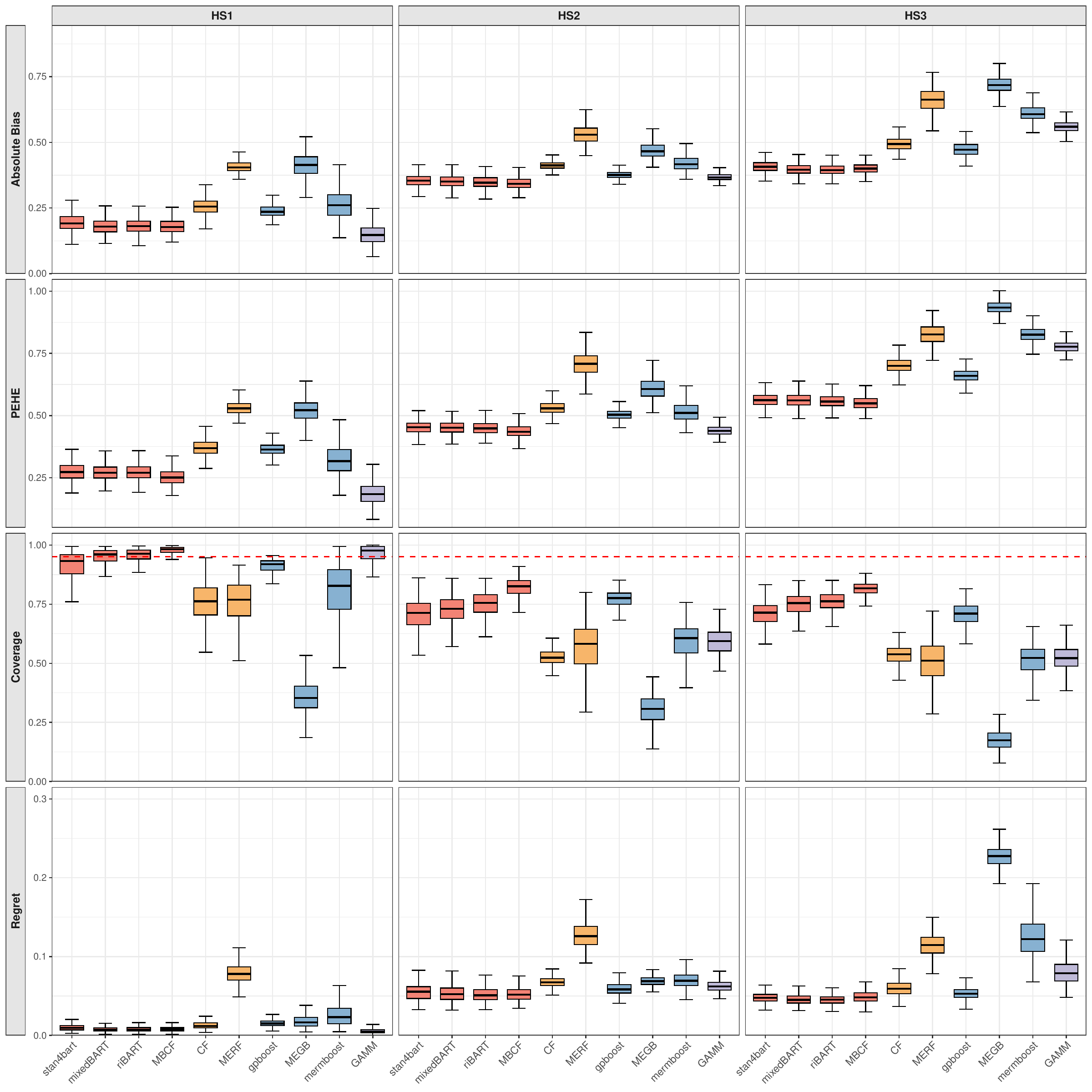}
  \caption{Absolute bias, PEHE, empirical coverage, and regret across methods for the setting with \(\mathrm{ICC}=0.1\) and \(I=100\) clusters. Columns correspond to scenarios HS1--HS3. Rows show absolute bias (top), PEHE (second), empirical coverage of nominal \(95\%\) intervals (third), and expected regret (bottom). Boxplots summarize Monte Carlo replicates, and the red dashed line marks the nominal 0.95 frequentist coverage level.}
  \label{fig:sim_i100}
\end{figure}

We conducted additional simulation experiments by varying the number of clusters to \(I=10\) and \(I=100\), while keeping all other data-generating mechanisms and analysis settings identical to those in the main text (\(\mathrm{ICC}=0.1\)). Figures~\ref{fig:sim_i10} and \ref{fig:sim_i100} summarize absolute bias, PEHE, empirical coverage, and regret across methods under HS1--HS3. These results are used to assess whether the relative ranking of methods is stable when the number of clusters changes.

\subsection{Simulation results for interval length} \label{supp:interval_length}

\begin{figure}[htbp!]
  \centering
  \includegraphics[width=\textwidth,keepaspectratio]{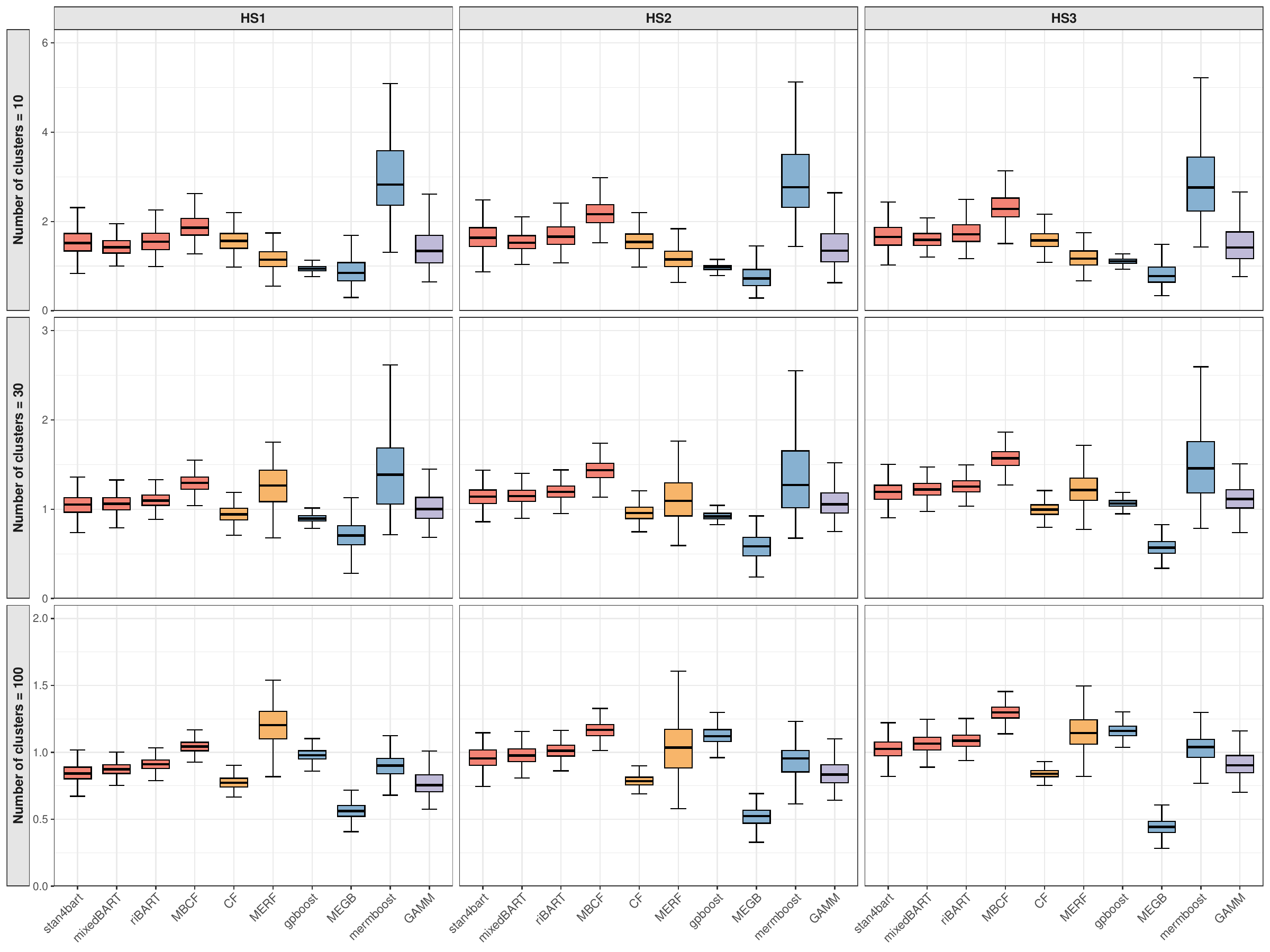}
  \caption{Average interval length across methods for the setting with \(\mathrm{ICC}=0.1\). Columns correspond to scenarios HS1--HS3, and panels summarize results for \(I=10\), \(I=30\), and \(I=100\) clusters, respectively. Boxplots summarize Monte Carlo replicates. Larger values indicate wider interval estimates.}
  \label{fig:sim_interval_length}
\end{figure}

To complement the empirical coverage results, we also examined average interval length across methods. Figure~\ref{fig:sim_interval_length} reports interval lengths for the main simulation setting with \(\mathrm{ICC}=0.1\), allowing comparisons of uncertainty quantification across methods beyond coverage alone.

\subsection{Simulation results for \(\mathrm{ICC}=0.01\) and \(0.05\)} \label{supp:icc_sensitivity}

\begin{figure}[!htbp]
  \centering
  \includegraphics[width=\textwidth,keepaspectratio]{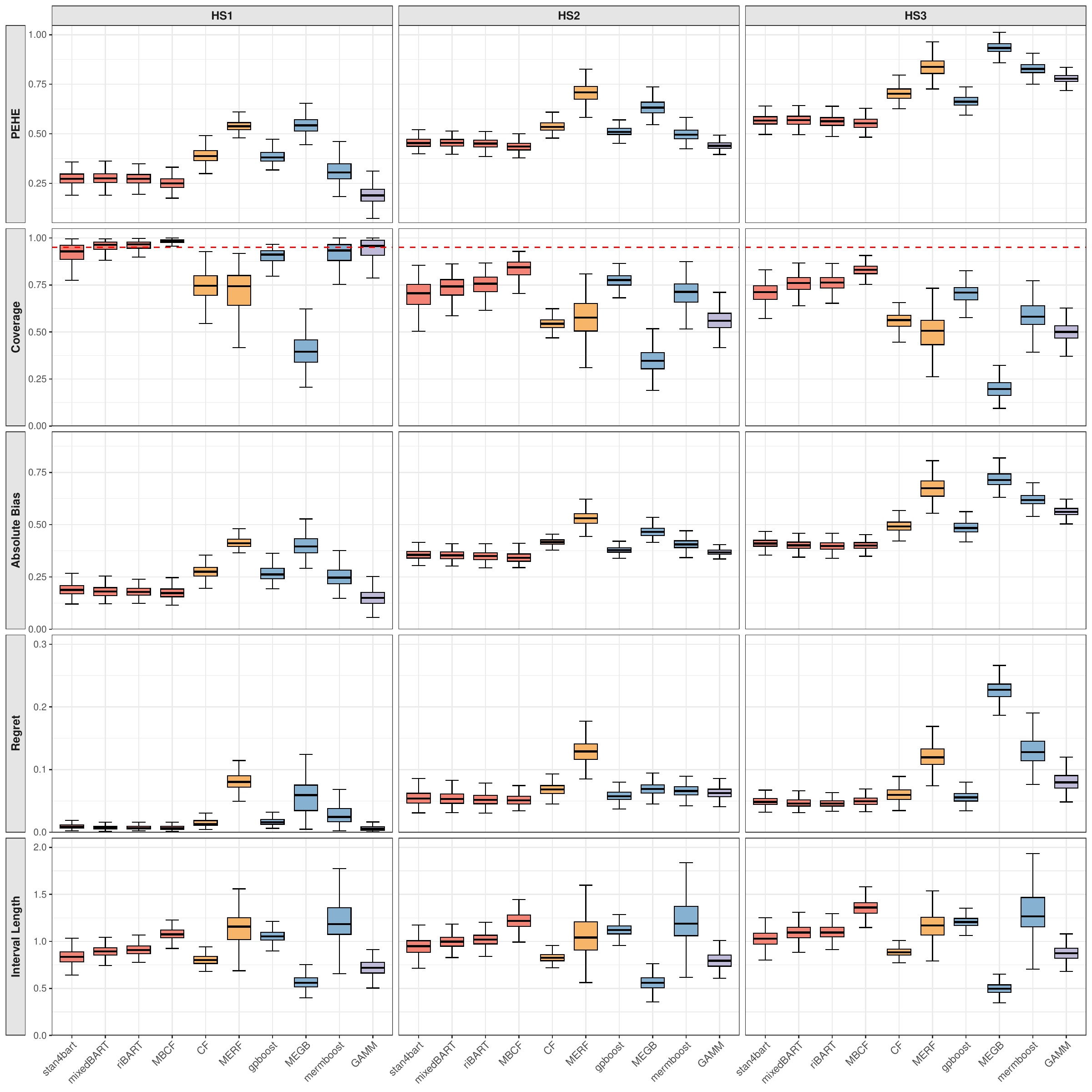}
  \caption{Absolute bias, PEHE, empirical coverage, regret, and average interval length across methods for the setting with \(\mathrm{ICC}=0.01\) and \(I=30\) clusters. Columns correspond to scenarios HS1--HS3. Rows summarize absolute bias, PEHE, empirical coverage, regret, and interval length, respectively. Boxplots summarize Monte Carlo replicates.}
  \label{fig:sim_icc001}
\end{figure}

\begin{figure}[!htbp]
  \centering
  \includegraphics[width=\textwidth,keepaspectratio]{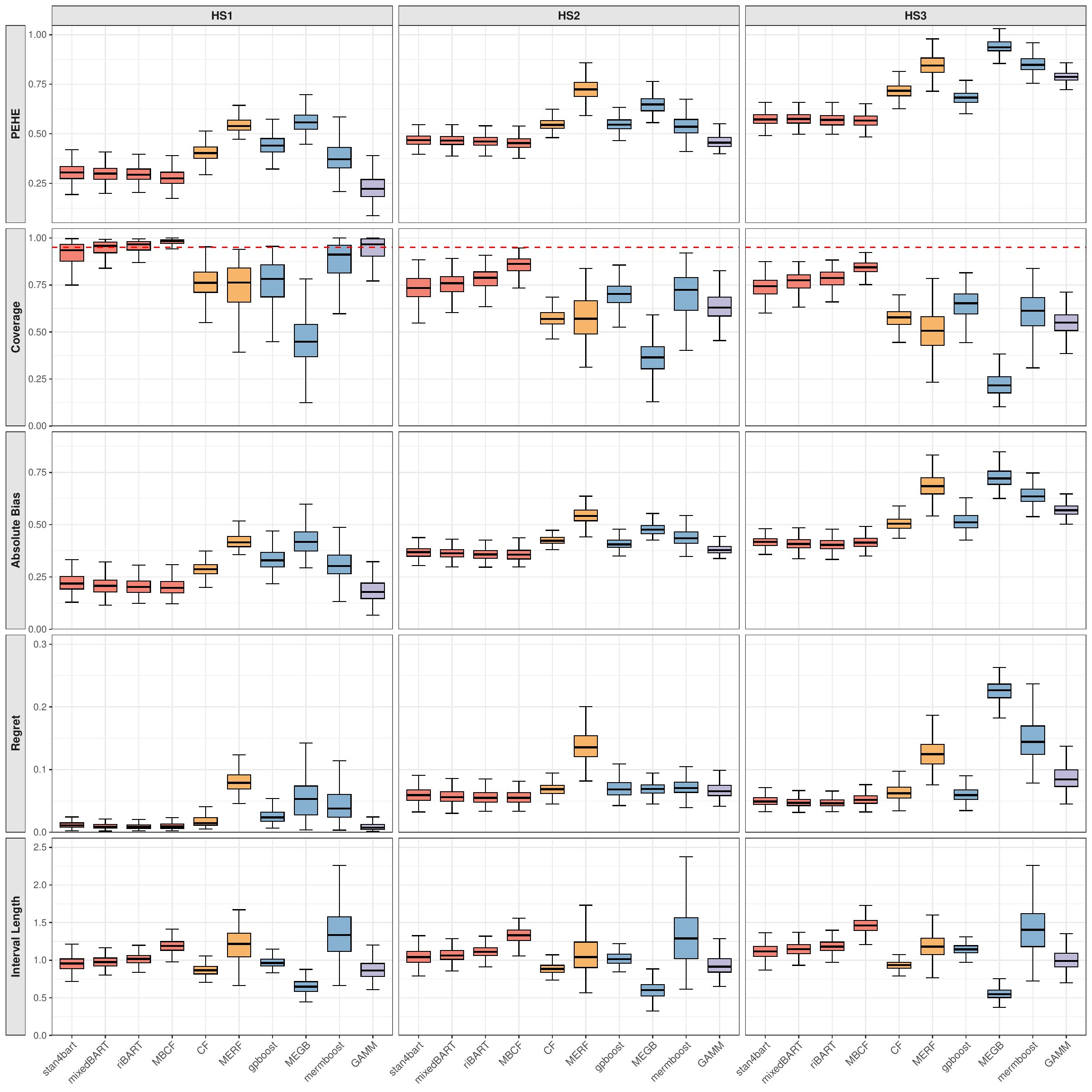}
  \caption{Absolute bias, PEHE, empirical coverage, regret, and average interval length across methods for the setting with \(\mathrm{ICC}=0.05\) and \(I=30\) clusters. Columns correspond to scenarios HS1--HS3. Rows summarize absolute bias, PEHE, empirical coverage, regret, and interval length, respectively. Boxplots summarize Monte Carlo replicates.}
  \label{fig:sim_icc005}
\end{figure}

We further assessed the sensitivity of the simulation findings to lower intraclass correlation settings. Figures~\ref{fig:sim_icc001} and \ref{fig:sim_icc005} present the corresponding results for \(\mathrm{ICC}=0.01\) and \(\mathrm{ICC}=0.05\), respectively. These figures allow us to evaluate whether the comparative performance of the methods changes materially when within-cluster dependence is weaker.

\FloatBarrier

\FloatBarrier

\section{Sensitivity to gamma-distributed random intercepts} \label{supp:sen_gamma}

\begin{figure}[!htbp]
  \centering
  \includegraphics[width=\textwidth,keepaspectratio]{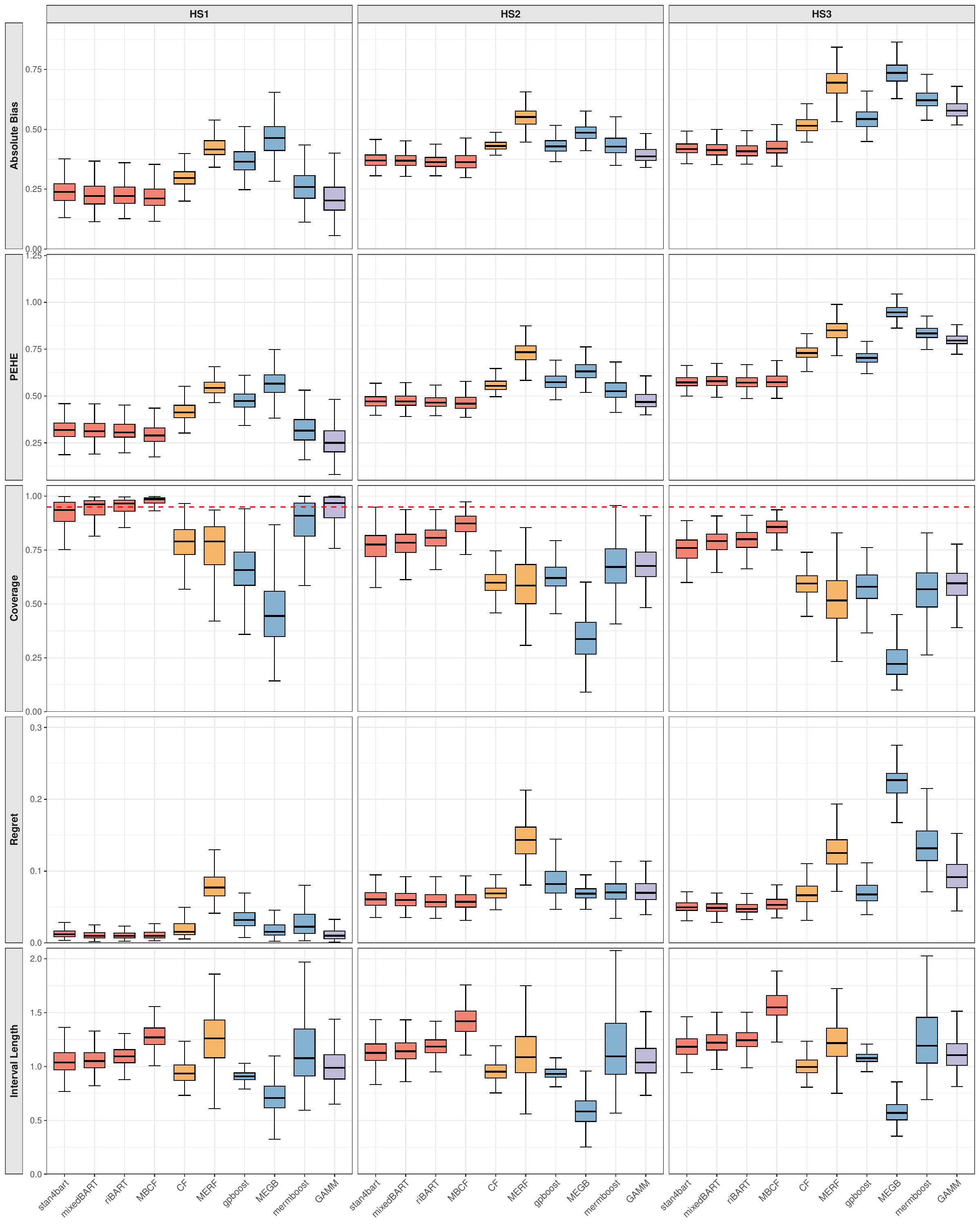}
  \caption{Absolute bias, PEHE, empirical coverage, regret, and average interval length across methods under gamma-distributed random intercepts for the setting with \(\mathrm{ICC}=0.1\) and \(I=30\) clusters. Columns correspond to scenarios HS1--HS3. Rows show absolute bias (top), PEHE (second), empirical coverage of nominal \(95\%\) intervals (third), expected regret (fourth), and average interval length (bottom). Boxplots summarize Monte Carlo replicates, and the red dashed line marks the nominal 0.95 frequentist coverage level.}
  \label{fig:sim_gamma}
\end{figure}

To assess robustness to departures from the normal random-intercept assumption, we repeated the main simulation setting with \(I=30\) clusters and \(\mathrm{ICC}=0.1\), but generated cluster-specific random intercepts from a gamma distribution. All other data-generating mechanisms and analysis settings remained identical to those in the main text. Figure~\ref{fig:sim_gamma} summarizes absolute bias, PEHE, empirical coverage, regret, and interval length across methods under HS1--HS3 in this sensitivity analysis.

\FloatBarrier

\section{Sensitivity to high-dimensional cases} \label{supp::sen::high}

We assess robustness of our findings under three high-dimensional data-generating mechanisms (DGMs) that vary the number and type of covariates while preserving the cluster-randomized structure and the functional forms of the baseline outcome \(f_0(\cdot)\) and the conditional treatment effect \(\tau(\cdot)\). In all DGMs, clusters are the unit of randomization and outcomes are continuous.

Across settings we fix the number of clusters to \(I=30\) and draw cluster sizes \(\{N_i\}_{i=1}^I\) from a discrete uniform distribution on integers \(\{n_{\min},\ldots,n_{\max}\}\) with \(n_{\min}=\lfloor 0.5\cdot 3000/I\rfloor\) and \(n_{\max}=\lceil 1.5\cdot 3000/I\rceil\), ensuring \(I\cdot \mathbb{E}[N_i]=3000\). Treatment is assigned at the cluster level with balanced \(1{:}1\) allocation.
Potential outcomes follow
\[
Y_{ij}(0) \;=\; f_0(\bm X_{ij},\bm V_i) + b_i + \varepsilon_{ij},
\qquad
Y_{ij}(1) \;=\; f_0(\bm X_{ij},\bm V_i) + \tau(\bm X_{ij},\bm V_i) + b_i + \varepsilon_{ij},
\]
where \(b_i \overset{\text{i.i.d.}}{\sim} \mathcal N(0,\sigma_b^2)\) denotes a cluster random effect and \(\varepsilon_{ij} \overset{\text{i.i.d.}}{\sim} \mathcal N(0,\sigma^2)\) an individual error. For a specified total variance \(\sigma_b^2+\sigma^2\) and intracluster correlation \(\mathrm{ICC}\), we set \(\sigma_b^2=\mathrm{ICC}\times(\sigma_b^2+\sigma^2)\) and \(\sigma^2=(1-\mathrm{ICC})\times(\sigma_b^2+\sigma^2)\).
\vspace{0.5em}

\paragraph{DGM with 20 covariates.}
We consider \(4\) cluster-level covariates \(\bm V_i=(V_{i1},\ldots,V_{i4})\) and \(16\) individual-level covariates \(\bm X_{ij}=(X_{ij1},\ldots,X_{ij16})\) with
\[
V_{i1},V_{i2}\overset{\text{i.i.d.}}{\sim}\mathcal N(0,1), \quad
V_{i3},V_{i4}\overset{\text{i.i.d.}}{\sim}\mathrm{Bernoulli}(1/2),
\]
\[
X_{ij1},\ldots,X_{ij8}\overset{\text{i.i.d.}}{\sim}\mathcal N(0,1), \quad
X_{ij9},\ldots,X_{ij16}\overset{\text{i.i.d.}}{\sim}\mathrm{Bernoulli}(1/2).
\]
We fix the total outcome variance to \(1.5\) with \(\mathrm{ICC}=0.1\). The baseline function and conditional treatment effect are
\[
f_0(\bm X_{ij},\bm V_i)
= \sin(\pi X_{ij1}X_{ij2})
+ \log\!\big(\lvert X_{ij5}+V_{i3}\rvert\big)
+ 0.3\,X_{ij10}
- 0.7\,X_{ij15},
\]
\[
\tau(\bm X_{ij},\bm V_i)
= 0.5
- \frac{0.8}{1+\exp\!\big(-(X_{ij1}+X_{ij2})\big)}
+ 0.4\,\sin(\pi X_{ij6}X_{ij8})
+ 0.3\,X_{ij10}
+ 0.7\,\log\!\big(\lvert 0.5 + X_{ij16}\rvert\big)
- 0.6\,V_{i2}
+ 0.2\,V_{i3}.
\]

\paragraph{DGM with 50 covariates.}
We consider \(8\) cluster-level covariates \(\bm V_i=(V_{i1},\ldots,V_{i8})\) and \(42\) individual-level covariates \(\bm X_{ij}=(X_{ij1},\ldots,X_{ij42})\) with
\[
V_{i1},\ldots,V_{i4}\overset{\text{i.i.d.}}{\sim}\mathcal N(0,1), \quad
V_{i5},\ldots,V_{i8}\overset{\text{i.i.d.}}{\sim}\mathrm{Bernoulli}(1/2),
\]
\[
X_{ij1},\ldots,X_{ij21}\overset{\text{i.i.d.}}{\sim}\mathcal N(0,1), \quad
X_{ij22},\ldots,X_{ij42}\overset{\text{i.i.d.}}{\sim}\mathrm{Bernoulli}(1/2).
\]
We fix the total outcome variance to \(2\) with \(\mathrm{ICC}=0.1\). The baseline function and conditional treatment effect are
\[
\begin{aligned}
f_0(\bm X_{ij},\bm V_i)
&= \sin(\pi X_{ij1}X_{ij2})
+ \log\!\big(\lvert X_{ij5}+V_{i6}\rvert\big)
+ 0.3\,X_{ij16}
- \sqrt{\lvert X_{ij20}-X_{ij30}\rvert}
+ 0.4\,X_{ij36}^2
- 0.5\,V_{i2},
\end{aligned}
\]
\[
\begin{aligned}
\tau(\bm X_{ij},\bm V_i)
&= 0.7
+ 0.5\,\sin(\pi X_{ij5}X_{ij6})
+ 0.2\,X_{ij13}
- 0.3\,X_{ij17}^2
- \frac{0.6}{1+\exp\!\big(-(X_{ij18}+X_{ij20})\big)}
- 0.4\,\cos(X_{ij21}) \\
&\qquad
+ 0.7\,X_{ij30}
+ 0.6\,\log\!\big(\lvert 0.5 + X_{ij42}\rvert\big)
- 0.6\,V_{i2}
+ 0.4\,V_{i6}.
\end{aligned}
\]

\paragraph{DGM with 100 covariates.}
We consider \(10\) cluster-level covariates \(\bm V_i=(V_{i1},\ldots,V_{i10})\) and \(90\) individual-level covariates \(\bm X_{ij}=(X_{ij1},\ldots,X_{ij90})\) with
\[
V_{i1},\ldots,V_{i5}\overset{\text{i.i.d.}}{\sim}\mathcal N(0,1), \quad
V_{i6},\ldots,V_{i10}\overset{\text{i.i.d.}}{\sim}\mathrm{Bernoulli}(1/2),
\]
\[
X_{ij1},\ldots,X_{ij50}\overset{\text{i.i.d.}}{\sim}\mathcal N(0,1), \quad
X_{ij51},\ldots,X_{ij90}\overset{\text{i.i.d.}}{\sim}\mathrm{Bernoulli}(1/2).
\]
We fix the total outcome variance to \(2.5\) with \(\mathrm{ICC}=0.1\). The baseline function and conditional treatment effect are
\[
\begin{aligned}
f_0(\bm X_{ij},\bm V_i)
&= \sin(\pi X_{ij1}X_{ij2})
+ \log\!\big(\lvert X_{ij5}+V_{i6}\rvert\big)
+ 0.3\,X_{ij10}
- 0.7\,X_{ij15}
- \sqrt{\lvert X_{ij20}-X_{ij60}\rvert} \\
&\qquad
+ 0.4\,X_{ij25}^2
+ 0.5\,X_{ij60}
+ 0.8\,X_{ij62}
- \log\!\big(1 + X_{ij70}\big),
\end{aligned}
\]
\[
\begin{aligned}
\tau(\bm X_{ij},\bm V_i)
&= 1.5
- \frac{0.8}{1+\exp\!\big(-(X_{ij1}+X_{ij2})\big)}
- 0.4\,X_{ij6}^2
- \cos(X_{ij22})
+ 0.3\,\sin(\pi X_{ij30}X_{ij34})
+ 0.1\,X_{ij36}
- 0.2\,X_{ij50}^2 \\
&\qquad
+ 0.7\,X_{ij60}
- 0.5\,X_{ij65}
+ 0.8\,\log\!\big(\lvert 1 + X_{ij80}\rvert\big)
- 0.6\,V_{i2}
- 0.2\,\exp\!\big(V_{i5}\big)
+ 0.5\,V_{i6}.
\end{aligned}
\]

\paragraph{Signal-to-total variance ratio.}
For each DGM we summarize the strength of the systematic signal relative to total outcome variability via
\[
\mathrm{STR}
= \frac{\operatorname{Var}\!\big(f_0(\bm X_{ij},\bm V_i) + A_i\,\tau(\bm X_{ij},\bm V_i) + b_i\big)}
       {\operatorname{Var}\!\big(f_0(\bm X_{ij},\bm V_i) + A_i\,\tau(\bm X_{ij},\bm V_i) + b_i + \varepsilon_{ij}\big)}\,.
\]
We control the STR around 0.65.

\begin{figure}[!htbp]
  \centering
  \includegraphics[width=\textwidth,keepaspectratio]{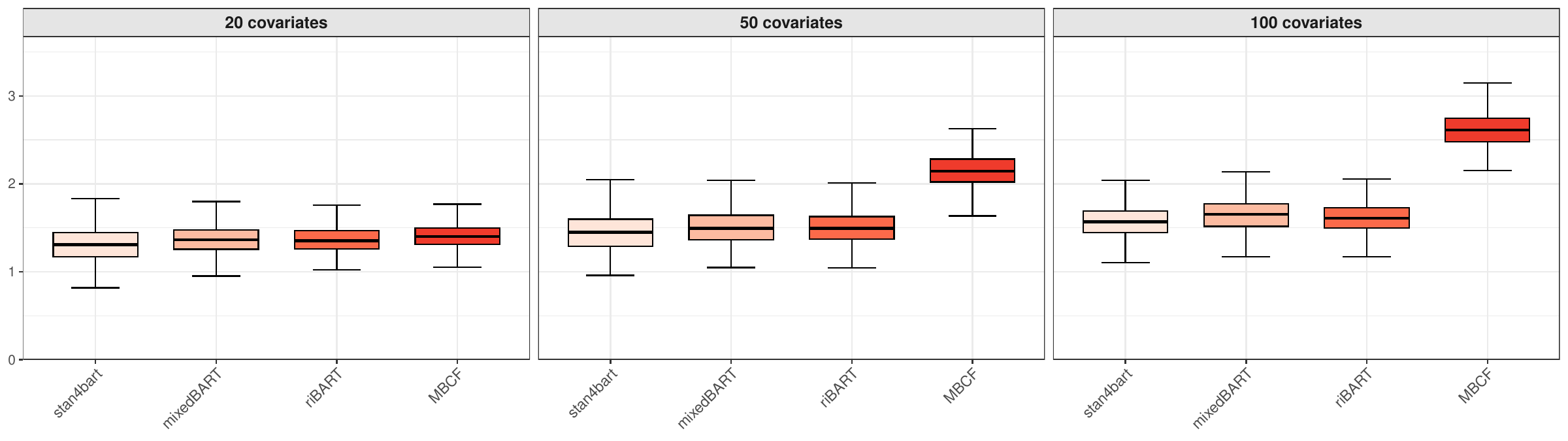}
  \caption{Average interval length across the four BART-based methods in the high-dimensional sensitivity analysis. Columns correspond to 20, 50, and 100 covariates. Boxplots summarize Monte Carlo replicates. Larger values indicate wider interval estimates.}
  \label{fig:hd_interval_length}
\end{figure}

The corresponding results for absolute bias, PEHE, empirical coverage, and regret are reported in the main text. Figure~\ref{fig:hd_interval_length} supplements those results by presenting interval length across the four BART-based methods under the three high-dimensional settings.

\FloatBarrier

\subsection{Comparison of mixedBART and mixedDART in high-dimensional settings}
\label{supp:mxdart}

To further examine the role of sparsity-aware regularization in high-dimensional settings, we conducted an additional comparison between mixedBART and mixedDART, where mixedDART denotes the mixedBART model fitted with the sparse Dirichlet prior on splitting probabilities. Across all three high-dimensional settings, mixedDART generally improved upon mixedBART in terms of absolute bias, PEHE, empirical coverage, and regret, with the improvement becoming more pronounced as the number of covariates increased from 20 to 50 and 100. In particular, mixedDART showed better protection against the inclusion of noise variables, leading to more accurate CATE estimation and coverage closer to the nominal level in the higher-dimensional settings. These gains were accompanied by somewhat wider interval lengths, especially when the number of covariates was large. Overall, these findings support the use of sparsity-aware priors, such as the Dirichlet prior implemented in \texttt{mxbart}, to improve the robustness of mixed-effects BART methods for high-dimensional CATE estimation in CRTs.

\begin{figure}[!htbp]
  \centering
  \includegraphics[width=\textwidth,keepaspectratio]{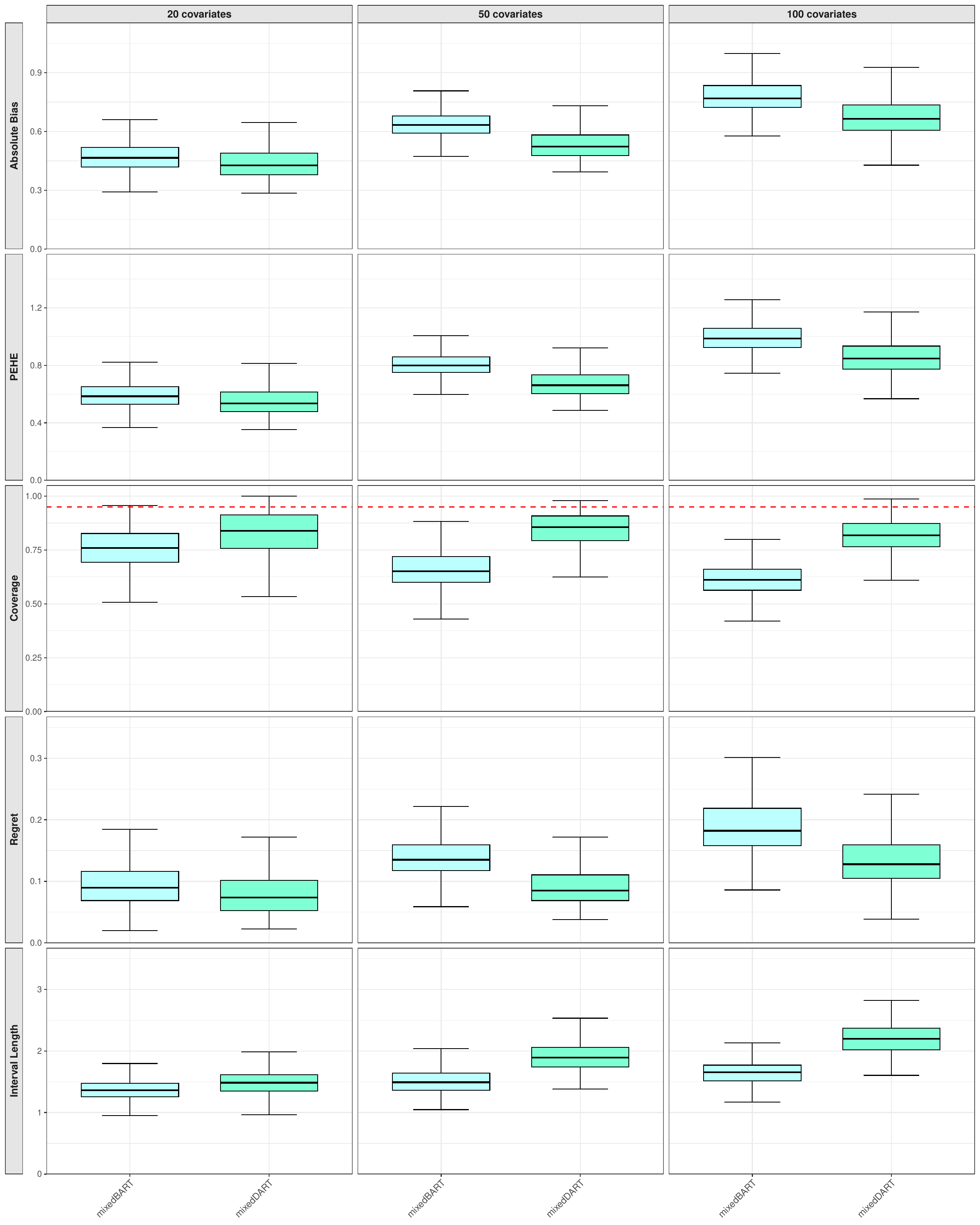}
  \caption{Comparison of mixedBART and mixedDART in the high-dimensional sensitivity analysis. Columns correspond to 20, 50, and 100 covariates. Rows show absolute bias (top), PEHE (second), empirical coverage of nominal \(95\%\) intervals (third), expected regret (fourth), and average interval length (bottom). Boxplots summarize Monte Carlo replicates, and the red dashed line marks the nominal 0.95 frequentist coverage level.}
  \label{fig:mixeddart_hd}
\end{figure}
\FloatBarrier

\section{Additional heterogeneous treatment effect analysis in TASSH} \label{supp:tassh}

This section provides supplementary results for the TASSH application using three additional BART-based methods, namely stan4bart, mixedBART, and riBART. For each method, we report the distribution of the estimated individual-level CATEs for 12-month DBP together with the corresponding leave-one-out TE-VIM results. We then compare the four BART-based methods, including MBCF, in terms of the distribution and agreement of their posterior mean CATE estimates. The estimation and analysis procedure was the same as that used for MBCF in the main text. Specifically, each method was fitted using 200 trees, 5{,}000 burn-in iterations, and 5{,}000 posterior draws. After estimating the CATEs, we computed the TE-VIM values and selected the six most important covariates according to TE-VIM to fit a classification and regression tree for exploring potential effect modifiers.

\subsection{Three-level CART for MBCF}
\label{supp:MBCF_cart_3}

As an additional exploratory analysis, we fitted a three-level classification and regression tree (CART) to the posterior mean CATE estimates from the MBCF model. Compared with the two-level tree reported in the main text, the three-level tree allows a more detailed subgroup structure. In this expanded tree, employment status enters as an additional splitting variable, although the resulting terminal-node summaries should still be interpreted cautiously as exploratory subgroup descriptions rather than confirmatory findings.

\begin{figure}[htbp]
  \centering
  \includegraphics[width=0.8\textwidth,keepaspectratio]{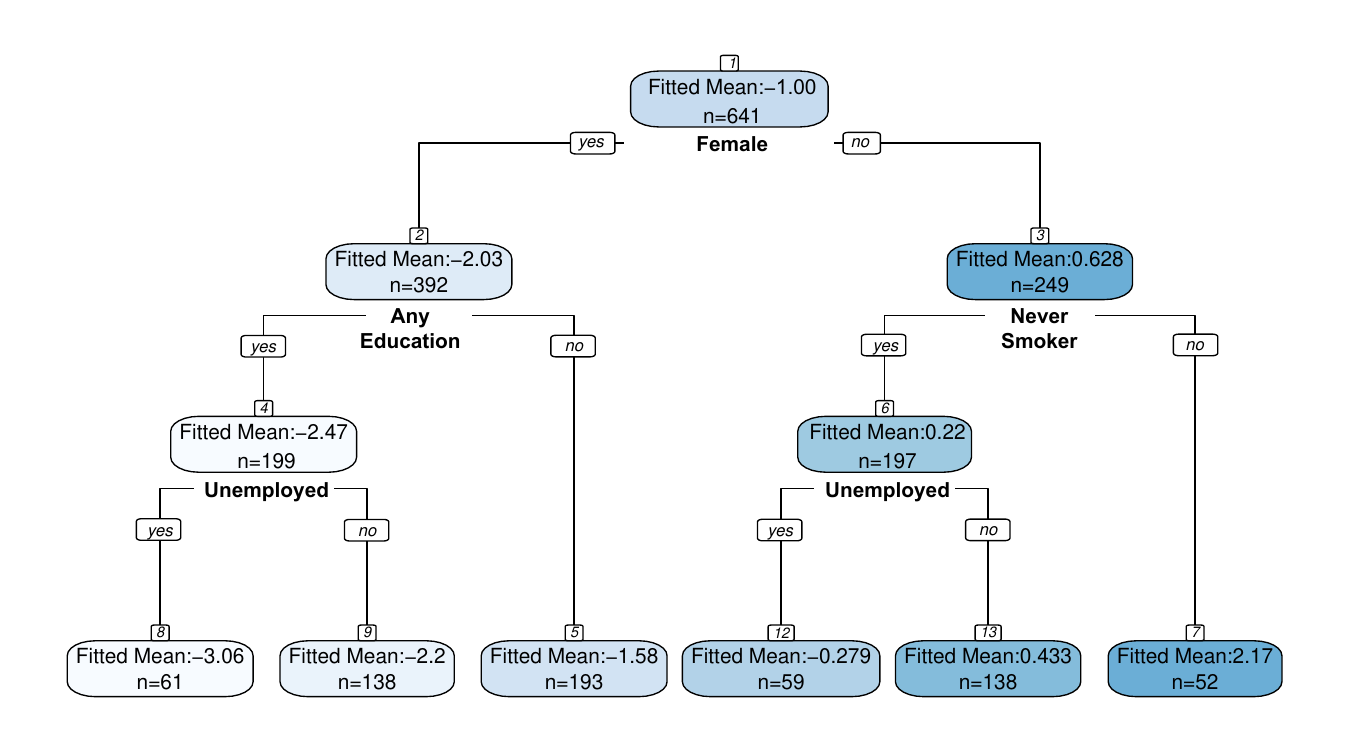}
  \caption{Three-level CART for posterior mean CATE on DBP at 12 months based on the MBCF model. The tree regresses participant-specific posterior mean CATE estimates on selected baseline covariates to provide a more detailed exploratory summary of treatment effect heterogeneity between TASSH plus HIC and HIC alone. Internal nodes show splitting variables and cut points, and terminal nodes report the node-specific mean CATE and sample size. Negative values indicate lower 12-month DBP under TASSH plus HIC relative to HIC alone. Compared with the two-level tree in the main text, this more elaborate tree additionally includes employment status as a splitting variable.}
  \label{fig:dbp_MBCF_cart2}
\end{figure}

\subsection{Results using stan4bart}
\label{supp:TASSH_stan4bart}

\begin{figure}[htbp]
  \centering
  \includegraphics[width=\textwidth,keepaspectratio]{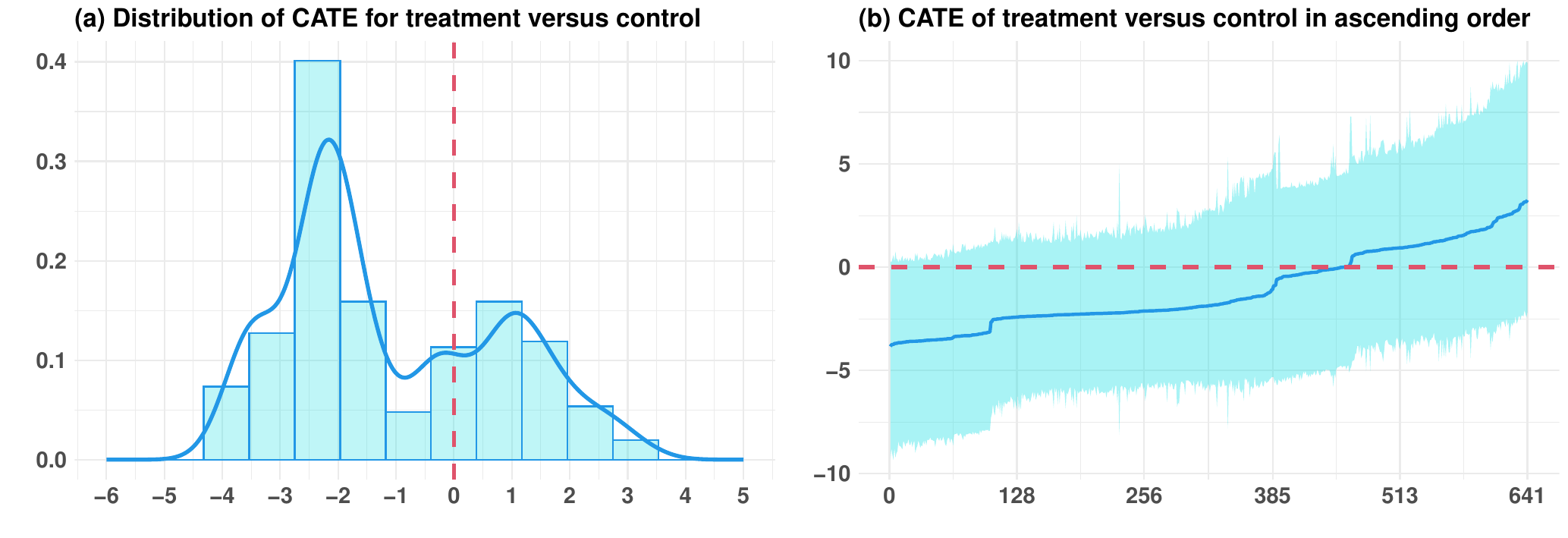}
  \caption{Estimated CATEs for DBP at 12 months using stan4bart.
  Panel (a) shows the posterior distribution of individual CATEs; values below zero indicate lower DBP under the intervention. Panel (b) plots participant-specific CATEs in ascending order; the solid line is the posterior mean and the shaded band is the corresponding 95\% credible interval.}
  \label{fig:dbp_hist_stan4bart}
\end{figure}

\begin{figure}[htbp]
  \centering
  \includegraphics[width=0.8\textwidth,keepaspectratio]{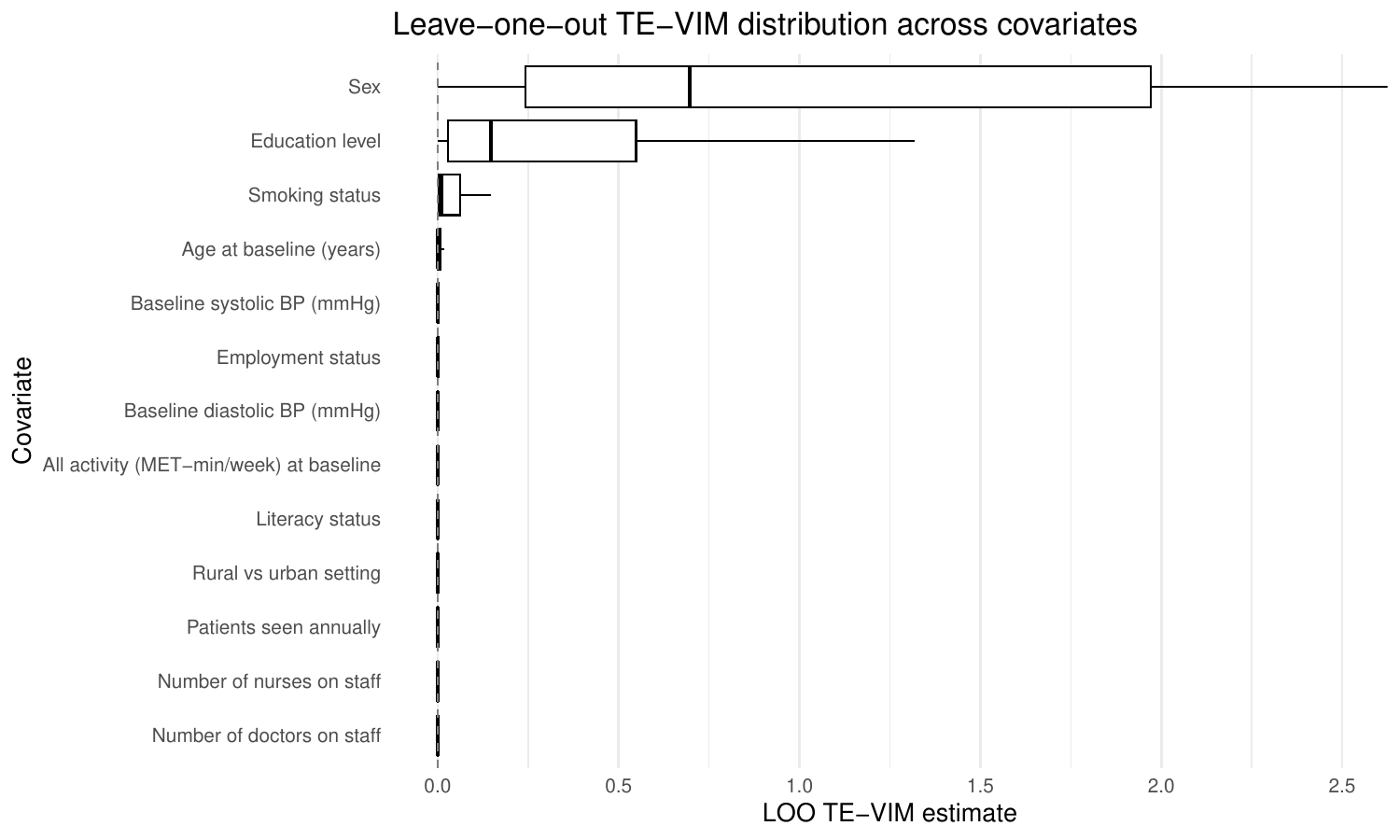}
  \caption{Leave-one-out TE-VIM across baseline covariates using stan4bart. For each covariate \(p\), we first obtained the full-model posterior mean CATE estimates \(\widehat{\tau}(\mathbf{x},\mathbf{v})\), and then regressed these estimates on the reduced covariate set excluding \(p\) using a BART projection model to obtain \(\widetilde{\tau}_{-p}(\mathbf{z}_{-p})\). Individual-level contributions to the TE-VIM were computed as \(\left\{\widehat{\tau}(\mathbf{x},\mathbf{v})-\widetilde{\tau}_{-p}(\mathbf{z}_{-p})\right\}^{2}\), and the boxplots summarize their distribution across participants. Larger values indicate that omitting the covariate leads to poorer prediction of the full-model CATEs and therefore greater importance for treatment-effect heterogeneity. Sex, education level, smoking status and age show the largest TE-VIM values, whereas most other covariates display comparatively small contributions.}
  \label{fig:dbp_tevim_stan4bart}
\end{figure}
\FloatBarrier

\begin{figure}[htbp]
  \centering
  \includegraphics[width=0.8\textwidth,keepaspectratio]{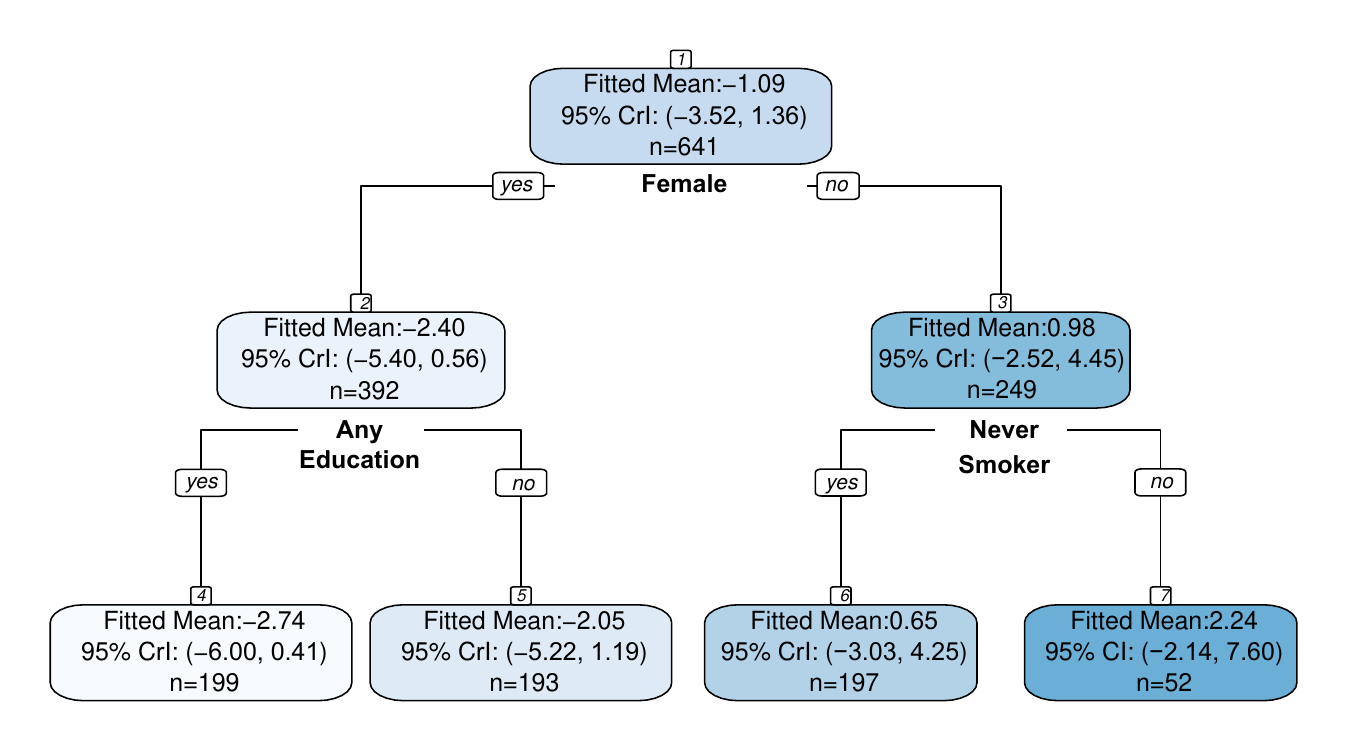}
  \caption{Two-level CART for posterior mean CATE on DBP at 12 months based on the stan4bart model. The tree regresses participant-specific posterior mean CATE estimates on selected baseline covariates to provide a more detailed exploratory summary of treatment effect heterogeneity between TASSH plus HIC and HIC alone.}
  \label{fig:dbp_stan4bart_cart}
\end{figure}

\subsection{Results using mixedBART}
\label{supp:TASSH_mxbart}

\begin{figure}[htbp]
  \centering
  \includegraphics[width=\textwidth,keepaspectratio]{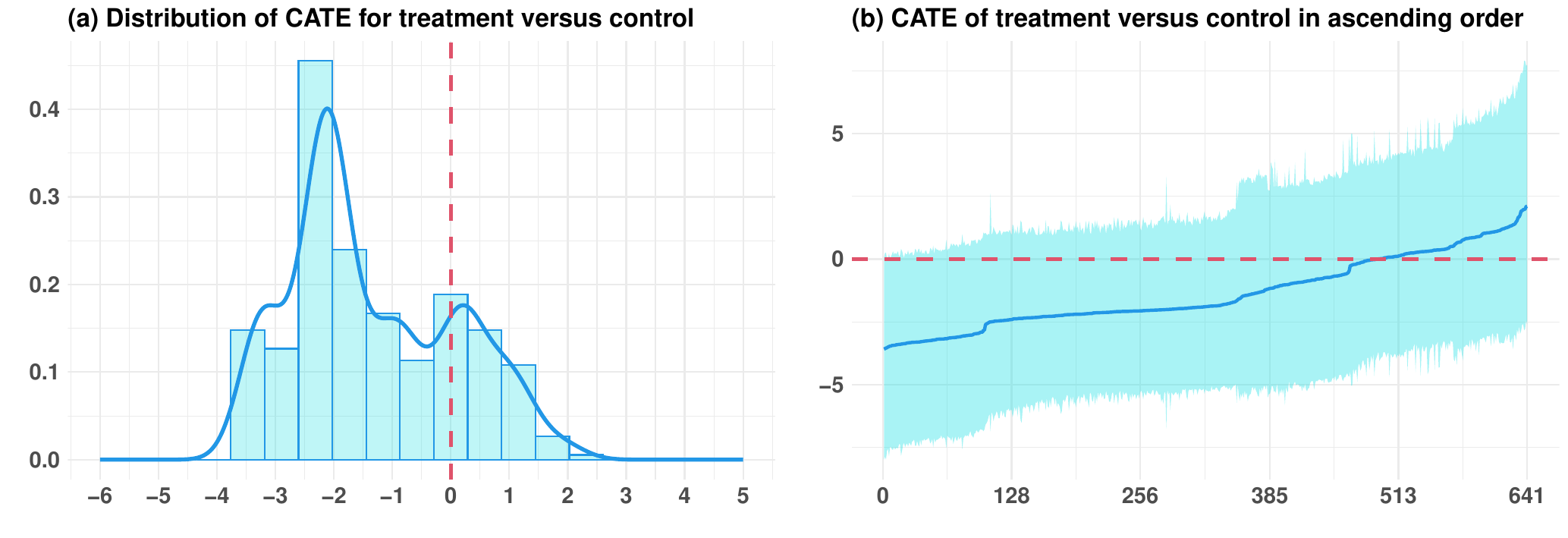}
  \caption{Estimated CATEs for DBP at 12 months using mixedBART.
  Panel (a) shows the posterior distribution of individual CATEs; values below zero indicate lower DBP under the intervention. Panel (b) plots participant-specific CATEs in ascending order; the solid line is the posterior mean and the shaded band is the corresponding 95\% credible interval.}
  \label{fig:dbp_hist_mixedbart}
\end{figure}

\begin{figure}[htbp]
  \centering
  \includegraphics[width=0.8\textwidth,keepaspectratio]{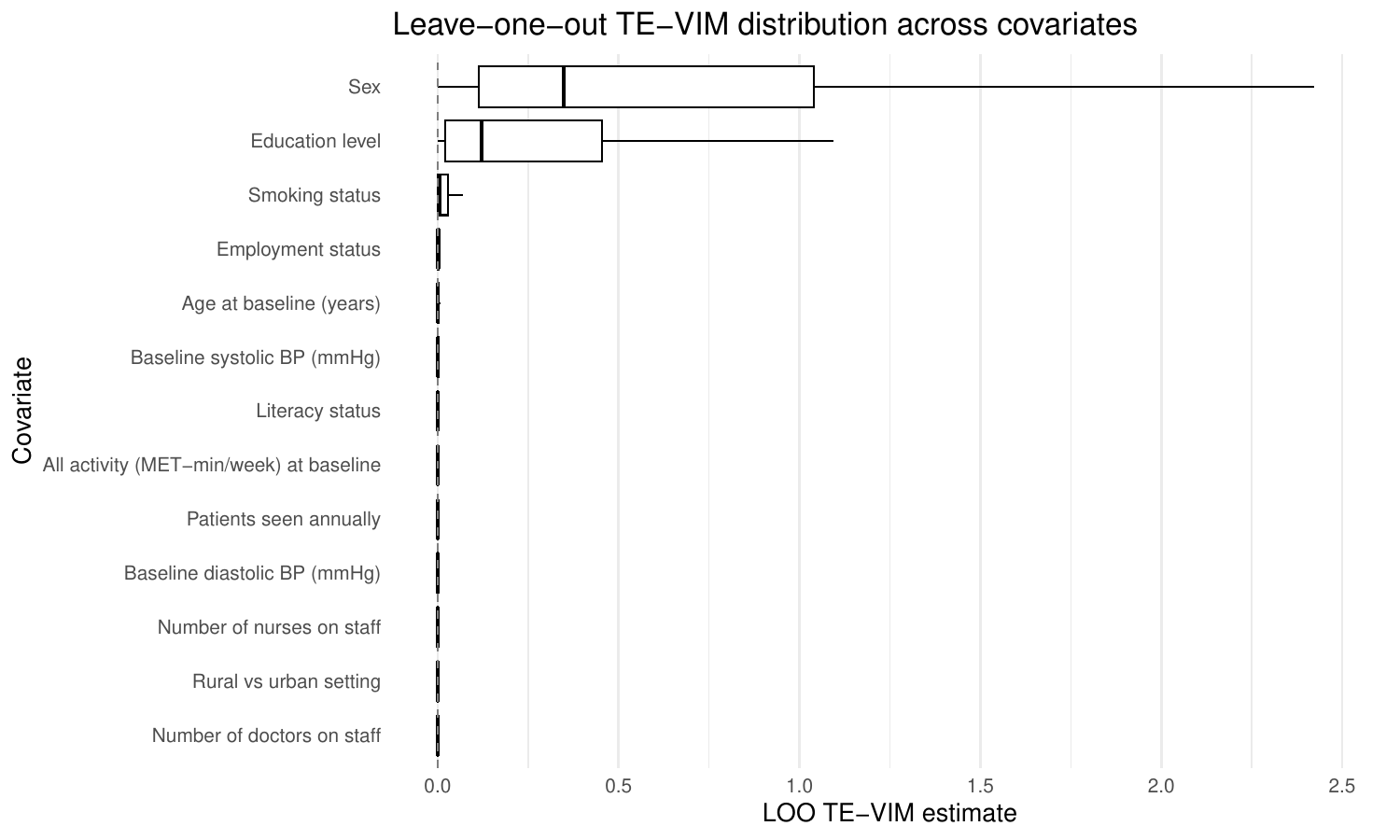}
  \caption{Leave-one-out TE-VIM across baseline covariates using mixedBART. For each covariate \(p\), we first obtained the full-model posterior mean CATE estimates \(\widehat{\tau}(\mathbf{x},\mathbf{v})\), and then regressed these estimates on the reduced covariate set excluding \(p\) using a BART projection model to obtain \(\widetilde{\tau}_{-p}(\mathbf{z}_{-p})\). Individual-level contributions to the TE-VIM were computed as \(\left\{\widehat{\tau}(\mathbf{x},\mathbf{v})-\widetilde{\tau}_{-p}(\mathbf{z}_{-p})\right\}^{2}\), and the boxplots summarize their distribution across participants. Larger values indicate that omitting the covariate leads to poorer prediction of the full-model CATEs and therefore greater importance for treatment-effect heterogeneity. Sex, education level, smoking status and employment status show the largest TE-VIM values, whereas most other covariates display comparatively small contributions.}
  \label{fig:dbp_tevim_mixedbart}
\end{figure}
\FloatBarrier

\begin{figure}[htbp]
  \centering
  \includegraphics[width=0.8\textwidth,keepaspectratio]{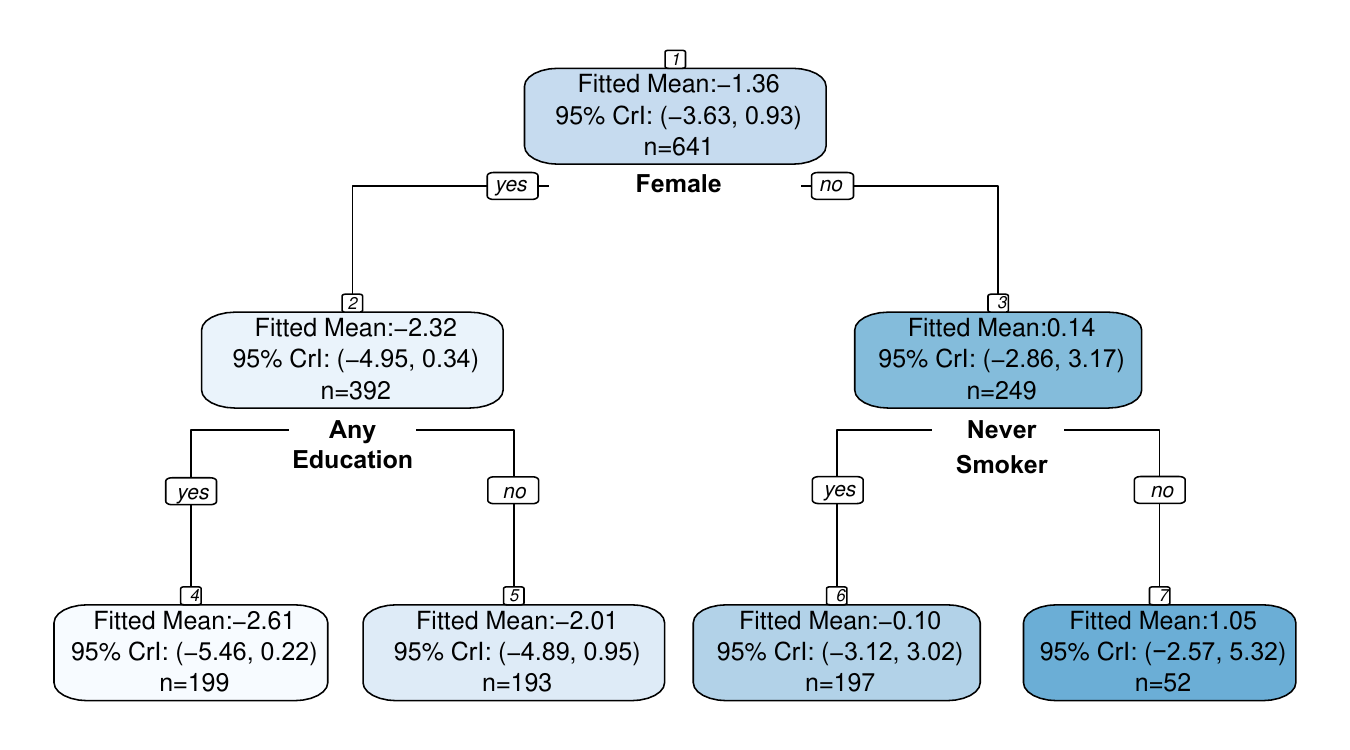}
  \caption{Two-level CART for posterior mean CATE on DBP at 12 months based on the mixedBART model. The tree regresses participant-specific posterior mean CATE estimates on selected baseline covariates to provide a more detailed exploratory summary of treatment effect heterogeneity between TASSH plus HIC and HIC alone.}
  \label{fig:dbp_mxbart_cart}
\end{figure}
\FloatBarrier

\subsection{Results using riBART}
\label{supp:TASSH_ribart}

\begin{figure}[htbp]
  \centering
  \includegraphics[width=\textwidth,keepaspectratio]{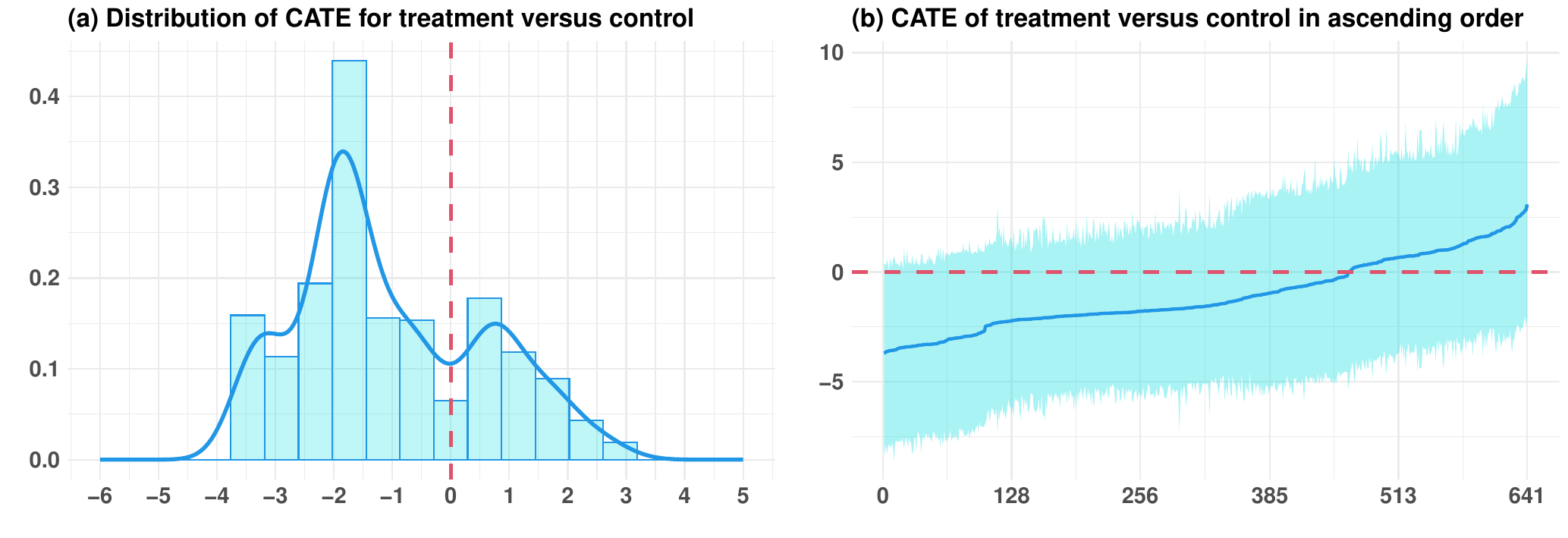}
  \caption{Estimated CATEs for DBP at 12 months using riBART.
  Panel (a) shows the posterior distribution of individual CATEs; values below zero indicate lower DBP under the intervention. Panel (b) plots participant-specific CATEs in ascending order; the solid line is the posterior mean and the shaded band is the corresponding 95\% credible interval.}
  \label{fig:dbp_hist_ribart}
\end{figure}

\begin{figure}[htbp]
  \centering
  \includegraphics[width=0.8\textwidth,keepaspectratio]{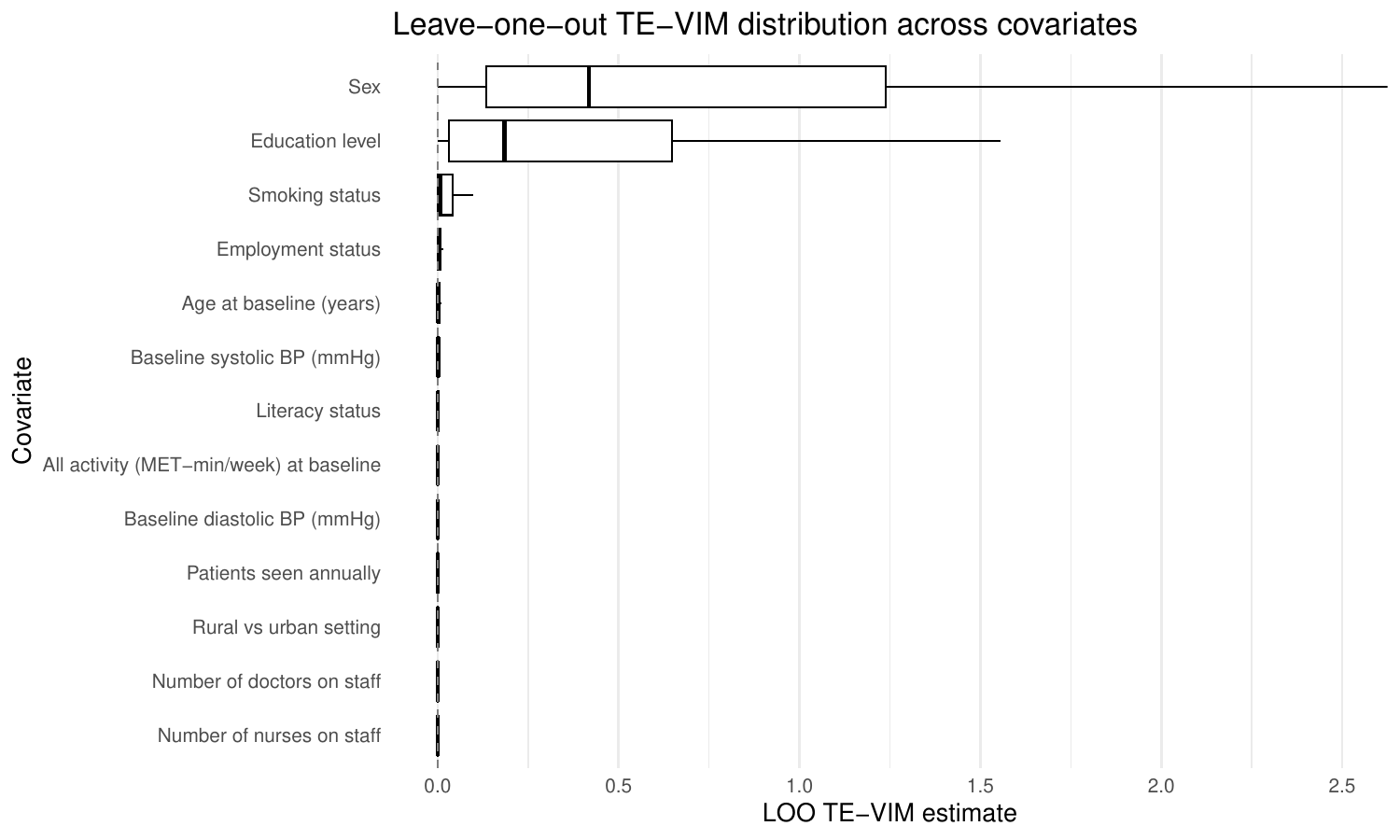}
  \caption{Leave-one-out TE-VIM across baseline covariates using riBART. For each covariate \(p\), we first obtained the full-model posterior mean CATE estimates \(\widehat{\tau}(\mathbf{x},\mathbf{v})\), and then regressed these estimates on the reduced covariate set excluding \(p\) using a BART projection model to obtain \(\widetilde{\tau}_{-p}(\mathbf{z}_{-p})\). Individual-level contributions to the TE-VIM were computed as \(\left\{\widehat{\tau}(\mathbf{x},\mathbf{v})-\widetilde{\tau}_{-p}(\mathbf{z}_{-p})\right\}^{2}\), and the boxplots summarize their distribution across participants. Larger values indicate that omitting the covariate leads to poorer prediction of the full-model CATEs and therefore greater importance for treatment-effect heterogeneity. Sex, education level, smoking status and employment status show the largest TE-VIM values, whereas most other covariates display comparatively small contributions.}
  \label{fig:dbp_tevim_ribart}
\end{figure}
\FloatBarrier

\begin{figure}[htbp]
  \centering
  \includegraphics[width=0.8\textwidth,keepaspectratio]{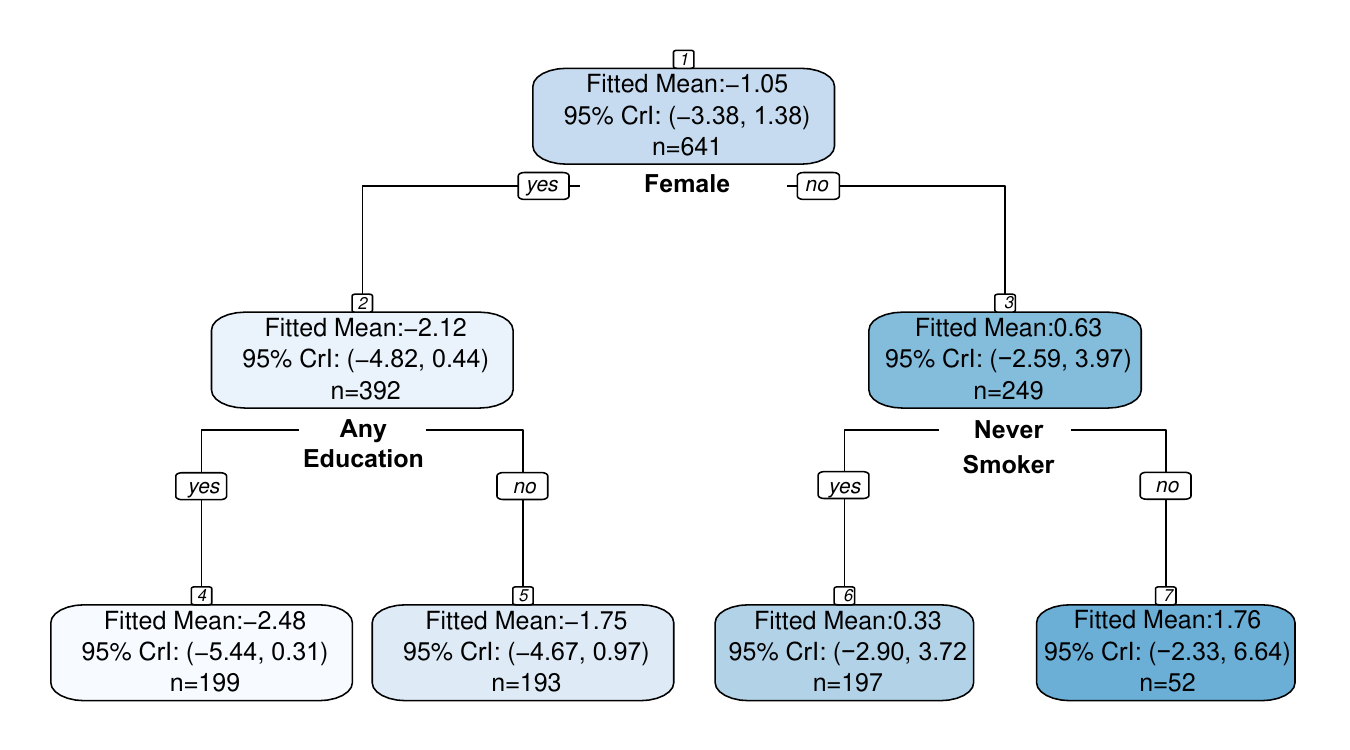}
  \caption{Two-level CART for posterior mean CATE on DBP at 12 months based on the riBART model. The tree regresses participant-specific posterior mean CATE estimates on selected baseline covariates to provide a more detailed exploratory summary of treatment effect heterogeneity between TASSH plus HIC and HIC alone.}
  \label{fig:dbp_rbart_cart}
\end{figure}
\FloatBarrier

\subsection{Comparison across the four BART-based methods}
\label{supp:TASSH_comparison}

\begin{figure}[htbp]
  \centering
  \includegraphics[width=\textwidth,keepaspectratio]{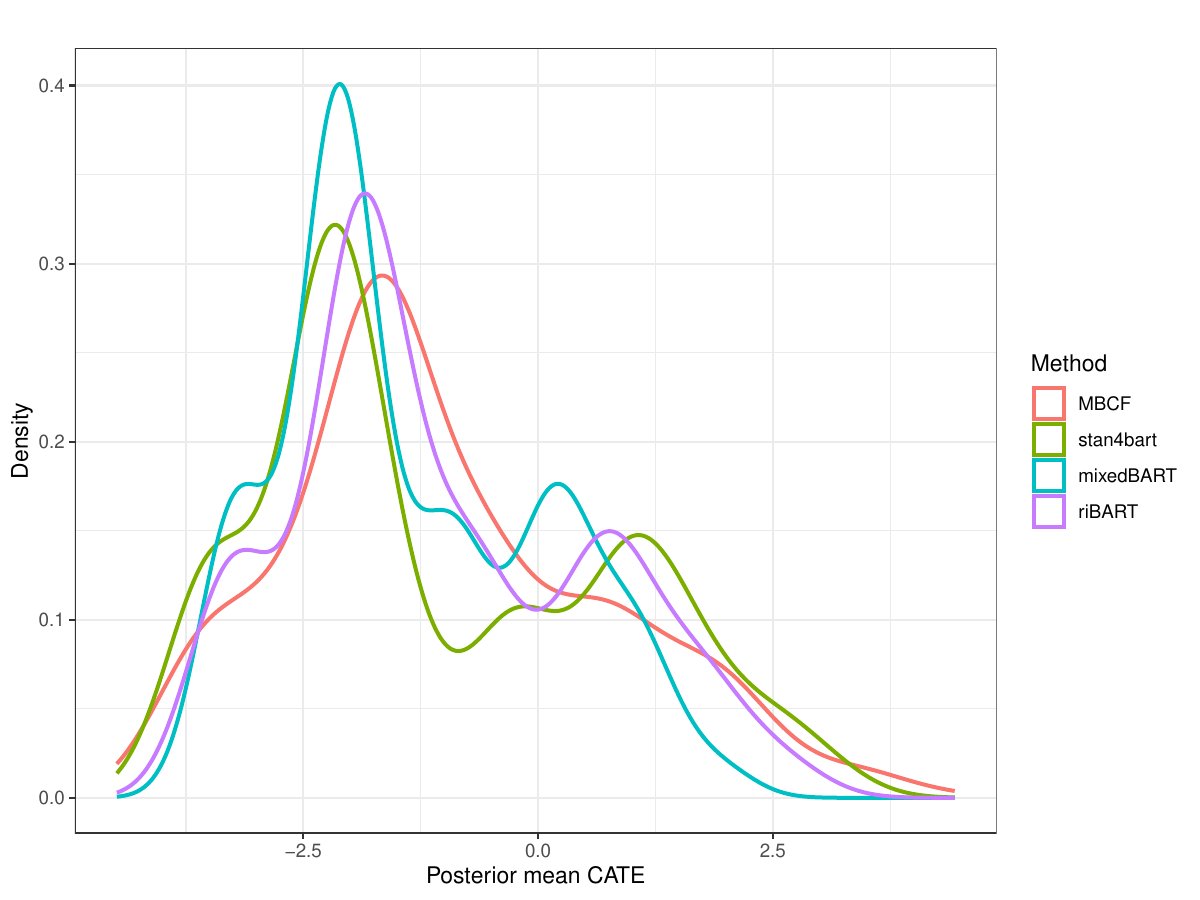}
  \caption{Distribution of participant-specific posterior mean CATE estimates across the four BART-based methods in the TASSH analysis for 12-month DBP. The overlaid density curves show that the marginal distributions of estimated CATEs are highly similar across methods.}
  \label{fig:cate_density_by_method}
\end{figure}

Table \ref{tab:corr_bart_methods_dbp} and Figure \ref{fig:cate_density_by_method} summarize the agreement of posterior mean CATE estimates across the four BART-based methods. The correlations quantify pairwise agreement, and the overlaid density plot compares the marginal distributions of the participant-specific posterior mean CATEs across methods. Overall, the four BART-based methods yielded very similar posterior mean CATE estimates in the TASSH analysis. As shown in Table \ref{tab:corr_bart_methods_dbp}, all pairwise Pearson and Spearman correlations exceeded 0.95, indicating strong agreement in both the magnitudes and rankings of the participant-specific posterior mean CATEs across methods. A notable pattern is that the three mixed-effects BART methods, namely stan4bart, mixedBART, and riBART, were even more similar to one another, with pairwise correlations around 0.99. By comparison, their correlations with MBCF were slightly lower, although still very high. Figure \ref{fig:cate_density_by_method} further shows that the marginal distributions of the estimated posterior mean CATEs were highly similar across the four methods. Taken together, these results suggest that the substantive conclusions from the TASSH application are highly robust across the four BART-based approaches.

\begin{table}[htbp]
\centering
\caption{Pairwise Pearson and Spearman correlations of participant-specific posterior mean CATE estimates across four BART-based methods in the TASSH analysis for 12-month DBP}
\label{tab:corr_bart_methods_dbp}
\begin{tabular}{lcccc}
\hline
\multicolumn{5}{c}{Panel A. Pearson correlation} \\
\hline
           & MBCF  & stan4bart & mixedBART & riBART \\
\hline
MBCF       & 1.000 & 0.951     & 0.967     & 0.979  \\
stan4bart  & 0.951 & 1.000     & 0.991     & 0.992  \\
mixedBART  & 0.967 & 0.991     & 1.000     & 0.993  \\
riBART     & 0.979 & 0.992     & 0.993     & 1.000  \\
\hline
\multicolumn{5}{c}{Panel B. Spearman correlation} \\
\hline
           & MBCF  & stan4bart & mixedBART & riBART \\
\hline
MBCF       & 1.000 & 0.955     & 0.967     & 0.977  \\
stan4bart  & 0.955 & 1.000     & 0.985     & 0.992  \\
mixedBART  & 0.967 & 0.985     & 1.000     & 0.991  \\
riBART     & 0.977 & 0.992     & 0.991     & 1.000  \\
\hline
\end{tabular}
\end{table}

\FloatBarrier 

\clearpage



\section*{Supplementary material}

# Appendix B

## Overview

This appendix provides reproducible R code and a worked data example for estimating conditional average treatment effects (CATEs) and the average treatment effect (ATE) in cluster-randomized trials (CRTs), along with associated uncertainty quantification (credible or confidence intervals, depending on the method). We implement the machine-learning approaches considered in our simulation study: stan4bart, mixedBART, riBART, MBCF, causal forest (CF), MERF, GPBoost, MEGB, mermboost, and GAMM.

For the full simulation design and evaluation metrics, see the main text. The complete codebase is available at https://github.com/changjun-li/crt-hte-ml-continuous.

### 1. stan4bart

Reference paper: https://www.mdpi.com/1099-4300/24/12/1782

R package: stan4bart

### Function inputs

- `df`
  A `data.frame` containing **all** variables used by the function.

  - **Required** columns:
    * Outcome (continuous)

    * Treatment indicator (0/1)

    * All covariates for the BART term

    * Cluster/group ID

- `outcome`
  *(character)* Name of the numeric outcome column in `df` (e.g. `"Y"`).

- `treatment`
  *(character)* Name of the binary treatment column in `df` (0/1, e.g. `"A"`).
  Included **inside** `bart(...)` so BART can learn heterogeneous treatment interactions.

- `covariates`
  *(character vector)* Names of the predictor columns to feed into `bart(...)`, e.g. `c("X1","X2",…,"X9")`.

- `id`
  *(character)* Name of the cluster/group identifier column in `df` (e.g. `"id"`).
  Used to fit a random intercept (`1 | id`).

- `iter`
  *(integer, default 5000)* Positive integer indicating the number of posterior samples to draw and return **per chain**. Includes warmup samples.

- **chains**
  *(integer, default 1)* Positive integer giving the number of Markov Chains to sample..

- **adapt_delta**
  *(numeric, default 0.95)* Target acceptance probability for Stan's HMC sampler.

- **max_treedepth**
  *(integer, default 10)* Maximum tree depth for the No-U-Turn sampler.

- **ntree**
  *(integer, default 200)* Number of trees in the BART ensemble.

```r
stan4bart_HTE <- function(df,
                          outcome,     # e.g. "Y"
                          treatment,   # e.g. "A"
                          covariates,  # e.g. c("X1","X2",...) list of covariates
                          id,          # e.g. "id"
                          iter = 5000L,
                          chains = 1L,
                          adapt_delta = 0.95,
                          max_treedepth = 10,
                          ntree = 200L) {

  stopifnot(all(c(outcome, treatment, id, covariates) %in% names(df)))
  df[[id]]        <- factor(df[[id]])
  df[[treatment]] <- as.integer(df[[treatment]])


  all_terms <- paste(c(treatment, covariates), collapse = " + ")
  fmla_text <- sprintf(
    "%s ~ bart(%s) + (1 | %s)",
    outcome,
    all_terms,
    id
  )
  fmla <- as.formula(fmla_text)

  environment(fmla) <- globalenv()


  args <- list(
    formula   = fmla,
    data      = df,
    test      = NULL,
    weights   = NULL,
    treatment = treatment,
    iter      = iter,
    warmup    = iter %/% 2L,
    chains    = chains,
    verbose   = -1,
    stan_args = list(
      adapt_delta   = adapt_delta,
      max_treedepth = max_treedepth
    ),
    bart_args = list(
```

```r
      ntree     = ntree,
      keepTrees = TRUE
    )
  )
  fit <- do.call(stan4bart, args)

  df0 <- df; df0[[treatment]] <- 0L
  df1 <- df; df1[[treatment]] <- 1L

  y0_post <- predict(fit, newdata = df0, type = "ev")
  y1_post <- predict(fit, newdata = df1, type = "ev")
  ite_post <- y1_post - y0_post

  iCATE    <- rowMeans(ite_post)
  CrI_i    <- t(apply(ite_post, 1, quantile, probs = c(0.025, 0.975)))

  ate_post <- colMeans(ite_post)
  ATE      <- mean(ate_post)
  CrI_ATE  <- quantile(ate_post, probs = c(0.025, 0.975))

  list(
    iCATE     = iCATE,
    iCATE_CrI = CrI_i,
    ATE       = ATE,
    ATE_CrI   = CrI_ATE,
    fit_object = fit,
    ite_post  = ite_post
  )
}


#### Example usage
library(stan4bart)
set.seed(20251229)
start <- Sys.time()
res <- stan4bart_HTE(
  df         = dat_bpi_output,
  outcome    = "Y",
  treatment  = "A",
  covariates = paste0("X", 1:9),
  id         = "id",
  iter       = 10000L,
  chains     = 1L,
  adapt_delta   = 0.95,
  max_treedepth = 50,
  ntree         = 200L
)
end <- Sys.time()
cat("Elapsed time:", as.numeric(end - start), "seconds\n")

res$ATE
res$ATE_CrI

iCATE_CrI_stan4bart <- cbind(
```

```
    iCATE      = res$iCATE,
    Lower_CrI  = res$iCATE_CrI[,1],
    Upper_CrI  = res$iCATE_CrI[,2])
head(iCATE_CrI_stan4bart)
```

## 2. mixedBART

Reference paper: https://pubmed.ncbi.nlm.nih.gov/33751659/

R package: mxbart, bart3, github:https://github.com/rsparapa/bnptools

### Function inputs

- **df**
  A `data.frame` containing in CRT data.
  Must include at least:

  - The **outcome** column (named by `outcome`)

  - The **treatment** indicator (named by `treatment`, values 0/1)

  - All **covariate** columns (named in `covariates`)

  - The **cluster ID** column (named by `id`)

- **outcome**
  *(character)*
  Name of the numeric continuous outcome column in **df**.

- **treatment**
  *(character)*
  Name of the binary treatment column in **df** (must be encoded as 0/1).

- **covariates**
  *(character vector)*
  Names of the predictor columns to feed into the nonparametric BART term.

- **id**
  *(character)*
  Name of the cluster/group identifier column in **df**.
  Used to fit a random intercept for each cluster.

- **ntree**
  *(integer, default 200)*
  Number of trees in the BART ensemble.

- **nskip**
  *(integer, default 5000)*
  Number of "burn-in" MCMC iterations to discard before collecting samples.

- **ndpost**
  *(integer, default 5000)*
  Number of posterior samples to collect **after** burn-in.

```

- **keepevery**
  *(integer, default 1)*
  Thinning interval: keep every **keepevery**-th draw (1 = no thinning).

```r
mxbart_HTE <- function(df,
                       outcome,         # outcome Y
                       treatment,       # treatment (character, 0/1)
                       covariates,      #covariates (character vector)
                       id,              # cluster ID column names (character)
                       ntree    = 200L, # BART trees
                       nskip    = 5000L, # MCMC burn-in
                       ndpost   = 5000L, # MCMC posterior sample
                       keepevery = 1L
) {
  stopifnot(all(c(outcome, treatment, id, covariates) %in% names(df)))

  df        <- df[order(df[[id]]), ]
  y.train   <- df[[outcome]]
  x.train   <- as.matrix(df[, c(treatment, covariates)])
  id.train  <- as.integer(df[[id]])
  n         <- nrow(df)

  df0       <- df; df0[[treatment]] <- 0L
  df1       <- df; df1[[treatment]] <- 1L
  x0        <- as.matrix(df0[, c(treatment, covariates)])
  x1        <- as.matrix(df1[, c(treatment, covariates)])
  x.test    <- rbind(x0, x1)

  fit <- mxbart(
    x.train   = x.train,
    y.train   = y.train,
    id.train  = id.train,
    x.test    = x.test,
    ntree     = ntree,
    nskip     = nskip,
    ndpost    = ndpost,
    keepevery = keepevery
  )

  posterior <- fit$fhat.test
  y0_post   <- posterior[,        1:n]
  y1_post   <- posterior[, (n+1):(2*n)]

  ite_post  <- y1_post - y0_post

  iCATE     <- colMeans(ite_post)
  CrI_i     <- t(apply(ite_post, 2, quantile, probs = c(0.025, 0.975)))

  ate_post  <- rowMeans(ite_post)
  ATE       <- mean(ate_post)
  CrI_ATE   <- quantile(ate_post, probs = c(0.025, 0.975))

  list(
    iCATE      = iCATE,
```

```
    iCATE_CrI  = CrI_i,
    ATE        = ATE,
    ATE_CrI    = CrI_ATE,
    fit_object = fit,
    ite_post   = ite_post
  )
}
```

```
#### Example usage
library(mxBART)
set.seed(20251229)
start <- Sys.time()
res <- mxbart_HTE(
    df         = dat_bpi_output,
    outcome    = "Y",
    treatment  = "A",
    covariates = paste0("X", 1:9),
    id         = "id",
    ntree      = 200L,
    nskip      = 5000L,
    ndpost     = 5000L,
    keepevery  = 1L
)

end <- Sys.time()
cat("Elapsed time:", as.numeric(end - start), "seconds\n")

print(res$ATE)
print(res$ATE_CrI)

head(
    data.frame(
        iCATE     = res$iCATE,
        lower_CrI = res$iCATE_CrI[,1],
        upper_CrI = res$iCATE_CrI[,2]
    )
)
```

## 3. riBART

Reference paper: https://intlpress.com/site/pub/files/_fulltext/journals/sii/2018/0011/0004/SII-2018-0011-0004-a001.pdf

R package: dbarts

**Function inputs**

- `df`
  A `data.frame` containing CRT data.
  Must include at least:

  - The **outcome** column (named by `outcome`)

- The **treatment** indicator (named by `treatment`, values 0/1)

- All **covariate** columns (named in `covariates`)

- The **cluster ID** column (named by `id`)

- `outcome`
  *(character)*
  Name of the numeric continuous outcome column in `df`.

- `treatment`
  *(character)*
  Name of the binary treatment column in `df` (must be encoded as 0/1).

- `covariates`
  *(character vector)*
  Names of the predictor columns to feed into the nonparametric BART term.

- `id`
  *(character)*
  Name of the cluster/group identifier column in `df`.
  Used to fit a random intercept for each cluster via `group.by`.

- `n.trees`
  *(integer, default 200L)*
  Number of trees in the BART ensemble.

- `n.burn`
  *(integer, default 5000L)*
  Number of "burn-in" MCMC iterations to discard before collecting samples.

- `n.samples`
  *(integer, default 5000L)*
  Number of posterior samples to collect **after** burn-in.

- `n.chains`
  *(integer, default 1L)*
  Number of MCMC chains.

- `keepTrees`
  *(logical, default TRUE)*
  Whether to keep the fitted tree structures in the returned model object.

- `verbose`
  *(logical, default FALSE)*
  Whether to print progress messages during MCMC.

```r
rbart_HTE <- function(df,
                      outcome,
                      treatment,
                      covariates,
                      id,
                      n.burn = 5000L,
                      n.samples = 5000L,
                      n.trees = 200L,
                      n.chains = 1L,
                      keepTrees = TRUE,
```

```r
                    verbose = FALSE) {

    stopifnot(all(c(outcome, treatment, id, covariates) %in% names(df)))

    Y <- df[[outcome]]
    A <- df[[treatment]]
    X <- df[, covariates, drop = FALSE]

    if (!all(na.omit(unique(A)) %in% c(0, 1))) stop("Treatment must be coded as 0/1.")

    dat <- data.frame(Y = Y, A = A, X)

    start_time <- Sys.time()

    fit <- rbart_vi(
        formula   = Y ~ A + .,
        data      = dat,
        group.by  = df[[id]],
        n.burn    = n.burn,
        n.samples = n.samples,
        n.trees   = n.trees,
        n.chains  = n.chains,
        keepTrees = keepTrees,
        verbose   = verbose
    )

    end_time <- Sys.time()

    new1 <- data.frame(A = 1, X)
    new0 <- data.frame(A = 0, X)

    mu1_draws <- predict(fit, newdata = new1, group.by = df[[id]])
    mu0_draws <- predict(fit, newdata = new0, group.by = df[[id]])

    ite_post <- mu1_draws - mu0_draws

    iCATE <- colMeans(ite_post)
    CrI_i <- t(apply(ite_post, 2, quantile, probs = c(0.025, 0.975)))

    ate_draws <- rowMeans(ite_post)
    ATE <- mean(ate_draws)
    CrI_ATE <- as.numeric(quantile(ate_draws, probs = c(0.025, 0.975)))

    list(
        iCATE     = iCATE,
        iCATE_CrI = CrI_i,
        ATE       = ATE,
        ATE_CrI   = CrI_ATE,
        fit_object = fit,
        ite_post  = ite_post
    )
}
```

```
#### Example usage
library(dbarts)
set.seed(20251229)
start <- Sys.time()
res <- rbart_HTE(
    df        = dat_bpi_output,
    outcome   = "Y",
    treatment = "A",
    covariates = paste0("X", 1:9),
    id        = "id"
)

end <- Sys.time()
cat("Elapsed time:", as.numeric(end - start), "seconds\n")

print(res$ATE)
print(res$ATE_CrI)

head(
    data.frame(
        iCATE     = res$iCATE,
        lower_CrI = res$iCATE_CrI[,1],
        upper_CrI = res$iCATE_CrI[,2]
    )
)
```

## 4. MBCF

Reference paper:

https://www.nature.com/articles/s41586-022-04907-7

https://journals.sagepub.com/doi/full/10.1177/09567976211028984

https://link.springer.com/chapter/10.1007/978-3-031-55548-0_25

Source: R package multibart

OSF: https://osf.io/3zmqc/

**Function inputs**

- `df`
  *data.frame* containing **all** variables used by the function.

  - **Required** columns:
    * **Outcome**: continuous numeric column (referenced by `outcome`).

    * **Treatment**: binary 0/1 indicator (referenced by `treatment`).

    * **Covariates**: all predictor columns (referenced by `covariates`).

    * **Cluster/Group ID**: integer or factor column (referenced by `id`).

9

- **id**
  *character*
  Name of the cluster/group ID column in `df`. Used to construct the random-intercept design matrix.

- **treatment**
  *character*
  Name of the binary treatment column in `df` (0/1).

- **outcome**
  *character*
  Name of the continuous outcome column in df.

- **covariates**
  *character vector*
  Names of the predictor columns to include in both the prognostic ($\mu(x)$) and moderation ($\tau(x)$) forests, e.g. `c("X1","X2",…,"X9")`.

- **nburn**
  *integer, default = 5000*
  Number of burn-in MCMC iterations to discard before saving.

- **nsim**
  *integer, default = 5000*
  Number of post-burn-in MCMC draws to retain.

- **nthin**
  *integer, default = 1*
  Thinning interval: save every $n_{\text{thin}}$-th draw (total iterations = `nburn` + `nsim` * `nthin`).

- **ntree_control**
  *integer, default = 200*
  Number of trees in the prognostic forest $\mu(x)$.

- **ntree_moderate**
  *integer, default = 200*
  Number of trees in the moderation forest $\tau(x)$.

- **base_control**
  *numeric, default = 0.95*
  Base hyperparameter $\alpha$ in the tree-splitting prior for $\mu(x)$.

- **power_control**
  *numeric, default = 2*
  Power hyperparameter $\beta$ in the tree-splitting prior for $\mu(x)$.

- **base_moderate**
  *numeric, default = 0.25*
  Base hyperparameter $\alpha$ in the tree-splitting prior for $\tau(x)$.

- **power_moderate**
  *numeric, default = 3*
  Power hyperparameter $\beta$ in the tree-splitting prior for $\tau(x)$.

- **nu**
  *integer, default = 3*
  Degrees of freedom of the chi-square prior on the residual variance $\sigma^2$.

- **sigq**
  *numeric, default = 0.9*
  Calibration quantile for the chisq prior on $\sigma^2$.

- **include_zhat**
  *character, default = **"none"***
  Whether to include an estimated propensity $\hat{z}$ in

  – `"none"` (no inclusion, typical for RCTs),

  – `"control"` (in $\mu(x)$),

  – `"moderate"` (in $\tau(x)$),

  – `"both"`.

```r
MBCF_HTE <- function(
    df,
    id,
    treatment,
    outcome,
    covariates,
    nburn            = 5000,
    nsim             = 5000,
    nthin            = 1,
    ntree_control    = 200,
    ntree_moderate   = 200,
    base_control     = 0.95,
    power_control    = 2,
    base_moderate    = 0.25,
    power_moderate   = 3,
    nu               = 3,
    sigq             = 0.9,
    include_zhat     = "none"
) {

  y <- df[[outcome]]
  z <- df[[treatment]]

  Xc_df <- df %>% select(all_of(covariates))
  is_fac_or_char <- vapply(Xc_df, function(x) is.factor(x) || is.character(x), logical(1))
  if (any(is_fac_or_char)) {
      Xc_df[is_fac_or_char] <- lapply(Xc_df[is_fac_or_char], factor)
  }
  Xm_df <- Xc_df
  Xc    <- dbarts::makeModelMatrixFromDataFrame(Xc_df)
  Xm    <- dbarts::makeModelMatrixFromDataFrame(Xm_df)

  random_intercepts = function(groups, original_colnames = TRUE) {
    dummies = get_dummies(groups, original_colnames = original_colnames)
    random_des = cbind(dummies)
    Q = matrix(1, nrow=ncol(random_des), ncol=1)
    dummies = get_dummies(groups, original_colnames)

    intercept_ix = get_group_indices(dummies)

    return(list(randeff_design = random_des,
                randeff_variance_component_design = as.matrix(Q),
```

```
            group_dummies = dummies,
            intercept_ix = intercept_ix
  ))
}


ri <- random_intercepts(groups = df[[id]])

fit <- suppressWarnings(bcf_continuous_linear(
    y              = y,
    z              = z,
    x_control      = Xc,
    x_moderate     = Xm,
    zhat           = rep(0.5, length(y)), ##for CRT
    # MCMC
    nburn          = nburn,
    nsim           = nsim,
    nthin          = nthin,
    # mu(x) forest prior
    ntree_control  = ntree_control,
    base_control   = base_control,
    power_control  = power_control,
    # tau(x) forest prior
    ntree_moderate = ntree_moderate,
    base_moderate  = base_moderate,
    power_moderate = power_moderate,
    # sigma^2 prior
    nu             = nu,
    sigq           = sigq,
    # random intercepts
    randeff_design                      = ri$randeff_design,
    randeff_variance_component_design   = ri$randeff_variance_component_design,
    randeff_scales                      = 1,
    randeff_df                          = nu,
    include_zhat                        = include_zhat
))


tau_fitted <- fit$moderate_fit
tau_samps  <- get_forest_fit(tau_fitted, Xm)

iCATE      <- colMeans(tau_samps)
iCATE_CrI <- t(apply(tau_samps, 2, quantile, probs = c(0.025, 0.975)))
ATE_samps <- rowMeans(tau_samps)
ATE        <- mean(ATE_samps)
ATE_CrI    <- quantile(ATE_samps, probs = c(0.025, 0.975))

list(
    iCATE     = iCATE,
    iCATE_CrI = iCATE_CrI,
    ATE       = ATE,
    ATE_CrI   = ATE_CrI,
    ite_post  = tau_samps,
    fit = fit
```

```
    )
}

#### Example usage
library(multibart)
library(dplyr)
set.seed(20251229)
start <- Sys.time()
res <- MBCF_HTE(
    df            = dat_bpi_output,
    id            = "id",
    treatment     = "A",
    outcome       = "Y",
    covariates    = paste0("X", 1:9),
    nburn         = 5000,
    nsim          = 5000,
    nthin         = 1,
    ntree_control = 200,
    ntree_moderate = 200
)

end <- Sys.time()
cat("Elapsed time:", as.numeric(end - start), "minutes\n")

cat(sprintf("ATE = %.3f  (95%% CrI: [%.3f, %.3f])\n",
            res$ATE, res$ATE_CrI[1], res$ATE_CrI[2]))

iCATE_mat <- cbind(
    iCATE     = res$iCATE,
    Lower_CrI = res$iCATE_CrI[,1],
    Upper_CrI = res$iCATE_CrI[,2]
)
head(iCATE_mat, 6)
```

## 5. CF

Reference paper:

https://projecteuclid.org/journals/annals-of-statistics/volume-47/issue-2/Generalized-random-forests/10.1214/18-AOS1709.full

https://www.tandfonline.com/doi/full/10.1080/01621459.2017.1319839

https://academic.oup.com/biomet/article/108/2/299/5911092

R package: grf

**Function inputs**

- `df`
  A `data.frame` containing CRT data.
  Must include at least:

  - The **outcome** column (named by `outcome`)

- The **treatment** indicator (named by `treatment`, values 0/1)

- All **covariate** columns (named in `covariates`)

- The **cluster ID** column (named by `id`)

- `outcome`
  *(character)*
  Name of the numeric continuous outcome column in `df`.

- `treatment`
  *(character)*
  Name of the binary treatment column in `df` (must be encoded as 0/1).

- `covariates`
  *(character vector)*
  Names of the predictor columns used as features in the causal forest.

- `id`
  *(character)*
  Name of the cluster/group identifier column in `df`.
  Used to specify clustered sampling/variance estimation via `clusters` in `grf::causal_forest()`.

- `num.trees`
  *(integer, default 2000)*
  Number of trees in the causal forest.

- `W.hat`
  *(numeric, default 0.5)*
  Propensity score used by the forest.
  Use `0.5` when treatment is randomized with equal assignment probability; otherwise provide an estimated propensity score.

- `estimate.variance`
  *(logical, default TRUE)*
  Whether to estimate prediction variance for individual treatment effects.
  If `TRUE`, the function returns 95% Wald confidence intervals for CATEs.

```r
CF_HTE <- function(df,
                   outcome,
                   treatment,
                   covariates,
                   id,
                   num.trees = 2000L,
                   W.hat = 0.5,
                   estimate.variance = TRUE) {

  stopifnot(all(c(outcome, treatment, covariates, id) %in% names(df)))

  Y <- as.numeric(df[[outcome]])
  W <- as.numeric(df[[treatment]])

  if (!all(na.omit(unique(W)) %in% c(0, 1))) stop("Treatment must be coded as 0/1.")

  X <- as.matrix(df[, covariates, drop = FALSE])
  clusters <- as.vector(df[[id]])
```

```r
  fit <- grf::causal_forest(
    X = X,
    Y = Y,
    W = W,
    W.hat = W.hat,
    num.trees = num.trees,
    clusters = clusters
  )

  pred <- stats::predict(fit, estimate.variance = estimate.variance)
  tau_hat <- as.numeric(pred$predictions)

  if (estimate.variance) {
    var_hat <- as.numeric(pred$variance.estimates)
    se_hat  <- sqrt(var_hat)

    iCATE_CI <- cbind(
      lower_95 = tau_hat - 1.96 * se_hat,
      upper_95 = tau_hat + 1.96 * se_hat
    )
  } else {
    iCATE_CI <- NULL
  }

  ate_res <- grf::average_treatment_effect(fit)
  ate_hat <- as.numeric(ate_res[["estimate"]])
  ate_se  <- as.numeric(ate_res[["std.err"]])

  ATE_CI <- c(
    lower_95 = ate_hat - 1.96 * ate_se,
    upper_95 = ate_hat + 1.96 * ate_se
  )

  list(
    iCATE      = tau_hat,
    iCATE_CI   = iCATE_CI,
    ATE        = ate_hat,
    ATE_CI     = ATE_CI,
    fit_object = fit
  )
}


#### Example usage
library(grf)
set.seed(20251229)
res <- CF_HTE(
  df        = dat_bpi_output,
  id        = "id",
  treatment = "A",
  outcome   = "Y",
  covariates = paste0("X", 1:9)
)
```

```
cat(sprintf("ATE = %.3f  (95%% CI: [%.3f, %.3f])\n",
            res$ATE, res$ATE_CI["lower_95"], res$ATE_CI["upper_95"]))

iCATE_mat <- cbind(
  iCATE    = res$iCATE,
  Lower_CI = res$iCATE_CI[, "lower_95"],
  Upper_CI = res$iCATE_CI[, "upper_95"]
)
head(iCATE_mat, 6)
```

**Function inputs**

## 6. MERF

Reference paper: https://journals.sagepub.com/doi/full/10.1177/0962280220946080

R package: LogituRF

**Function inputs**

- `df`
  A `data.frame` containing CRT data.
  Must include at least:

  - The **outcome** column (named by `outcome`, numeric)

  - The **treatment** indicator (named by `treatment`, values 0/1 or coercible to 0/1)

  - All **covariate** columns (named in `covariates`)

  - The **cluster ID** column (named by `id`)

- `id`
  *(character)*
  Name of the cluster/group identifier column in `df`.
  Used for cluster-level resampling in the bootstrap and as the grouping variable in `LongituRF::MERF()`.

- `treatment`
  *(character)*
  Name of the binary treatment column in `df`.

- `outcome`
  *(character)*
  Name of the numeric continuous outcome column in `df`.

- `covariates`
  *(character vector)*
  Names of the predictor columns used as features in the mixed-effect random forest mean function.

- `B`
  *(integer, default 200)*
  Number of **cluster bootstrap** replicates used to obtain uncertainty intervals for CATE and ATE.

- **ntree**
  *(integer, default 500)*
  Number of trees in the random forest component of `LongituRF::MERF()`.

- **mtry**
  *(integer, default 3)*
  Number of variables randomly sampled as candidates at each split. $p/3$ is recommended.

- **sto**
  *(character, default `"none"`)*
  Defines the covariance function of the stochastic process in `LongituRF::MERF()`.
  Can be one of:

  - `"none"`: no stochastic process

  - `"BM"`: Brownian motion

  - `"OrnUhl"`: standard Ornstein–Uhlenbeck process

  - `"BBridge"`: Brownian bridge

  - `"fbm"`: fractional Brownian motion
    It can also be a user-defined covariance function.

```r
LogituRF_HTE <- function(df,
                         id,
                         treatment,
                         outcome,
                         covariates,
                         B     = 200,
                         ntree = 500,
                         mtry  = 3,
                         sto   = "none") {

  if (!requireNamespace("LongituRF", quietly = TRUE)) {
    stop("Package 'LongituRF' is required. Please install it first.")
  }
  MERF <- LongituRF::MERF

  clusters <- unique(df[[id]])
  n_obs    <- nrow(df)

  X_main_global <- as.matrix(df[, covariates, drop = FALSE])
  Z_global      <- matrix(1, n_obs, 1)
  time_global   <- rep(1, n_obs)

  trt_by_cl <- tapply(df[[treatment]], df[[id]], function(x) {
    ux <- unique(x)
    if (length(ux) > 1L) {
      warning("Cluster has multiple treatment values; using the first.")
    }
    ux[1]
  })
  as01 <- function(x) {
    if (is.logical(x)) return(as.integer(x))
```

```r
    if (is.factor(x))  x <- as.character(x)
    if (is.character(x)) return(ifelse(x %in% c("1","TRUE","T","true","t"), 1L, 0L))
    as.integer(x)
}
cluster_trt <- as01(trt_by_cl)
cl_levels   <- names(trt_by_cl)

if (length(unique(cluster_trt[!is.na(cluster_trt)])) < 2L) {
  stop("Original data appears to have only one treatment arm across clusters.")
}

.all_finite_rows <- function(M) {
  apply(M, 1L, function(r) all(is.finite(r)))
}

bootstrap_rep <- function(b) {
  max_tries <- 1000L
  tries <- 0L
  repeat {
    sampled <- sample(clusters, length(clusters), replace = TRUE)
    trt_vec <- cluster_trt[match(sampled, cl_levels)]
    if (length(unique(trt_vec)) < 2L) {
      tries <- tries + 1L
      if (tries >= max_tries) stop("Failed to get both arms after 1000 resamples.")
      next
    }

    boot_df <- do.call(
      rbind,
      lapply(sampled, function(cl) df[df[[id]] == cl, , drop = FALSE])
    )

    keep <- is.finite(boot_df[[outcome]]) &
      .all_finite_rows(as.matrix(boot_df[, covariates, drop = FALSE]))
    if (!all(keep)) boot_df <- boot_df[keep, , drop = FALSE]
    if (nrow(boot_df) == 0L) {
      tries <- tries + 1L
      if (tries >= max_tries) stop("All rows non-finite after 1000 resamples.")
      next
    }

    x_mat <- as.matrix(boot_df[, covariates, drop = FALSE])
    A_vec <- as01(boot_df[[treatment]])
    y_vec <- as.numeric(boot_df[[outcome]])

    if (length(unique(A_vec)) < 2L || isTRUE(all.equal(stats::var(y_vec), 0))) {
      tries <- tries + 1L
      if (tries >= max_tries) stop("Both arms/var(Y) failed after 1000 resamples.")
      next
    }

    Xb_main <- x_mat
    Xb      <- cbind(A = A_vec, Xb_main)
```

```r
    Zb        <- matrix(1, nrow(boot_df), 1)
    time_b    <- rep(1, nrow(boot_df))
    id_b      <- boot_df[[id]]

    mtry_use <- min(mtry, ncol(Xb))

    fit_b <- try(
      MERF(
        X     = Xb,
        Y     = y_vec,
        Z     = Zb,
        id    = id_b,
        time  = time_b,
        ntree = ntree,
        mtry  = mtry_use,
        sto   = sto
      ),
      silent = TRUE
    )
    if (inherits(fit_b, "try-error")) {
      tries <- tries + 1L
      if (tries >= max_tries) stop("MERF failed repeatedly on bootstrap samples.")
      next
    }

    new0 <- cbind(A = 0, X_main_global)
    new1 <- cbind(A = 1, X_main_global)

    dummy_id <- boot_df[[id]][1]
    id_pred  <- rep(dummy_id, n_obs)

    pred0 <- try(predict(fit_b, X = new0, Z = Z_global,
                         id = id_pred, time = time_global), silent = TRUE)
    if (inherits(pred0, "try-error") || any(!is.finite(pred0))) {
      tries <- tries + 1L
      if (tries >= max_tries) stop("Prediction (A=0) failed repeatedly.")
      next
    }

    pred1 <- try(predict(fit_b, X = new1, Z = Z_global,
                         id = id_pred, time = time_global), silent = TRUE)
    if (inherits(pred1, "try-error") || any(!is.finite(pred1))) {
      tries <- tries + 1L
      if (tries >= max_tries) stop("Prediction (A=1) failed repeatedly.")
      next
    }

    return(drop(pred1 - pred0))
  }
}

tau_mat <- matrix(NA_real_, nrow = B, ncol = n_obs)
```

```r
  pb <- utils::txtProgressBar(min = 0, max = B, style = 3)
  on.exit(try(close(pb), silent = TRUE), add = TRUE)

  for (b in seq_len(B)) {
    tau_mat[b, ] <- bootstrap_rep(b)
    utils::setTxtProgressBar(pb, b)
  }

  iCATE_hat <- colMeans(tau_mat)
  iCATE_ci  <- t(apply(tau_mat, 2, stats::quantile, probs = c(0.025, 0.975)))
  colnames(iCATE_ci) <- c("lower", "upper")

  ATE_boot <- rowMeans(tau_mat)
  ATE_hat  <- mean(ATE_boot)
  ATE_ci   <- stats::quantile(ATE_boot, probs = c(0.025, 0.975))

  list(
    iCATE    = iCATE_hat,
    iCATE_CI = iCATE_ci,
    ATE      = ATE_hat,
    ATE_CI   = setNames(as.numeric(ATE_ci), c("lower", "upper")),
    raw_boot = tau_mat
  )
}
```

```r
#### Example usage
library(LongituRF)
set.seed(20251229)
start <- Sys.time()
res <- LogituRF_HTE(
    df              = dat_bpi_output,
    id              = "id",
    treatment       = "A",
    outcome         = "Y",
    covariates      = paste0("X", 1:9),
    B               = 5,
    ntree           = 500,
    mtry            = 3,
    sto             = "none"
)
end <- Sys.time()
cat("Elapsed time:", as.numeric(end - start), "minutes\n")

head(res$iCATE)
head(res$iCATE_CI)
res$ATE
res$ATE_CI
```

## 7. GPBoost

Reference paper:

https://www.jmlr.org/papers/v23/20-322.html

https://ieeexplore.ieee.org/abstract/document/9759834

R package: gpboost

## Function inputs

- `df`
  A `data.frame` containing CRT data.
  Must include at least:

    - The **outcome** column (named by `outcome`, numeric)

    - The **treatment** indicator (named by `treatment`, values 0/1)

    - All **covariate** columns (named in `covariates`)

    - The **cluster ID** column (named by `id`)

- `id`
  *(character)*
  Name of the cluster/group identifier column in `df`.

- `treatment`
  *(character)*
  Name of the binary treatment column in `df` (must be encoded as 0/1).

- `outcome`
  *(character)*
  Name of the numeric continuous outcome column in `df`.

- `covariates`
  *(character vector)*
  Names of the predictor columns used as fixed-effect features in GPBoost.

- `B`
  *(integer, default 200)*
  Number of **cluster bootstrap** replicates used to obtain uncertainty intervals for CATE and ATE.

- `params`
  *(list)*
  A list of LightGBM hyperparameters passed to `gpb.train()`.
  Common entries include:

    - `objective` (default `"regression_l2"`)

    - `learning_rate` (default `0.1`)

    - `max_depth` (default `4`)

    - `min_data_in_leaf` (default `100`)

    - `lambda_l2` (default `1`)

    - `num_threads` (default `1`)

- **nrounds**
  *(integer, default 2000)*
  Number of boosting iterations (trees) used in `gpb.train()`.

```
GPBoost_HTE <- function(df, id, treatment, outcome, covariates,
                        B = 200,
                        params = list(objective = "regression_l2",
                                      learning_rate = 0.1,
                                      max_depth = 4,
                                      min_data_in_leaf = 100,
                                      lambda_l2 = 1,
                                      num_threads = 1),
                        nrounds = 2000) {

clusters <- unique(df[[id]])
n <- nrow(df)

gid_o_num <- as.integer(factor(df[[id]], levels = unique(df[[id]])))

if (is.null(params$force_col_wise)) params$force_col_wise <- TRUE

tau_boot <- matrix(NA_real_, nrow = B, ncol = n)

bootstrap_rep <- function(b) {
  samp_cl <- sample(clusters, size = length(clusters), replace = TRUE)
  boot_list <- lapply(seq_along(samp_cl), function(i) {
    cl  <- samp_cl[i]
    tmp <- df[df[[id]] == cl, , drop = FALSE]
    tmp$boot_gid_chr <- paste0(tmp[[id]], "__", i)
    tmp
  })
  boot_df <- do.call(rbind, boot_list)

  boot_df$boot_gid <- as.integer(factor(boot_df$boot_gid_chr,
                                 levels = unique(boot_df$boot_gid_chr)))
  boot_df$boot_gid_chr <- NULL

  Xb <- boot_df[, covariates, drop = FALSE]
  Ab <- boot_df[[treatment]]
  if (!is.numeric(Ab)) Ab <- as.numeric(Ab)
  Db <- cbind(Xb, A = Ab)

  stopifnot(all(Ab %in% c(0, 1)))
  stopifnot(!anyNA(Db))
  stopifnot(all(is.finite(as.matrix(Db))))
  stopifnot(all(is.finite(boot_df[[outcome]])))

  dtrain <- gpb.Dataset(data = as.matrix(Db), label = boot_df[[outcome]])
  gp_mod <- GPModel(group_data = boot_df$boot_gid, likelihood = "gaussian")

  bst <- gpb.train(data = dtrain, gp_model = gp_mod,
                   params = params, nrounds = nrounds, verbose = 0)

  Xo <- df[, covariates, drop = FALSE]
```

```r
    D0 <- cbind(Xo, A = 0)
    D1 <- cbind(Xo, A = 1)

    p0 <- predict(bst, data = as.matrix(D0),
                  group_data_pred = gid_o_num, pred_latent = TRUE)
    p1 <- predict(bst, data = as.matrix(D1),
                  group_data_pred = gid_o_num, pred_latent = TRUE)

    ic <- as.numeric(p1$fixed_effect - p0$fixed_effect)

    rm(dtrain, gp_mod, bst); gc(FALSE)

    ic
  }

  for (b in seq_len(B)) {
    tau_boot[b, ] <- bootstrap_rep(b)
  }

  iCATE_hat <- colMeans(tau_boot)
  iCATE_lb  <- apply(tau_boot, 2, quantile, probs = 0.025)
  iCATE_ub  <- apply(tau_boot, 2, quantile, probs = 0.975)

  ATE_boot  <- rowMeans(tau_boot)
  ATE_hat   <- mean(ATE_boot)
  ATE_lb    <- quantile(ATE_boot, 0.025)
  ATE_ub    <- quantile(ATE_boot, 0.975)

  list(
    iCATE    = iCATE_hat,
    iCATE_CI = cbind(lower = iCATE_lb, upper = iCATE_ub),
    ATE      = ATE_hat,
    ATE_CI   = c(lower = ATE_lb, upper = ATE_ub),
    tau_boot = tau_boot,
    ATE_boot = ATE_boot
  )
}

#### Example usage
library(gpboost)
start <- Sys.time()
set.seed(20251229)
result <- GPBoost_HTE(
  df         = dat_bpi_output,
  id         = "id",
  treatment  = "A",
  outcome    = "Y",
  covariates = paste0("X", 1:9),
  B          = 200,
  nrounds    = 2000
)
end <- Sys.time()
cat("Elapsed time:", as.numeric(end - start), "minutes\n")
```

```
cat(sprintf(
  "ATE = %.3f  (95%% CrI: [%.3f, %.3f])\n",
  result$ATE,
  result$ATE_CI[1],
  result$ATE_CI[2]
))

iCATE_mat <- cbind(
  iCATE     = result$iCATE,
  Lower_CrI = result$iCATE_CI[, 1],
  Upper_CrI = result$iCATE_CI[, 2]
)
head(iCATE_mat, 6)
```

## 8. MEGB

Reference paper: https://www.nature.com/articles/s41598-025-16526-z

R package: MEGB

### Function inputs

- `df`
  A `data.frame` containing CRT data.
  Must include at least:

  - The **outcome** column (named by `outcome`, numeric)

  - The **treatment** indicator (named by `treatment`, values 0/1 or coercible to 0/1)

  - All **covariate** columns (named in `covariates`)

  - The **cluster ID** column (named by `id`)

- `id`
  *(character)*
  Name of the cluster/group identifier column in `df`.
  Used as the trajectory identifier (`id`) in `MEGB::MEGB()` and for **cluster-level resampling** in the bootstrap.

- `treatment`
  *(character)*
  Name of the binary treatment column in `df`.

- `outcome`
  *(character)*
  Name of the numeric continuous outcome column in `df`.

- `covariates`
  *(character vector)*
  Names of the predictor columns used as fixed-effect features in MEGB.

- `B`
  *(integer, default 200)*
  Number of **cluster bootstrap** replicates used to construct uncertainty intervals for CATE and ATE.

- **ntree**
  *(integer, default 500)*
  Number of trees to grow in the boosting expansion in MEGB.

- **shrinkage**
  *(numeric, default 0.05)*
  Learning rate / step-size reduction applied to each tree in the boosting expansion.

- **interaction.depth**
  *(integer, default 2)*
  Maximum depth of variable interactions in the boosting model.

  - 1 fits an additive model

  - 2 allows up to two-way interactions, etc.

- **n.minobsinnode**
  *(integer, default 5)*
  Minimum number of observations allowed in a terminal node of each tree.

```r
MEGB_HTE <- function(df,
                     id,
                     treatment,
                     outcome,
                     covariates,
                     B                 = 200,
                     ntree             = 500,
                     shrinkage         = 0.05,
                     interaction.depth = 2,
                     n.minobsinnode    = 5) {
  if (!requireNamespace("MEGB", quietly = TRUE)) {
    stop("Package 'MEGB' is required. Please install it first.")
  }
  MEGB_fit <- MEGB::MEGB

  clusters <- unique(df[[id]])
  n_obs    <- nrow(df)

  X_main <- as.matrix(df[, covariates, drop = FALSE])
  time   <- rep(1, n_obs)

  as01 <- function(x) {
    if (is.logical(x)) return(as.integer(x))
    if (is.factor(x))  x <- as.character(x)
    if (is.character(x)) return(ifelse(x %in% c("1","TRUE","T","true","t"), 1L, 0L))
    as.integer(x)
  }

  trt_by_cl <- tapply(df[[treatment]], df[[id]], function(x) {
    ux <- unique(x)
    if (length(ux) > 1L) {
      warning("Cluster has multiple treatment values; using the first.")
    }
    ux[1]
  })
```

```
cluster_trt <- as01(trt_by_cl)
cl_levels  <- names(trt_by_cl)
if (length(unique(cluster_trt[!is.na(cluster_trt)])) < 2L) {
  stop("Original data appears to have only one treatment arm across clusters.")
}

bootstrap_rep <- function(b) {
  max_tries <- 1000L
  tries <- 0L
  repeat {
    sampled <- sample(clusters, length(clusters), replace = TRUE)
    trt_vec <- cluster_trt[match(sampled, cl_levels)]
    if (length(unique(trt_vec)) < 2L) {
      tries <- tries + 1L
      if (tries >= max_tries) stop("Failed to get both arms after 1000 resamples.")
      next
    }

    boot_df <- do.call(
      rbind,
      lapply(sampled, function(cl) df[df[[id]] == cl, , drop = FALSE])
    )

    Xb_main <- as.matrix(boot_df[, covariates, drop = FALSE])
    Ab      <- as01(boot_df[[treatment]])
    id_b    <- boot_df[[id]]
    Zb      <- matrix(1, nrow(boot_df), 1)
    time_b  <- rep(1, nrow(boot_df))
    Xb_simple <- cbind(A = Ab, Xb_main)

    fit_b <- try(
      MEGB_fit(
        X                = Xb_simple,
        Y                = boot_df[[outcome]],
        id               = id_b,
        Z                = Zb,
        time             = time_b,
        ntree            = ntree,
        shrinkage        = shrinkage,
        interaction.depth = interaction.depth,
        n.minobsinnode   = n.minobsinnode
      ),
      silent = TRUE
    )
    if (inherits(fit_b, "try-error")) {
      tries <- tries + 1L
      if (tries >= max_tries) stop("MEGB failed repeatedly on bootstrap samples.")
      next
    }

    new0 <- cbind(A = 0, X_main)
    new1 <- cbind(A = 1, X_main)
```

```
    Z0         <- matrix(1, n_obs, 1)
    dummy_id <- sampled[1]
    id_pred   <- rep(dummy_id, n_obs)

    pred0 <- try(predict(fit_b, X = new0, Z = Z0, id = id_pred, time = time), silent = TRUE)
    if (inherits(pred0, "try-error") || any(!is.finite(pred0))) {
      tries <- tries + 1L
      if (tries >= max_tries) stop("Prediction (A=0) failed repeatedly.")
      next
    }

    pred1 <- try(predict(fit_b, X = new1, Z = Z0, id = id_pred, time = time), silent = TRUE)
    if (inherits(pred1, "try-error") || any(!is.finite(pred1))) {
      tries <- tries + 1L
      if (tries >= max_tries) stop("Prediction (A=1) failed repeatedly.")
      next
    }

    return(drop(pred1 - pred0))
  }
}

tau_mat <- matrix(NA_real_, nrow = B, ncol = n_obs)
pb <- utils::txtProgressBar(min = 0, max = B, style = 3)
on.exit(try(close(pb), silent = TRUE), add = TRUE)

for (b in seq_len(B)) {
  tau_mat[b, ] <- bootstrap_rep(b)
  utils::setTxtProgressBar(pb, b)
}

iCATE_hat <- colMeans(tau_mat)
iCATE_ci   <- t(apply(tau_mat, 2, stats::quantile, probs = c(0.025, 0.975)))
colnames(iCATE_ci) <- c("lower", "upper")

ATE_boot <- rowMeans(tau_mat)
ATE_hat  <- mean(ATE_boot)
ATE_ci    <- stats::quantile(ATE_boot, probs = c(0.025, 0.975))

list(
  iCATE     = iCATE_hat,
  iCATE_CI = iCATE_ci,
  ATE       = ATE_hat,
  ATE_CI    = setNames(as.numeric(ATE_ci), c("lower", "upper")),
  raw_boot = tau_mat
)
}


#### Example usage
library(MEGB)
set.seed(20251229)
start <- Sys.time()
res <- MEGB_HTE(
```

```
  df                  = dat_bpi_output,
  id                  = "id",
  treatment           = "A",
  outcome             = "Y",
  covariates          = paste0("X", 1:9),
  B                   = 200,
  ntree               = 500,
  shrinkage           = 0.05,
  interaction.depth   = 2,
  n.minobsinnode      = 100
)

end <- Sys.time()
cat("Elapsed time:", as.numeric(end - start), "minutes\n")

head(res$iCATE)
head(res$iCATE_CI)
res$ATE
res$ATE_CI
```

## 9. mermboost

Reference paper: https://link.springer.com/article/10.1007/s11222-025-10612-y

package: mermboost

### Function inputs

- **data**
  A **data.frame** containing CRT data.
  Must include at least:

  - The **outcome** column (named by outcome, numeric)

  - The **treatment** indicator (named by treatment, values 0/1 or coercible to 0/1)

  - All **covariate** columns (named in covariates)

  - The **cluster ID** column (named by id)

- **outcome**
  *(character)*
  Name of the numeric continuous outcome column in **data**.

- **treatment**
  *(character)*
  Name of the binary treatment column in **data**.

- **id**
  *(character)*
  Name of the cluster/group identifier column in **data**.
  Converted to a factor and used as the random-intercept grouping term (1 | id) in the mixed-effects boosting model.
  Also used for **cluster-level bootstrap resampling**.

- **covariates**
  *(character vector)*
  Names of candidate baseline covariates.
  The function builds main-effect terms and treatment interaction terms using `bols()` / `bbs()` depending on the number of unique values and whether the covariate varies within both treatment arms.

- **mstop_max**
  *(integer, default 2000)*
  Maximum number of boosting iterations considered for the initial model fit.
  The function fits an initial `mermboost` model with `mstop = mstop_max`, then selects an optimal stopping iteration `best_m` via cross-validation.

- **folds**
  *(integer, default 5)*
  Number of cross-validation folds used in `mer_cvrisk()` to select the optimal boosting stopping iteration `best_m`.

- **B**
  *(integer, default 200)*
  Number of **cluster bootstrap** replicates used to obtain uncertainty intervals for CATE and ATE.

- **mc.cores**
  *(integer, default `detectCores() - 1`)*
  Number of CPU cores used for the bootstrap.

```r
mermboost_HTE <- function(data,
                          outcome,
                          treatment,
                          id,
                          covariates,
                          mstop_max = 2000,
                          folds     = 5,
                          B         = 200,
                          mc.cores  = detectCores() - 1) {

  df <- data

  MIN_UNIQUES_FOR_SPLINE <- 10L

  .to_num <- function(x) {
    if (is.logical(x)) return(as.integer(x))
    if (is.factor(x))  x <- as.character(x)
    suppressWarnings(as.numeric(x))
  }
  as01 <- function(x) {
    if (is.logical(x)) return(as.integer(x))
    if (is.factor(x))  x <- as.character(x)
    if (is.character(x)) return(ifelse(x %in% c("1","TRUE","T","true","t"), 1L, 0L))
    as.integer(x)
  }

  covariates <- intersect(covariates, colnames(df))
  covariates <- setdiff(covariates, c(id, treatment))
  if (!length(covariates)) stop("No valid covariates after filtering id/treatment.")
```

```r
A01 <- as01(df[[treatment]])
is_dup_of_A <- function(v) {
  xv <- .to_num(df[[v]]); ok <- is.finite(xv) & is.finite(A01)
  u <- unique(xv[ok]); if (length(u) != 2L) return(FALSE)
  mapped <- ifelse(xv == max(u, na.rm=TRUE), 1L, 0L)
  all(mapped[ok] == A01[ok])
}
if (length(covariates)) {
  dup_mask <- vapply(covariates, is_dup_of_A, logical(1))
  if (any(dup_mask)) covariates <- covariates[!dup_mask]
}
if (!length(covariates)) stop("No valid covariates after filtering equivalence to treatment.")

cont   <- covariates[sapply(covariates,
                            function(v) is.numeric(df[[v]]) && length(unique(df[[v]])) > 2)]
binary <- covariates[sapply(covariates,
                            function(v) length(unique(df[[v]])) == 2)]
others <- covariates[sapply(covariates,
                            function(v) is.factor(df[[v]]) && length(unique(df[[v]])) > 2)]

#if (length(cont) > 0) df[cont] <- scale(df[cont], center = TRUE, scale = FALSE)
df[[id]]   <- factor(df[[id]])
df$A_num   <- A01

has_var_both <- function(v, dat=df) {
  xv <- .to_num(dat[[v]])
  s0 <- stats::sd(xv[dat$A_num == 0], na.rm = TRUE)
  s1 <- stats::sd(xv[dat$A_num == 1], na.rm = TRUE)
  is.finite(s0) && is.finite(s1) && s0 > 0 && s1 > 0
}
ok_by_cont <- if (length(cont)) cont[ sapply(cont,
                                             function(v) has_var_both(v, df)) ]
              else character(0)
ok_by_binary <- if (length(binary)) binary[sapply(binary,
                                             function(v) has_var_both(v, df)) ]
              else character(0)
ok_by_others <- if (length(others)) {
  keep <- vapply(others, function(v) {
    tb0 <- length(unique(df[df$A_num==0, v, drop=TRUE]))
    tb1 <- length(unique(df[df$A_num==1, v, drop=TRUE]))
    tb0 >= 2 && tb1 >= 2
  }, logical(1))
  others[keep]
} else character(0)

nuniq <- setNames(sapply(cont, function(v) length(unique(df[[v]]))), cont)
mk_bbs_main <- function(v) {
  u <- nuniq[[v]]
  if (is.na(u) || u < MIN_UNIQUES_FOR_SPLINE) {
    sprintf("bols(%s, intercept=FALSE, lambda=1e-4)", v)
  } else {
    df_use <- min(5L, u - 1L)
    k_use  <- min(8L, max(6L, u - 2L))
```

```r
      sprintf("bbs(%s)", v)
  }
}
mk_bbs_by <- function(v) {
  u <- nuniq[[v]]
  if (is.na(u) || u < MIN_UNIQUES_FOR_SPLINE) {
    sprintf("bols(%s, by=A_num, lambda=1e-4)", v)
  } else {
    df_use <- min(5L, u - 1L)
    k_use  <- min(8L, max(6L, u - 2L))
    sprintf("bbs(%s, by=A_num)", v)
  }
}
main <- c(
  "bols(A_num, intercept=FALSE, lambda=1e-4)",
  if (length(cont))   sapply(cont, mk_bbs_main),
  if (length(binary)) sapply(binary,
                             function(v) sprintf("bols(%s, intercept=FALSE, lambda=1e-4)", v)),
  if (length(others)) sapply(others,
                             function(v) sprintf("bols(%s, intercept=FALSE, lambda=1e-4)", v))
)
inter <- c(
  if (length(ok_by_cont)) sapply(ok_by_cont, mk_bbs_by),
  if (length(ok_by_binary)) sapply(ok_by_binary,
                             function(v) sprintf("bols(%s, by=A_num, lambda=1e-4)", v)),
  if (length(ok_by_others)) sapply(ok_by_others,
                             function(v) sprintf("bols(%s, by=A_num, lambda=1e-4)", v))
)
rand <- sprintf("(1 | %s)", id)
fmla <- as.formula(paste0(outcome, " ~ ", paste(c(main, inter, rand), collapse = " + ")))

best_m <- tryCatch({
  fit0 <- suppressWarnings(suppressMessages(
    mermboost(fmla, data = df, family = gaussian(),
              control = boost_control(mstop = mstop_max))
  ))
  cv <- suppressWarnings(suppressMessages(mer_cvrisk(fit0, no_of_folds = folds)))
  mstop(cv)
}, error = function(e) {
  warning("initial CV choose mstop fail, use mstop_max = ", mstop_max,
          " reason ", e$message)
  mstop_max
})

## Bootstrap
new0 <- df; new0$A_num <- 0
new1 <- df; new1$A_num <- 1
clusters <- unique(df[[id]])

one_boot <- function(i) {
  tryCatch({
    suppressWarnings(suppressMessages({
      tries <- 0L
```

```r
repeat {
  tries <- tries + 1L
  samp_cl <- sample(clusters, length(clusters), replace = TRUE)
  df_b <- do.call(rbind, lapply(seq_along(samp_cl), function(k) {
    tmp <- df[df[[id]] == samp_cl[k], , drop = FALSE]
    tmp[[id]] <- factor(sprintf("%s__boot%04d", as.character(tmp[[id]]), k))
    tmp
  }))
  df_b[[id]] <- droplevels(df_b[[id]])

  keep <- is.finite(.to_num(df_b[[outcome]]))
  num_cols <- names(which(vapply(df_b[, covariates, drop = FALSE], is.numeric, logical(1))))
  if (length(num_cols)) {
    keep <- keep & apply(is.finite(as.matrix(df_b[, num_cols, drop = FALSE])), 1L, all)
  }
  if (!all(keep)) df_b <- df_b[keep, , drop = FALSE]
  if (nrow(df_b) == 0L) { if (tries >= 1000L)
    return(list(iCATE=rep(NA_real_, nrow(df)), ATE=NA_real_)); next }

  if (length(unique(df_b$A_num)) == 2L &&
      !isTRUE(all.equal(stats::var(.to_num(df_b[[outcome]])), 0))) break
  if (tries >= 1000L) return(list(iCATE=rep(NA_real_, nrow(df)), ATE=NA_real_))
}

has_var_both_b <- function(v) has_var_both(v, dat = df_b)
ok_by_cont_b <- if (length(cont)) cont[ sapply(cont, has_var_both_b) ] else character(0)
ok_by_binary_b <- if (length(binary)) binary[sapply(binary, has_var_both_b) ] else character(0)
ok_by_others_b <- if (length(others)) {
  keep2 <- vapply(others, function(v) {
    tb0 <- length(unique(df_b[df_b$A_num==0, v, drop=TRUE]))
    tb1 <- length(unique(df_b[df_b$A_num==1, v, drop=TRUE]))
    tb0 >= 2 && tb1 >= 2
  }, logical(1))
  others[keep2]
} else character(0)

nuniq_b <- setNames(sapply(cont, function(v) length(unique(df_b[[v]]))), cont)
mk_bbs_main_b <- function(v) {
  u <- nuniq_b[[v]]
  if (is.na(u) || u < MIN_UNIQUES_FOR_SPLINE) {
    sprintf("bols(%s, intercept=FALSE, lambda=1e-4)", v)
  } else {
    df_use <- min(5L, u - 1L)
    k_use  <- min(8L, max(6L, u - 2L))
    sprintf("bbs(%s)", v)
  }
}
mk_bbs_by_b <- function(v) {
  u <- nuniq_b[[v]]
  if (is.na(u) || u < MIN_UNIQUES_FOR_SPLINE) {
    sprintf("bols(%s, by=A_num, lambda=1e-4)", v)
  } else {
    df_use <- min(5L, u - 1L)
  }
}
```

```
          k_use  <- min(8L, max(6L, u - 2L))
          sprintf("bbs(%s, by=A_num", v)
        }
      }

      main_b <- c(
        "bols(A_num, intercept=FALSE, lambda=1e-4)",
        if (length(cont))   sapply(cont,   mk_bbs_main_b),
        if (length(binary)) sapply(binary,
                                   function(v)
                                   sprintf("bols(%s, intercept=FALSE, lambda=1e-4)", v)),
        if (length(others)) sapply(others,
                                   function(v)
                                   sprintf("bols(%s, intercept=FALSE, lambda=1e-4)", v))
      )
      inter_b <- c(
        if (length(ok_by_cont_b))   sapply(ok_by_cont_b, mk_bbs_by_b),
        if (length(ok_by_binary_b)) sapply(ok_by_binary_b,
                                           function(v)
                                           sprintf("bols(%s, by=A_num, lambda=1e-4)", v)),
        if (length(ok_by_others_b)) sapply(ok_by_others_b,
                                           function(v)
                                           sprintf("bols(%s, by=A_num, lambda=1e-4)", v))
      )

      fmla_b <- as.formula(paste0(outcome, " ~ ", paste(c(main_b, inter_b, rand), collapse = " + ")))

      fit_b <- mermboost(
        fmla_b, data = df_b, family = gaussian(),
        control = boost_control(mstop = best_m)
      )

      mu0_b <- predict(fit_b, newdata = new0, type = "response", RE = FALSE)
      mu1_b <- predict(fit_b, newdata = new1, type = "response", RE = FALSE)
      iC_b  <- mu1_b - mu0_b
      list(iCATE = iC_b, ATE = mean(iC_b))
    }))
  }, error = function(e) {
    warning("bootstrap ", i, " fail: ", e$message)
    list(iCATE = rep(NA_real_, nrow(df)), ATE = NA_real_)
  })
}


if (mc.cores <= 1L) {
  boot_list <- lapply(seq_len(B), one_boot)
} else if (.Platform$OS.type == "unix") {
  boot_list <- parallel::mclapply(seq_len(B), one_boot, mc.cores = mc.cores, mc.set.seed = TRUE)
} else {
  cl <- parallel::makeCluster(mc.cores)
  parallel::clusterEvalQ(cl, library(mermboost))
  parallel::clusterExport(
    cl,
    c("df","fmla","new0","new1","id","treatment","best_m",
```

```
        "covariates","clusters","outcome","one_boot",".to_num","as01",
        "MIN_UNIQUES_FOR_SPLINE","cont","binary","others"),
     envir = environment()
  )
  boot_list <- parallel::parLapply(cl, seq_len(B), one_boot)
  parallel::stopCluster(cl)
}

ite_mat <- do.call(cbind, lapply(boot_list, `[[`, "iCATE"))
ate_vec <- sapply(boot_list, `[[`, "ATE")

iCATE_point <- rowMeans(ite_mat, na.rm = TRUE)
iCATE_CI    <- t(apply(ite_mat, 1, quantile, probs = c(0.025, 0.975), na.rm = TRUE))
ATE_point   <- mean(ate_vec, na.rm = TRUE)
ATE_CI      <- quantile(ate_vec, c(0.025, 0.975), na.rm = TRUE)

list(
   best_m    = best_m,
   ATE       = ATE_point,
   ATE_CI    = ATE_CI,
   iCATE     = iCATE_point,
   iCATE_CI  = iCATE_CI,
   ate_vec   = ate_vec,
   ite_mat   = ite_mat
)
}
```

```
#### Example usage
library(mermboost)
res <- mermboost_HTE(
   data       = dat_bpi_output,
   outcome    = "Y",
   treatment  = "A",
   id         = "id",
   covariates = paste0("X", 1:9),
   mstop_max  = 5000,
   folds      = 5,
   B          = 5,
   mc.cores   = detectCores() - 1
)

res$ATE
res$ATE_CI
head(res$iCATE)
head(res$iCATE_CI)
```

## 10. GAMM

Reference paper: https://academic.oup.com/jrsssb/article/73/1/3/7034726

R package: mgcv

**Function inputs**

- `df`
  A `data.frame` containing CRT data.
  Must include at least:

  - The **outcome** column (named by `outcome`)

  - The **treatment** indicator (named by `treatment`, values 0/1)

  - All **covariate** columns (named in `covariates`)

  - The **cluster ID** column (named by `cluster`)

- `outcome`
  *(character)*
  Name of the outcome column in `df`.

- `treatment`
  *(character)*
  Name of the binary treatment column in `df` (must be encoded as 0/1).

- `covariates`
  *(character vector)*
  Names of baseline covariates to include in the model.
  The function automatically:

  - keeps only valid column names in `df`
  - classifies covariates into continuous / binary (0/1) / multi-level factors
  - chooses smooth vs linear terms for continuous covariates depending on support within both treatment arms

- `cluster`
  *(character)*
  Name of the cluster/group identifier column in `df`.

- `family`
  *(family object, default `gaussian()`)*
  Outcome distribution and link function used for model fitting (e.g., `gaussian()`, `binomial()`).

- `nsim`
  *(integer, default 2000)*
  Number of coefficient draws used to simulate uncertainty intervals.

- `use_bam`
  *(logical, default FALSE)*
  If `TRUE`, fits the model using `mgcv::bam()` with a penalized random effect term `s(cluster, bs="re")`.
  If `FALSE`, attempts `mgcv::gamm()` first; if `gamm()` fails, falls back to `mgcv::gam()` with `s(cluster, bs="re")`.

- `select`
  *(logical, default TRUE)*
  Passed to `mgcv` to enable additional smoothness/term selection penalization.

- `min_unique`
  *(integer, default 5)*
  Minimum number of unique values **within each treatment arm** required to allow a continuous covariate to be modeled with smooth terms.
  If the support is insufficient, the covariate is included as a linear main effect plus a linear treatment interaction instead.

```r
GAMM_HTE <- function(df, outcome, treatment, covariates, cluster,
                     family    = gaussian(),
                     nsim      = 2000,
                     use_bam   = FALSE,
                     select    = TRUE,
                     min_unique = 5) {

  stopifnot(requireNamespace("mgcv", quietly = TRUE))
  stopifnot(requireNamespace("MASS", quietly = TRUE))

  df[[cluster]] <- factor(df[[cluster]])
  df[[cluster]] <- droplevels(df[[cluster]])
  if (!all(df[[treatment]] %in% c(0, 1))) stop("`treatment` must be coded 0/1.")
  df$A_num <- as.integer(df[[treatment]])

  covariates <- unique(as.character(covariates))
  covariates <- intersect(covariates, colnames(df))
  if (!length(covariates)) stop("No valid covariates.")

  is_binary01 <- function(x) {
    if (is.logical(x)) return(TRUE)
    u <- unique(x[!is.na(x)])
    if (is.factor(u) || is.character(u)) u <- suppressWarnings(as.numeric(as.character(u)))
    is.numeric(u) && length(u) <= 2 && all(u %in% c(0, 1))
  }

  allow_smooth <- function(x, A, min_unique = 5) {
    if (!is.numeric(x)) return(FALSE)
    ok <- is.finite(x)
    x0 <- unique(x[ ok & A == 0 ])
    x1 <- unique(x[ ok & A == 1 ])
    nu <- length(unique(x[ok]))
    length(x0) >= min_unique && length(x1) >= min_unique && nu >= 4
  }


  A <- df[[treatment]]

  cov_bin  <- covariates[vapply(covariates, function(v) is_binary01(df[[v]]), logical(1))]
  cov_cont <- covariates[vapply(
    covariates,
    function(v) is.numeric(df[[v]]) && !is_binary01(df[[v]]),
    logical(1)
  )]
  used     <- union(cov_bin, cov_cont)
  cov_fac  <- setdiff(covariates, used)
  if (length(cov_fac)) {
    cov_fac <- cov_fac[vapply(cov_fac, function(v) is.factor(df[[v]]), logical(1))]
    if (length(cov_fac)) {
      df[cov_fac] <- lapply(df[cov_fac], droplevels)
      cov_fac <- cov_fac[vapply(cov_fac, function(v) nlevels(df[[v]]) >= 2, logical(1))]
    }
  }
```

```r
## build formula: A + X + A_num:X
terms <- c("A_num")

if (length(cov_cont)) {
  for (m in cov_cont) {
    if (allow_smooth(df[[m]], A, min_unique)) {
      terms <- c(terms,
                 sprintf("s(%s, bs='cr')", m),
                 sprintf("s(%s, by=A_num, bs='cr')", m))
    } else {
      terms <- c(terms, m, sprintf("A_num:%s", m))
    }
  }
}

if (length(cov_bin)) {
  terms <- c(terms, cov_bin, paste0("A_num:", cov_bin))
}

if (length(cov_fac)) {
  terms <- c(terms, cov_fac, paste0("A_num:", cov_fac))
}

if (!length(terms)) stop("No terms left for the model.")
fmla <- as.formula(sprintf("%s - %s", outcome, paste(terms, collapse = " + ")))


fit_source <- "gamm"
fit_lme <- NULL
if (use_bam) {
  re_term <- sprintf("s(%s, bs='re')", cluster)
  fmla_bam <- as.formula(paste(paste(deparse(fmla), collapse = " "), "+", re_term))
  fit_gam <- mgcv::bam(fmla_bam, data = df, family = family,
                       method = "fREML", discrete = TRUE, select = select)
  fmla_final <- fmla_bam
  fit_source <- "bam"
} else {
  ctrl <- nlme::lmeControl(
    apVar       = FALSE,
    returnObject = TRUE,
    msMaxIter   = 200,
    msMaxEval   = 400,
    niterEM     = 0,
    tolerance   = 1e-6,
    msVerbose   = FALSE
  )
  gamm_fit_try <- try(
    mgcv::gamm(fmla, data = df, family = family,
               random = stats::setNames(list(~1), cluster),
               method = "REML", control = ctrl),
    silent = TRUE
  )
  if (inherits(gamm_fit_try, "try-error")) {
    re_term   <- sprintf("s(%s, bs='re')", cluster)
```

```
      fmla_gam  <- as.formula(paste(paste(deparse(fmla), collapse = " "), "+", re_term))
      fit_gam   <- mgcv::gam(fmla_gam, data = df, family = family,
                             method = "REML", select = select)
      fit_lme   <- NULL
      fit_source<- "gam_fallback"
      fmla_final<- fmla_gam
    } else {
      fit_gam   <- gamm_fit_try$gam
      fit_lme   <- gamm_fit_try$lme
      fmla_final<- fmla
      fit_source<- "gamm"
    }
  }

  df0 <- df; df0[[treatment]] <- 0; df0$A_num <- 0L
  df1 <- df; df1[[treatment]] <- 1; df1$A_num <- 1L

  mu0_hat <- as.numeric(predict(fit_gam, newdata = df0, type = "response"))
  mu1_hat <- as.numeric(predict(fit_gam, newdata = df1, type = "response"))
  iCATE   <- mu1_hat - mu0_hat
  ATE     <- mean(iCATE, na.rm = TRUE)

  X0 <- predict(fit_gam, newdata = df0, type = "lpmatrix")
  X1 <- predict(fit_gam, newdata = df1, type = "lpmatrix")
  beta_hat  <- stats::coef(fit_gam)
  V         <- mgcv::vcov.gam(fit_gam, unconditional = TRUE)
  beta_sims <- MASS::mvrnorm(n = nsim, mu = beta_hat, Sigma = V)

  eta0 <- X0 %*% t(beta_sims)
  eta1 <- X1 %*% t(beta_sims)
  linkinv <- family$linkinv; if (is.null(linkinv)) linkinv <- function(z) z
  mu0_sim <- linkinv(eta0)
  mu1_sim <- linkinv(eta1)

  iCATE_sim_mat <- mu1_sim - mu0_sim
  ATE_sim <- colMeans(iCATE_sim_mat, na.rm = TRUE)

  iCATE_CI <- t(apply(iCATE_sim_mat, 1, stats::quantile,
                      probs = c(0.025, 0.975), na.rm = TRUE))
  ATE_CI   <- as.numeric(stats::quantile(ATE_sim, probs = c(0.025, 0.975), na.rm = TRUE))

  list(
    iCATE    = iCATE,
    iCATE_CI = iCATE_CI,
    ATE      = ATE,
    ATE_CI   = ATE_CI,
    fit_gam  = fit_gam,
    fit_lme  = fit_lme,
    pieces   = list(
      cov_cont  = cov_cont,
      cov_bin   = cov_bin,
      cov_fac   = cov_fac,
      formula   = fmla_final,
```

```
        fit_source= fit_source
    )
  )
}
```

```
#### Example usage
library(mgcv)
res <- GAMM_HTE(
    df        = dat_bpi_output,
    outcome   = "Y",
    treatment = "A",
    cluster   = "id",
    covariates = paste0("X", 1:9),
    min_unique = 6
)

res$ATE
res$ATE_CI
head(res$iCATE)
head(res$iCATE_CI)
```



\end{document}